%% file: main.tex
\pdfoutput=1
\documentclass[a4paper]{jpconf}

\input{zzzzz_head_packs_hyphs_iop}

\newcounter{pageatref}			
\newcounter{mastereqcounterfigtab}		
\newcounter{mastereqcountertheo}		
\numberwithin{equation}{section} 		
\numberwithin{figure}{section}  		
\numberwithin{table}{section}  		
					
\numberwithin{theoremcounter}{section}	

\theoremstyle{plain}						
\theorembodyfont{\rmfamily}					
\input{__new_commands_general}

\input{__new_commands_special}

\begin{document}				
\input{zzzzz_title_iop}

%
\vspace{-6mm}
\section{Introduction}
The General Boundary Formulation (GBF) is a new framework for studying quantum theories
\cite{oeckl:_gbf_found_prob}.
The respective theory is considered in some spacetime region,
and the quantum states live in a Hilbert space associated to the region's boundary.
Amplitudes and a probability interpretation
can be assigned to the states.
Not needing a fixed metric, the GBF is a promising framework
for quantum theories treating the metric as a dynamical object.
Using Sch\"odinger-Feynman quantization, in
\cite{colosi_oeckl:_spat_asymp_smatrix_gbf}
the GBF reproduces the usual S-matrix for real
Klein-Gordon theory on Minkowski spacetime,
plus an equivalent S-matrix of different type.
In \cite{colosi:_smatrix_desitter_long} 
the same has been achieved for deSitter spacetime.
The techniques used for both types of S-matrices
are generalized for a wide class of spacetimes in
\cite{colosi_dohse:_structure_smatrix_gbqft}
and now applied here for Anti-deSitter spacetime (AdS),
where due to the lack of (temporal) asymptotically free states
the usual S-matrix cannot be defined.
The paper is structured as follows:
After short reviews of the standard S-matrix (Section 2),
AdS (3), the GBF (4) and Schrödinger-Feynman quantization (5),
we apply in (6) the GBF for constructing the different type of S-matrix
for real Klein-Gordon (KG) theory. 
\vspace{-3mm}
\section{Standard construction of the S-matrix}
In standard QFT there is \emph{one} Hilbert space $\hilb$ of free states.
The S-matrix $\scatmat: \hilb \rightarrow \hilb$ is a unitary operator.
Matrix elements are obtained as large-time limit
$\scatmat_{\et,\ze}  \!\sim\!\liminfty{t}\!
\braopketl{t}{\ze\!}{\mc U\lints{-t,t}}{\!\et}{-t}$,
\bfig{\igx[width=0.28\linewidth]{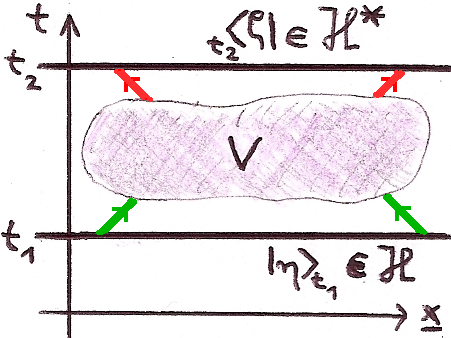}\;\,
	\begin{minipage}[b]{0.7\linewidth}
		with $\mc U\lints{-t,t}$ being the time-evolution operator.
		\vspace{-0.5mm}
		\caption{Slice of spacetime between the two equal-time 
			hypersurfaces at $t_1$ and $t_2$.
			Usually it is assumed that the interaction is "switched on"
			only between times $t_1$ and $t_2$,
			then $\ket[t_1]{\et}$ and $\bra[t_2]{\ze}$ are free states.
			Setting $t_1 = -t$ and $t_2 = + t$ and taking the limit $t\to \infty$
			the states become asymptotically free while the interaction
			can now be "switched on" on all of spacetime.
			}
	\label{fig:_gbf_standard_particles}
	\end{minipage}
	}
There is however a physically more convincing way to describe this.
Physical interactions are not switched on and off by \emph{time},
but decrease with the \emph{distance} between interacting particles.
Consider interacting massive particles leaving the interaction region V
in \ref{fig:_gbf_standard_particles}.
When their separation becomes large, the state can be considered free
and the particles as moving along timelike geodesics.
For later times the states remain free iff
spacetime geometry is such that the separation remains large,
which in curved spacetime cannot be taken for granted,
see figure \ref{fig:_ads_geodesics}.
%
\vspace{-6mm}
\section{Anti-deSitter spacetime (AdS)}
We coordinatize (the universal covering version of)
Lorentzian AdS of dimension $(d\pn1)$ by
$x \eq (t,\ro,\Om)$, with time $t\elof (-\infty,+\infty)$,
radial coordinate $\ro \elof [0,\frac\piu2)$, and angles
$\Om \eq (\te_1,\ldotsn,\te_{d\mn1})$ such that $\sphere[d\mn1]$
is covered.
The metric is $\dif s^2\lads  \eq \Rads^2\cosn{-2} \rho \,
 \biglrr{ \dif t^2 - \dif  \rho^2 - \sinn2 \rho \; \dif s^2_{\sphere[d\mn1]}}$,
with the curvature radius $\Rads$.
For large $\Rads$ the AdS metric approximates the
Minkowski metric.
The free Lagrangian of real KG theory
$\mc L = \frac12 \ruutabs g \biglrs{
 g^{\mu\nu} (\del_\mu\ph)(\del_\nu\ph) - m^2 \ph^2}$
yields as equation of motion the KG equation
$(\dalembertian \pn m^2  ) \, \ph = 0$.
Its solutions are the modes
$\ph_{\om l\vc m}\ar x$, with 
frequency $\om \elof (-\infty,+\infty)$
and angular quantum numbers $(l,\vc m)$
such that the spherical harmonics $\spherharmonicar{\vc m}{l}\Om$
form an orthonormal base (ONB) on $\sphere[d\mn1]$.
A separation ansatz \cite{breitfreed:_stability_gauged_ext_sugra}
gives $\ph_{\om l\vc m}\ar x \eq \eu^{-\iu \om t}  \,
\spherharmonicar{\vc m}{l}\Om\, S_{\om,l}\ar\ro$.
We consider only the case of odd $d$, where
$S_{\om l}$ can be any linear combination
of the two linear independent solutions of the radial part of the
KG equation given by
$\oS{\om,l} \ar\ro = \sinn l \ro\, \cosn{\timp}\!\ro\;
\hypergeo{\alp}{\bep}{\gaS}{\sinsq\ro}$ and
$\tSodd{\om,l}  \ar\ro = (\sin \ro)^{2\mn l\mn d}\,
\cosn{\timp}\!\ro\; \hypergeo{\alp\mn\gaS\pn1}
{\bep\mn\gaS\pn1}{2\mn\gaS}{\sinsq\ro}$.
\vspace{-2mm}
\bfig{\igx[width=0.23\linewidth]{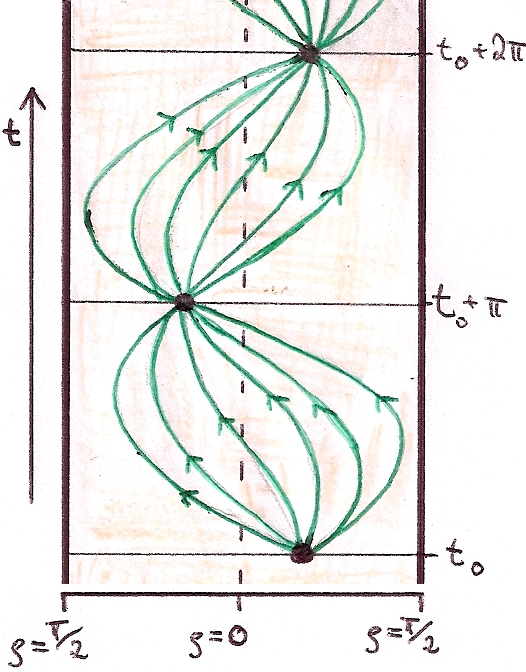}\;
	\begin{minipage}[b]{0.75\linewidth}
		$F$ is the hypergeometric function with
		$\al_+,\be_+,\ga^\txS$ depending on $\om,l$, mass $m$
		and spatial dimension $d$.
		\vspace{-0.5mm}
		\caption{Penrose diagram of AdS.
			Only temporal and radial directions are drawn.
			The negative curvature makes the timelike geodesics emanating
			from an arbitrary point at $t_0$ reconverge
			in periods of $\piu$, see \cite{avis_isham_storey:_qft_ads}.
			Thus there exists no time after which
			two particles leaving a scattering process
			are forever separated by large distances,
			i.e., after which interactions
			can be neglected forever.
			Therefore on AdS temporally asymptotic states
			do not exist and the standard S-matrix cannot be defined.
			}
		\label{fig:_ads_geodesics}
	\end{minipage}
	}
\vspace{-15.5mm}
\section{General Boundary Formulation (GBF)}
To see how another type of S-matrix can be constructed for AdS
now we outline the GBF \cite{oeckl:_gbf_found_prob}. 
To each hypersurface $\Si$
(of codimension one) in a spacetime is associated its quantum state space
$\hilb[\Si]$. Each region M (of top dimension)
has associated its boundary state space $\hilb[\del\txM]$,
see fig.~\ref{fig:_gbf_general}.
Amplitudes for the quantum states
are calculated with the amplitude map
$\ro_\txM: \hilb[\del\txM] \rightarrow \complex$.
\vspace{-2.5mm}
\bfig{\begin{minipage}[c]{0.49\linewidth}
		\igx[width=0.8\linewidth]{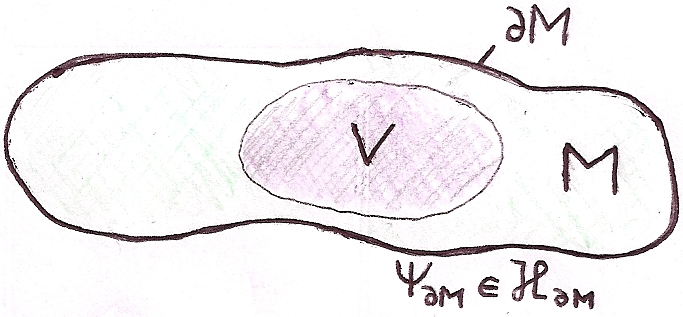}
		\caption{}
		\label{fig:_gbf_general}
	\end{minipage}
	\begin{minipage}[c]{0.49\linewidth}
		\igx[width=0.85\linewidth]{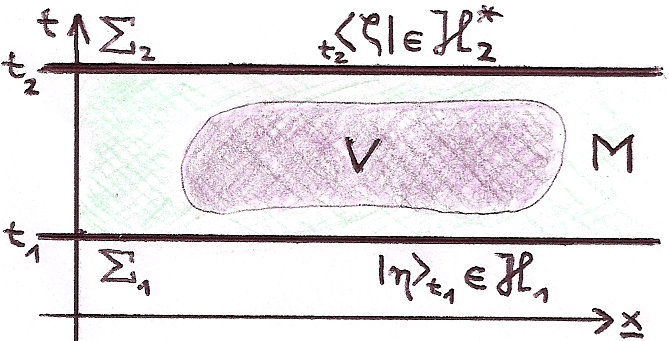}
		\caption{}
		\label{fig:_gbf_reform}
	\end{minipage}
	}
\vspace{-7.5mm}\\
Figure \ref{fig:_gbf_reform} shows how this general description
includes the standard setting
\cite{oeckl:_gbf_found_prob,oeckl:_observables_gbf}. 
There, M is the spacetime slice
$[t_1,t_2] \cross \reals[3]$ with boundary
$\del\txM \eq \Si_1 \disunion \Si_2$, a disjoint union of
two hypersurfaces. The boundary state space is
the tensor product 
$\hilb[\del\txM] \eq \hilb[1] \tensored \hilbdual[2]$.
The amplitude map $\ro_\txM\maps \hilb[1] \tensored \hilbdual[2]
\to \complex$ is now such that it is related to time evolution by
$\ro_\txM \biglrr{\ket[t_1]{\!\et} \!\!\tensored\! \bra[t_2]{\!\ze\!}} \eq 
\braopketl{2}{\ze}{\mc U\lint{t_1,t_2}}{\et}{1}$.
The tensor product $\ket[t_1]{\!\et} \!\!\tensored\! \bra[t_2]{\!\ze\!}$
also writes as \emph{one} boundary state
$\ps_{\del\txM} \elof \hilb[\del\txM]$ (dropping bra-ket notation).
\bfig{
		\igx[width=0.22\linewidth]{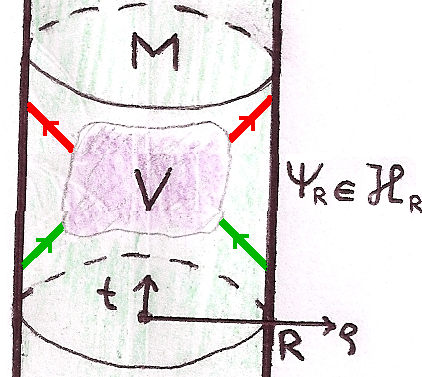}
	\;
	\begin{minipage}[b]{0.75\linewidth}
		\caption{Hypercylinder region $\txM = \reals\cross\ball[d]$:
			a solid ball of radius $R$ in space
			times all of time \cite{colosi_oeckl:_spat_asymp_smatrix_gbf}.
			The boundary $\del\txM$ is the \emph{single}
			hypersurface $\Si_R \defeq \reals \cross \sphere[d\mn1]$.
			An incoming (outgoing) particle is now one
			 that enters (leaves) M at any time.
			The boundary state space decomposes
			$\hilb[\del\txM] = \hilb[\text{in}]
			\tensored \hilb[\text{out}]$ into spaces of purely
			incoming and outgoing states.
			The amplitude $\ro_\txM$ of \emph{one} boundary state
			$\ps_{\del\txM} \eq \ps\ltx{in} \tensored \ps\ltx{out}$
			is thus related to observing
			its outcoming part $\ps\ltx{out}$
			given $\ps\ltx{in}$ was prepared.
			}
		\label{fig:_gbf_hypercylinder_particles}
	\end{minipage}
	}
\vspace{-8.5mm}\\
If we assume spacetime geometry being such that large radius $R$ 
implies large distance for two points on $\Si_R$,
then for $R\to\infty$ interactions can be neglected
and the boundary state is (radially) asymptotically free.
This is the case for AdS and so the problem
of no (temporally) asymptotic states can be avoided.
The large-$R$ limit of $\ro_\txM (\ps_{\del\txM})$
can thus be interpreted as a new type of S-matrix.
%
\vspace{-3mm}
\section{Schr\"odinger-Feynman quantization}
In the Schr\"odinger representation \cite{jackiw:_schrod_rep}
the states are complex function(al)s
on field configurations on a hypersurface.
Let $\txK_\Si$ the space of field configurations on $\Si$,
then $\ps_\Si \maps \txK_\Si \to \complex$.
Feynman quantization means that amplitudes are calculated as a path-integral
with $Z_\txM$ called field propagator:
\vspace{-4mm}
\bals{\ro_\txM(\ps_{\del\txM}) \eq \intliml{\txK_{\del\txM}}
	 \fundif \vph \ps_{\del\txM}(\vph) \, Z_\txM(\vph)
	\pquad 4
	Z_\txM(\vph) \eq
	 \intliml{\txK_\txM,\ph|_{\del\txM}\eq \vph}\,\fundif\ph
	 \eu^{\iu S_\txM(\ph)}\;.
	}
\vspace{-4mm}\\
The vacuum $\ps_{R,0}$ for a hypercylinder $\Si_R$ is a Gaussian
\cite{colosi:_structure_vacuum_gbqft,
colosi_oeckl:_unitary_evolution_qft_curved_spacetime},
with $\opA_R$ called vacuum operator:
\vspace{-5mm}
\bals{\ps_{R,0}(\vph) \sim \exp \biiiglrr{
		- \fracss{1}{2} \intliml{\Si_R} \dif t \volOm 
		\vph\arvc x (\op\txA_{\Si_R} \vph)\arvc x}
	\pquad 4
	\opA_{R} = - \iu \, \ruutabs{g_{\Si_R}g^{\ro\ro}} \, 
			\biiigl. \frac{\del_\ro \opUp\ar \ro}{\opUp\ar \ro}
			\biiigr|_{\ro=R} \;.
	}
\vspace{-4mm}\\
An ONS on $\Si_R$ is given by the functions
$u_{\om l\vc m} \artOm \sim\eu^{-\iu \om t}\, \spherharmonicar{\vc m}l\Om$,
and $\op\Up$ is defined by its eigenvalues:
$\opUp\ar\ro \, u_{\om l\vc m} \artOm \eq \biglrs{c^a_{\om l\vc m} \oS{\om,l} \ar\ro 
+ c^b_{\om l\vc m} \tSodd{\om,l}  \ar\ro }\,u_{\om l\vc m} \artOm$.
The vacuum is thus determined by fixing the coefficients $c^{a,b}$
(a,b are labels, not indices). We shall choose
$c^a_{\om l\vc m} \eq (p\Rads)^l/(2l\pn1)!!$ and $c^b_{\om l\vc m} \eq -\iu\,(2l\mn1)!!/(p\Rads)^{l\pn1}$
with $p\defeq \ruutabs{\om^2-m^2}$, because
in the limit $\Rads\to\infty$ this choice yields the well known Minkowski vacuum.

Next we can define coherent states. A coherent state is
determined by its complex characterising function $\et$
and usually written with creators $\opa^\dagger_{\om l\vc m}$ as
$\ps_{R, \et}(\vph) \sim \biglrs{\exp \int\dif\om \sum\,\!_{l\vc m}\;
\et_{\om l\vc m}\, \opa^\dagger_{\om l\vc m}}\;\ps_{R,0}(\vph)$.
In the interaction picture it can equivalently be written as
$\ps_{R, \xi}(\vph) \sim \biglrs{\exp \int\dif t\volOm 
\xi\artOm\, \opUp^{-1}\ar\ro\,\vph\artOm}\;\ps_{R,0}(\vph)$,
see \cite{colosi_dohse:_structure_smatrix_gbqft}.
%
\section{Radial S-matrix for AdS}
Labeling all quantities of the free theory by a zero,
calculating the free amplitude results in
\bals{\ro_{R,0}(\ps_{R,\xi}) 
	\eq \intliml{} \fundif \vph \ps_{R,\xi}(\vph) \, Z_{R,0}(\vph)
	\eq \exp \biiiglrs{\fracs{1}{2} \int_{\Si_R} \!\!\dif t \, \volOm \xi\artOm \,
		\Op{\mc{B}} \left( \frac{\coco{c_b}}{c_b} \, \xi\artOm 
		+ \coco{\xi\artOm}  \right)}\;,
	}
\vspace{-4mm}\\
with $\Op{\mc B}$ defined by the eigenvalues
$p/(2\Rads^{d\mn2})\cdot (2l\pn1)/(2l\pn d\mn2)$
when acting on the $u_{\om l \vc m} \artOm$.
Since the free amplitude is independent of $R$ (not $\Rads$),
taking the large-$R$ limit is easy and the amplitude
can then be interpreted as radial S-matrix.
Now we add a source field $\mu$ to include interactions:
$S_{R,\mu}(\ph) \eq S_{R,0}(\ph)
+\int_{\ro<R} \volx[d\pn1] \ruutabs g\, \mu\ar x\, \ph\ar x$.
At this point we assume the source field to vanish for $\ro\geq R$.
The resulting amplitude for a coherent state is
\bals{\ro_{R,\mu}(\ps_{R,\xi}) = \ro_{R,0}(\ps_{R,\xi}) \,
	\biiiglrs{ \exp\!\! \intliml{\ro<R}\, \vol[d\pn1]{\!\!x}\! \ruutabs{g} \,
	 \mu\, \Op{\mc C}\opX_a\xi}
	\biiiglrs{ \exp \fracs{\iu}{2} \!\!\intliml{\ro<R} \,
		\vol[d\pn1]{\!\!x}\!\vol[d\pn1]{\!\!x'}\! \ruutabs{g\ar x g\ar{x'}} \,
		\mu\ar x \, G\ltx F\ar{x,x'} \, \mu\ar{x'} }.
	}
The operators are again defined by their eigenvalues:
for $\opX_a\ar\ro$ by $\oS{\om,l} \ar\ro$
and for $\op{\mc C}$
by $\iu(p\Rads)^{l\pn1} [\Rads^{d\mn1}(2l\pn d\mn2)(2l\mn1)!!]^{-1}$.
The Feynman propagator $G_\txF$ fulfills the inhomogeneous KG equation
$(\beltrami\lads\pn m^2)\,G\ltx F\ar{x,x'} \eq \dirdelta[d]{x\mn x'}$
and is given by
\bals{G\ltx F\ar{x,x'} = \int\!\!\dif\om \sumliml{l,\vc m}\;
	\biiiglrc{ \te\ar{\ro\mn\ro'} 
	\left( \Op{\mc C}\opX_a\ar{\ro'}\,\coco{u_{\om l\vc m}}\right) 
	\!\ar{t',\Om'}\, \left(\opUp\ar \ro u_{\om l\vc m} \right)\!\ar{t,\Om}
	\;+ \;(x\leftrightarrow x') }\;.
	}
The amplitude is thus independent of $R$,
so we can take the limit $R\to\infty$
(such that the source field can have support on the whole spacetime)
and consider the amplitude as an S-matrix.
For a field interaction with potential $V$ and action
$S_{R,V}(\ph) \eq S_{R,0}(\ph)
+\int_{\ro<R} \volx[d\pn1] \ruutabs g\, V(x,\ph)$
we can use the functional derivative technique to obtain
\bals{\exp \iu S_{R,V}(\ph)
	\eq  \biiiiglrs{ \exp\,\iu \!\int_{\ro<R}\,\volx[d\pn1]
 	\ruutabs{g\ar x} \,
 	\opV \biiglrr{ x, -\iu \fracss{\delta}{\delta \mu\ar x}}}\,
 	\exp \iu S_{R,\mu}(\ph) \biiigl.\biiigr|_{\mu \eq0}\,
	}
leading to the $R$-independent amplitude below whose
large-$R$ limit is an S-matrix:
\bals{\ro_{R,V}(\ps) 
	\eq \biiiiglrs{ \exp \,\iu \intliml{\ro<R}\, \volx[4] \ruutabs{g\ar x} \,
	 \opV \biiglrr{ x, -\iu \fracss{\delta}{\delta \mu\ar x}} }\,
	 \ro_{R,\mu}(\ps) \biiigl.\biiigr|_{\mu \eq0}\;.
	}
%
\section{Conclusions}
Using the GBF, an S-matrix for AdS can be constructed
by calculating (generalized) amplitudes for the hypercylinder region.
This circumvents the problem of not having available (temporally)
asymptotic states on AdS, needed for the S-matrix of standard QFT.
Future projects consist in applying holomorphic quantization
\cite{oeckl:_holomorphic_quant_lin_field_gbf,oeckl:_affine_holo_quant,
oeckl:_schrorep_holorep} within the GBF framework,
studying properties of this S-matrix and connections
with the AdS/CFT correspondence \cite{witten:_ads_holo},
and relating our S-matrix to the one found by Giddings
\cite{giddings:_boundary_s_matrix_ads_cft_dictionary}.
%
\section*{Acknowledgments}
M.D. wishes to thank the organizers of the LOOPS 11 conference
for the possibility of presenting this work,
which was supported in part by CONACyT grant 49093
and scholarship 213531.
%
\section*{References}
\bibliography{zzzzz_references}
%
%
\end{document}

%% file: __new_commands_general.tex
\newcommand{\nc}{\newcommand}
\nc{\rnc}{\renewcommand}

\nc{\bosym}{\boldsymbol}
\nc{\enmat}{\ensuremath}

\nc{\TOM}{\TextOrMath}
\nc{\TOMt}[1]{\TOM {#1\xspace} {\text{#1}}}
\nc{\TOMm}[1]{\TOM {$#1$\xspace} {#1}}

\DeclareMathAlphabet{\mathpzc}{OT1}{pzc}{m}{it}

\nc{\noi}{\noindent}
\nc{\yesi}{\indent}
\nc{\whitei}[1]{\white{x}\hspace{#1}\negphantom{x}}

\nc{\sma}[1]{\TOM{{\small #1}}{\text{\small$#1$}}}
\nc{\smaa}[1]{\TOM{{\footnotesize #1}}{\text{\footnotesize$#1$}}}
\nc{\smaaa}[1]{\TOM{{\scriptsize #1}}{{\scriptstyle #1}}}
\nc{\smaaaa}[1]{\TOM{{\tiny #1}}{{\scriptscriptstyle #1}}}

\nc{\ssiz}{\fontsize{7pt}{9pt}\selectfont}
\nc{\sssiz}{\fontsize{5pt}{7pt}\selectfont}

\nc{\notafootnote}[1]{{\noindent\ssiz{#1}\par}\noindent\hspace{-3.5pt}}

\nc{\inv}[1][1]{^{-#1}}

\nc{\hs}[1]{^{\smaaaa{#1}}}
\nc{\hb}[1]{^{(#1)} }
\nc{\hbs}[1]{^{\smaaaa{(#1)}}}
\nc{\htx}[1]{^{\text{#1}}}
\nc{\htxs}[1]{^{\text{\tiny#1}}}
\nc{\hbtx}[1]{^{\text{(#1)}}}
\nc{\hbtxs}[1]{^{\text{\tiny(#1)}}}

\nc{\ls}[1]{_{\smaaaa{#1}}}
\nc{\lb}[1]{_{(#1)} }
\nc{\lbs}[1]{_{\smaaaa{(#1)}}}
\nc{\ltx}[1]{_{\text{#1}}}
\nc{\ltxs}[1]{_{\text{\tiny#1}}}
\nc{\lbtx}[1]{_{\text{(#1)}}}
\nc{\lbtxs}[1]{_{\text{\tiny(#1)}}}
\nc{\lint}[1]{_{[#1]}}
\nc{\lints}[1]{_{\scriptscriptstyle[#1]}}

\nc{\hvc}[2][]{^{\smaaaa{#1}\vc #2 }}
\nc{\hvcs}[2][]{^{\smaaaa{#1\vc #2 }}}

\nc{\lvc}[2][]{_{\smaaaa{#1}\vc #2 }}
\nc{\lvcs}[2][]{_{\smaaaa{#1\vc #2 }}}

\nc{\hi}[1]{\hs{#1} }	
\nc{\hii}[2]{\hi{#1#2} }	
\nc{\hiii}[3]{\hi{#1#2#3} }	
\nc{\hiiii}[4]{\hi{#1#2#3#4} }	

\nc{\ovl}{\overline}
\nc{\unl}{\underline}
\nc{\unlh}[1]{\settowidth{\negphantomlength}{{#1}}%
	\unl{\hspace{\negphantomlength}}\negphantom{#1}#1}

\nc{\undash}{\dashuline}				
\nc{\undot}{\dotuline}

\nc{\ovb}{\overbrace}
\nc{\unb}{\underbrace}
\nc{\ovs}[2]{\overset{#2}{#1}}			
\nc{\uns}[2]{\underset{#2}{#1}}		
\nc{\ovla}[1]{\overleftarrow{#1}}
\nc{\ovra}[1]{\overrightarrow{#1}}

\nc{\contraction}[1]{\overbracket{#1}}

\nc{\ti}[1]{\TOMm{\tilde{#1}}}
\nc{\tivc}[1]{\TOMm{\tilde{\vc{#1}}}}
\nc{\titi}[1]{\TOMm{\tilde{\tilde{#1}}}}

\nc{\Ti}[1]{\TOMm{\widetilde{#1}}}
\nc{\TTi}[1]{\TOMm{\!\widetilde{\,#1\,}\!}}	
\nc{\Tivc}[1]{\TOMm{\widetilde{\vc{#1}}}}

\nc{\Wi}[1]{\TOMm{\widehat{#1}}}
\nc{\WWi}[1]{\TOMm{\!\widehat{\,#1\,}\!}}	

\nc{\janusdel}{\partial}
\nc{\janusdelslashed}{\slashdel}
\declareslashed{}{\smaaa{\shortleftrightarrow}}{0}{1.6}{\janusdel}
\declareslashed{}{\smaaa{\shortleftrightarrow}}{0.07}{1.7}{\janusdelslashed}

\nc{\janus}{\slashed{\janusdel}}
\nc{\januslo}[1]{\janus\,\!\lo{\!#1}}
\nc{\janushi}[1]{\janus\,\!\hi{#1}}
\nc{\janusslashed}{\slashed{\janusdelslashed}}

\nc{\degree}{\TOMm{^\circ}}
\nc{\composed}{\circ}
\nc{\fish}{\textproto{\Adaleth}}
\nc{\helmet}{\textproto{\Ahelmet}}

\nc{\cross}{\mathchoice{\hspace{0.7pt}\sma\times\hspace{0.7pt}}
	{\hspace{0.7pt}\sma\times\hspace{0.7pt}}
	{\hspace{0.4pt}\times\hspace{0.4pt}}
	{\hspace{0.2pt}\times\hspace{0.2pt}} }

\nc{\dirsum}{\mathchoice{\hspace{1pt}\sma\oplus\hspace{1pt}}
	{\hspace{1pt}\sma\oplus\hspace{1pt}}
	{\hspace{0.5pt}\oplus\hspace{0.5pt}}
	{\hspace{0.3pt}\oplus\hspace{0.3pt}} }

\nc{\tensored}{\mathchoice{\hspace{1pt}\sma\otimes\hspace{1pt}}
	{\hspace{1pt}\sma\otimes\hspace{1pt}}
	{\hspace{0.5pt}\otimes\hspace{0.5pt}}
	{\hspace{0.3pt}\otimes\hspace{0.3pt}} }

\nc{\facwedgeslashed}{\wedge}
\declareslashed{}{\smaaaa!}{0}{-0.5}{\facwedgeslashed}
\nc{\facwedge}{\slashed{\facwedgeslashed}}
\nc{\ordwedgeslashed}{\wedge}
\declareslashed{}{\smaaaa\langle}{-0.08}{-0.5}{\ordwedgeslashed}
\nc{\ordwedge}{\slashed{\ordwedgeslashed}}

\nc{\wedged}{\mathchoice{\hspace{1pt}\sma\wedge\hspace{1pt}}
	{\hspace{1pt}\sma\wedge\hspace{1pt}}
	{\hspace{0.5pt}\wedge\hspace{0.5pt}}
	{\hspace{0.3pt}\wedge\hspace{0.3pt}}
	}
\nc{\unnormwedged}{\hspace{1pt}\sma\facwedge\hspace{1pt}}
\nc{\ordwedged}{\hspace{1pt}\sma\ordwedge\hspace{1pt}}

\nc{\ltridot}%
	{\TOM{$.\hspace{0.08ex}.\hspace{0.08ex}.\hspace{0.12ex}$\xspace}
	{\mathchoice{.\hspace{0.08ex}.\hspace{0.08ex}.\hspace{0.12ex}}
		{.\hspace{0.08ex}.\hspace{0.08ex}.\hspace{0.12ex}}
		{\ldots} {\ldots} }%
	}
\nc{\ctridot}%
	{\TOM{$\!\cdot\mhspace{-0.3}{x}\cdot\mhspace{-0.3}{x}\cdot\!$\xspace}
	{\mathchoice{\!\cdot\mhspace{-0.3}{x}\cdot\mhspace{-0.3}{x}\cdot\!}
		{\!\cdot\mhspace{-0.3}{x}\cdot\mhspace{-0.3}{x}\cdot\!}
		{\cdots} {\cdots} }%
	}
\nc{\ltridotw}{\,\ltridot\,}
\nc{\ctridotw}{\,\ctridot\,}

\nc{\aso}{\TOM{$,\hspace{0.5pt}\ltridot,\hspace{0.5pt}$}
	{\mathchoice{,\hspace{0.5pt}\ltridot,\hspace{0.5pt}}%
		{,\hspace{0.5pt}\ltridot,\hspace{0.5pt}}%
		{,\,\ldots,\,} {,\,\ldots,\,}} }
\nc{\ason}{\TOM{$,\hspace{-1.5pt}\ltridot\hspace{-0.5pt},\hspace{-1pt}$}
	{\mathchoice{,\hspace{-1.5pt}\ltridot\hspace{-0.5pt},\hspace{-1pt}}%
		{,\hspace{-1.5pt}\ltridot\hspace{-0.5pt},\hspace{-1pt}}%
		{,\ldots,} {,\ldots,}} }

\nc{\cdotw}{\,\cdot\,}
\nc{\ldotsn}{{\scriptstyle\ldots}}
\nc{\Euro}{\TOM {\euro\xspace} {\text{\euro}}}

\newcounter{puadcounter}
\nc{\pquad}[1]{\setcounter{puadcounter}{#1}%
			\ifte{#1>0}{\quad\addtocounter{puadcounter}{-1}%
				\pquad{\value{puadcounter}}}{}%
		   }

\nc{\puad}[1]{\setcounter{puadcounter}{#1}%
			\ifte{#1>0}{\,\addtocounter{puadcounter}{-1}%
				\puad{\value{puadcounter}}}{}%
		   }

\nc{\muad}[1]{\setcounter{puadcounter}{#1}%
			\ifte{#1>0}{\!\addtocounter{puadcounter}{-1}%
				\muad{\value{puadcounter}}}{}%
		   }

\nc{\bgl}{\left}
\nc{\biigl}{\Bigl}
\nc{\biiigl}{\biggl}
\nc{\biiiigl}{\Biggl}

\nc{\bgr}{\right}
\nc{\biigr}{\Bigr}
\nc{\biiigr}{\biggr}
\nc{\biiiigr}{\Biggr}

\nc{\bgm}{\middle}
\nc{\biigm}{\Bigm}
\nc{\biiigm}{\biggm}
\nc{\biiiigm}{\Biggm}

\nc{\bglrr}[1]{\bgl( #1 \bgr)}
\nc{\biglrr}[1]{\bigl( #1 \bigr)}
\nc{\biiglrr}[1]{\biigl( #1 \biigr)}
\nc{\biiiglrr}[1]{\biiigl( #1 \biiigr)}
\nc{\biiiiglrr}[1]{\biiiigl( #1 \biiiigr)}

\nc{\bglrs}[1]{\bgl[ #1 \bgr]}
\nc{\biglrs}[1]{\bigl[ #1 \bigr]}
\nc{\biiglrs}[1]{\biigl[ #1 \biigr]}
\nc{\biiiglrs}[1]{\biiigl[ #1 \biiigr]}
\nc{\biiiiglrs}[1]{\biiiigl[ #1 \biiiigr]}

\nc{\bglrc}[1]{\bgl\{ #1 \bgr\}}
\nc{\biglrc}[1]{\bigl\{ #1 \bigr\}}
\nc{\biiglrc}[1]{\biigl\{ #1 \biigr\}}
\nc{\biiiglrc}[1]{\biiigl\{ #1 \biiigr\}}
\nc{\biiiiglrc}[1]{\biiiigl\{ #1 \biiiigr\}}

\nc{\bglra}[1]{\bgl\langle #1 \bgr\rangle}
\nc{\biglra}[1]{\bigl\langle #1 \bigr\rangle}
\nc{\biiglra}[1]{\biigl\langle #1 \biigr\rangle}
\nc{\biiiglra}[1]{\biiigl\langle #1 \biiigr\rangle}
\nc{\biiiiglra}[1]{\biiiigl\langle #1 \biiiigr\rangle}

\nc{\bglrrn}[2]{\bgl( #1 \bgr)^{\!#2}}
\nc{\biglrrn}[2]{\bigl( #1 \bigr)^{\!#2}}
\nc{\biiglrrn}[2]{\biigl( #1 \biigr)^{\!#2}}
\nc{\biiiglrrn}[2]{\biiigl( #1 \biiigr)^{\!#2}}
\nc{\biiiiglrrn}[2]{\biiiigl( #1 \biiiigr)^{\!#2}}

\nc{\bglrsn}[2]{\bgl[ #1 \bgr]^{#2}}
\nc{\biglrsn}[2]{\bigl[ #1 \bigr]^{#2}}
\nc{\biiglrsn}[2]{\biigl[ #1 \biigr]^{#2}}
\nc{\biiiglrsn}[2]{\biiigl[ #1 \biiigr]^{#2}}
\nc{\biiiiglrsn}[2]{\biiiigl[ #1 \biiiigr]^{#2}}

\nc{\bglrcn}[2]{\bgl\{ #1 \bgr\}^{\!#2}}
\nc{\biglrcn}[2]{\bigl\{ #1 \bigr\}^{\!#2}}
\nc{\biiglrcn}[2]{\biigl\{ #1 \biigr\}^{\!#2}}
\nc{\biiiglrcn}[2]{\biiigl\{ #1 \biiigr\}^{\!#2}}
\nc{\biiiiglrcn}[2]{\biiiigl\{ #1 \biiiigr\}^{\!#2}}

\nc{\bglran}[2]{\bgl\langle #1 \bgr\rangle^{\!#2}}
\nc{\biglran}[2]{\bigl\langle #1 \bigr\rangle^{\!#2}}
\nc{\biiglran}[2]{\biigl\langle #1 \biigr\rangle^{\!#2}}
\nc{\biiiglran}[2]{\biiigl\langle #1 \biiigr\rangle^{\!#2}}
\nc{\biiiiglran}[2]{\biiiigl\langle #1 \biiiigr\rangle^{\!#2}}

\nc{\comcon}{\TOM{c.\hspace{-0.5pt}c.\hspace{-0.5pt}\xspace}%
			{\text{ c.c. }}}

\nc{\etc}{etc.\hspace{-0.5pt}\xspace}
\nc{\exgra}{e.\hspace{-0.5pt}g.\hspace{-0.5pt}\xspace}
\nc{\idest}{i.\hspace{-0.5pt}e.\hspace{-0.5pt}\xspace}
\nc{\resp}{resp.\hspace{-0.5pt}\xspace}

\nc{\cyrillic}[1]{{\fontencoding{OT2}\fontfamily{cmr}\selectfont#1}}

\nc{\icyr}{\TOMt {\cyrillic{i}}}
\nc{\shecyr}{\TOMt {\cyrillic{sh}}}
\nc{\shchecyr}{\TOMt {\cyrillic{shch}}}
\nc{\yucyr}{\TOMt {\cyrillic{yu}}}
\nc{\yacyr}{\TOMt {\cyrillic{ya}}}

\nc{\Icyr}{\TOMt {\cyrillic{I}}}
\nc{\Shecyr}{\TOMt {\cyrillic{SH}}}
\nc{\Shchecyr}{\TOMt {\cyrillic{SHCH}}}
\nc{\Yucyr}{\TOMt {\cyrillic{YU}}}
\nc{\Yacyr}{\TOMt {\cyrillic{YA}}}

\declareslashed{}{\iota}{0.2}{0}{\pi}
\nc{\piu}{\TOMm{\slashed{\pi}}}
\nc{\twopi}[1][]{\TOMm{ (2\piu)\unemptynumberone{#1}{^{#1}} }}

\nc{\iu}{\TOMt{i}}

\nc{\eu}{\TOMt{e}}

\nc{\Ampere}{\text{A}}
\nc{\Coulomb}{\text{C}}
\nc{\Evolt}{\TOMm{\echarge\ls{\!}\txV}}	
\nc{\Joule}{\text{J}}
\nc{\Kelvin}{\text{K}}
\nc{\Kilogram}{\text{kg}}
\nc{\Meter}{\text{m}}
\nc{\Newton}{\text{N}}
\nc{\Second}{\text{s}}
\nc{\Volt}{\text{V}}

\nc{\Mega}{\text{M}}
\nc{\Femto}{\text{f}}

\nc{\Avogadro}{\TOMm{\text{N}\ltxs A}}	
\nc{\Boltzmann}{\TOMm{\text{k}\ltxs B}}
\nc{\echarge}{\TOMm{e}}			
\nc{\finestruc}{\TOMm{\alpha}}		
\nc{\lambdadebro}{\TOMm{\lambda\ltxs{Bro}}}
\nc{\lightc}{\TOMm{c}}				
\nc{\newtonG}{\TOMm{G\ltxs N}}		
\nc{\Gnobar}{G}					
\declareslashed{}{\smaaa\smallsetminus}{-0.18}{0.48}{\Gnobar}
\nc{\newtonGbar}{\TOMm{\slashed{\Gnobar}}}
\nc{\vacepsi}{\TOMm{\ep\ls0}}		
\nc{\vacmu}{\TOMm{\mu\ls0}}			

\nc{\lelec}{\ltxs{el}}					
\nc{\helec}{\htxs{el}}
\nc{\lgrav}{\ltxs{gr}}					
\nc{\hgrav}{\htxs{gr}}
\nc{\linert}{\ltxs{in}}					
\nc{\hinert}{\htxs{in}}

\nc{\massgr}{\TOMm{m\lgrav}}		
\nc{\massin}{\TOMm{m\linert}}		

\nc{\hfirst} {\TOMm{\htx{st}}}
\nc{\hsecond}{\TOMm{\htx{nd}}}
\nc{\hthird} {\TOMm{\htx{rd}}}
\nc{\hnth}   {\TOMm{\htx{th}}}

\def\DHLhooksqrt#1#2{\setbox0=\hbox{$#1\sqrt#2$}%
				\dimen0=\ht0\advance\dimen0-0.2\ht0%
				\setbox2=\hbox{\vrule height\ht0 depth -\dimen0}%
				{\box0\lower0.4pt\box2}%
				}
\newcommand{\hooksqrt}[2][]{\mathpalette\DHLhooksqrt{[#1]{#2}}}

\nc{\ruut}[2][]{\TOM{$\hspace{1pt}\hooksqrt[#1]{#2\hspace{1pt}}$}%
		{\hspace{2pt}\hooksqrt[#1]{#2\hspace{1pt}}\hspace{1pt}}%
		}

\nc{\ruutabs}[2][]{\ruut[#1]{%
	\hspace{-1pt}\abs{\hspace{0.5pt}#2\hspace{0.5pt}}\hspace{-1pt}}%
	}
\nc{\ruutabsn}[2][]{\ruut[#1]{%
	\hspace{-1pt}\abs{#2}\hspace{-1pt}}%
	}

\nc{\ruuts}[2][]{\sma{\ruut[#1]{#2}}}
\nc{\ruutss}[2][]{\smaa{\ruut[#1]{#2}}}
\nc{\ruutsss}[2][]{\smaaa{\ruut[#1]{#2}}}
\nc{\ruutssss}[2][]{\smaaaa{\ruut[#1]{#2}}}

\nc{\ruutabss}[2][]{\sma{\ruutabs[#1]{#2}}}
\nc{\ruutabsss}[2][]{\smaa{\ruutabs[#1]{#2}}}
\nc{\ruutabssss}[2][]{\smaaa{\ruutabs[#1]{#2}}}
\nc{\ruutabsssss}[2][]{\smaaaa{\ruutabs[#1]{#2}}}

\nc{\ruutabssn}[2][]{\sma{\ruutabsn[#1]{#2}}}
\nc{\ruutabsssn}[2][]{\smaa{\ruutabsn[#1]{#2}}}
\nc{\ruutabssssn}[2][]{\smaaa{\ruutabsn[#1]{#2}}}
\nc{\ruutabsssssn}[2][]{\smaaaa{\ruutabsn[#1]{#2}}}

\nc{\fracs}[2]{\sma{\frac{#1}{#2}} }
\nc{\fracss}[2]{\smaa{\frac{#1}{#2}} }
\nc{\fracsss}[2]{\smaaa{\frac{#1}{#2}} }
\nc{\fracssss}[2]{\smaaaa{\frac{#1}{#2}} }

\nc{\fracwspace}{\hspace{0.2ex}}

\nc{\fracw}[2]{\TOMm{\, \frac{\fracwspace #1 \fracwspace}%
			{\fracwspace #2 \fracwspace} \,}%
			}
\nc{\fracws}[2]{\sma{\fracw{#1}{#2}} }
\nc{\fracwss}[2]{\smaa{\fracw{#1}{#2}} }
\nc{\fracwsss}[2]{\smaaa{\fracw{#1}{#2}} }
\nc{\fracwssss}[2]{\smaaaa{\fracw{#1}{#2}} }

\nc{\fracn}[2]{\TOMm{\frac{\fracwspace #1 \fracwspace}%
			{\fracwspace #2 \fracwspace}}%
			}
\nc{\fracns}[2]{\sma{\fracn{#1}{#2}} }
\nc{\fracnss}[2]{\smaa{\fracn{#1}{#2}} }
\nc{\fracnsss}[2]{\smaaa{\fracn{#1}{#2}} }
\nc{\fracnssss}[2]{\smaaaa{\fracn{#1}{#2}} }

\nc{\onehalf}{\fracn{1}{2}}
\nc{\onehalfs}{\sma{\onehalf}}
\nc{\onehalfss}{\smaa{\onehalf}}
\nc{\onehalfsss}{\smaaa{\onehalf}}
\nc{\onehalfssss}{\smaaaa{\onehalf}}

\nc{\piuhalf}{\fracn{\piu}{2}}
\nc{\piuhalfs}{\sma{\piuhalf}}
\nc{\piuhalfss}{\smaa{\piuhalf}}
\nc{\piuhalfsss}{\smaaa{\piuhalf}}
\nc{\piuhalfssss}{\smaaaa{\piuhalf}}

\nc{\piufourth}{\fracn{\piu}{4}}
\nc{\piufourths}{\sma{\piufourth}}
\nc{\piufourthss}{\smaa{\piufourth}}
\nc{\piufourthsss}{\smaaa{\piufourth}}
\nc{\piufourthssss}{\smaaaa{\piufourth}}

\nc{\dual}[1]{\TOMm{\,^* #1}}   
\nc{\dualspace}[1]{\TOMm{#1^*}}	

\nc{\coco}[1]{\TOMm{\ovl{#1}}}	
 	       	      
\nc{\hodge}{\TOMm{\star}}		

\nc{\pushfwd}[1]{\TOMm{#1 _{\smaaaa\rhd} }}	
\nc{\pushfwdat}[2]{\TOMm{#1 _{\smaaaa\rhd} ^{#2}}}
\nc{\pullback}[1]{\TOMm{#1 ^{\smaaaa\lhd} }}	
\nc{\pullbackat}[2]{\TOMm{#1 ^{\smaaaa\lhd} _{#2}}}

\nc{\Trans}[1]{\TOMm{#1^\top}}
\nc{\adjoint}[1]{\TOMm{#1^{\dagger}}}

\declareslashed{}{\circ}{0}{0.37}{\dagger}
\nc{\adgaz}[1]{\TOMm{#1^{\slashed{\dagger}} }}

\declareslashed{}{\smaaa{\shortrightarrow}}{0.2}{0}{\shortleftarrow}
\nc{\shortleftrightarrow}{\slashed{\shortleftarrow}}

\newlength{\bigcornerheight}
\newlength{\bigcornerwidth}
\newlength{\bigcornerthick}
\nc{\biglrcorner}{\mathrel{\hbox{\rule{\bigcornerwidth}{\bigcornerthick}%
	\rule{\bigcornerthick}{\bigcornerheight}}}}
\nc{\bigllcorner}{\mathrel{\hbox{\rule{\bigcornerthick}{\bigcornerheight}%
	\rule{\bigcornerwidth}{\bigcornerthick}}}}
\nc{\bigurcorner}{\mathrel{\hbox{\rule[\bigcornerwidth]{\bigcornerwidth}%
	{\bigcornerthick}\rule{\bigcornerthick}{\bigcornerheight}}}}
\nc{\bigulcorner}{\mathrel{\hbox{\rule{\bigcornerthick}{\bigcornerheight}%
	\rule[\bigcornerwidth]{\bigcornerwidth}{\bigcornerthick}}}}

\nc{\inserted}{\biglrcorner}

\declareslashed{}{/}{0.05}{0}{\del}
\nc{\slashdel}{\slashed{\del}}

\nc{\flowone}{I}
\declareslashed{}{\thicksim}{0.47}{0.6}{\flowone}
\nc{\flowtwo}{\slashed{\flowone}}
\declareslashed{}{\thicksim}{0.27}{0.12}{\flowtwo}
\nc{\flowthree}{\slashed{\flowtwo}}
\declareslashed{}{\thicksim}{0.23}{-0.27}{\flowthree}
\nc{\flowF}{\slashed{\flowthree}}

\declareslashed{}{\aquarius}{0}{0.545}{\aquarius}
\nc{\flowriver}{\sma{\slashed{\aquarius}}}

\nc{\lstar}{_{\smaaa *}}
\nc{\hstar}{^{\smaaa *}}

\nc{\lsharp}{_{\smaaa \sharp}}
\nc{\hsharp}{^{\smaaa \sharp}}

\nc{\lflat}{_{\smaaa \flat}}
\nc{\hflat}{^{\smaaa \flat}}

\nc{\opspace}[1]{\text{Op}\Arg{#1}}
\nc{\linopspace}[1]{\text{Lin}\Arg{#1}}

\rnc{\emptyset}{\TOMm{\varnothing}}

\nc{\union}{\mathop{\cup}}
\nc{\unionn}{\mathop{\bigcup}}
\nc{\unionnl}[2]{\mathop{\bigcup} \limits_{#1}^{#2}}
\nc{\disunion}{\mathop{\sqcup}}
\nc{\disunionn}{\mathop{\bigsqcup}}
\nc{\disunionnl}[2]{\mathop{\bigsqcup} \limits_{#1}^{#2}}
\nc{\intsec}{\mathop{\cap}}
\nc{\intsecc}{\mathop{\bigcap}}
\nc{\intseccl}[2]{\mathop{\bigcap} \limits_{#1}^{#2}}
\nc{\intseccu}[2]{\mathop{\bigcap} \limits_{#1}^{}}
\nc{\without}{\setminus}

\nc{\Else}{\text{else}}

\nc{\sign}{\text{sign}\,}
\nc{\diag}{\text{diag}\,}
\nc{\const}{\text{const.}}

\nc{\vcnabla}{\,\vc{\!\nabla\!}\,}
\nc{\grad}{\text{grad}\,}					
\nc{\gradvc}{\vc{\text{grad}}\;}
\rnc{\div}{\text{div}\;}						

\nc{\Id}[1][]{\TOMm{\text{Id} \ltx{#1} \,\!}}
\nc{\Idop}[1][]{\TOMm{\op{\text{Id}} \ltx{#1} \,\!}}
\nc{\One}[1][]{\TOMm{\mathbbm{1} \ltx{#1} \,\!}}
\nc{\Oneop}[1][]{\TOMm{\op{\mathbbm{1}} \ltx{#1} \,\!}}
\nc{\trace}{\text{tr }}
\nc{\rank}{\text{rank }}
\nc{\card}{\text{card }}
\rnc{\det}[1][]{\text{det}\unemptynumberone{#1}{#1}\;}
\nc{\Repart}{\mathbb{R}\text{e}\:}
\nc{\Impart}{\mathbb{I}\text{m}\:}

\nc{\interior}[1]{\TOMm{\text{int } #1}}
\nc{\closure}[1]{\TOMm{\ovl{#1}}}
\nc{\kernel}{\text{Ker}\;}
\nc{\image}{\text{Im}\;}
\nc{\supp}{\text{Supp}\;}
\nc{\domain}{\text{Dom}\;}

\nc{\eq}{\, = \,}
\nc{\neqq}{\, \neq \,}
\nc{\defeq}{\, := \,}
\nc{\eqdef}{\, =: \,}
\nc{\eqos}[2][]{\,\uns{\ovs{=}{#2}}{#1}\,}
\nc{\eqostx}[2][]{\eqos[\text{\tiny#1}]{\text{\tiny#2}} }
\nc{\eqosref}[2][]{\emptynumberone{#1}{\eqostx{\eqref{#2}}}{\eqostx[\eqref{#1}]{\eqref{#2}}} }
\nc{\eqx}{\eqos{!}}
\nc{\eqq}{\eqos{?}}

\nc{\equivw}{\,\equiv\,}
\nc{\nequiv}{\slashed{\equiv}}
\nc{\nequivw}{\,\nequiv\,}

\nc{\eqapprox}{\simeq}
\nc{\verysmall}{\smaa{\ll}}
\nc{\verylarge}{\smaa{\gg}}

\nc{\skewperp}{\enmat{\mathrel{\vcenter{\offinterlineskip\hbox{$\smaaa{\:\slash}$}\vskip.05ex\hbox{$\smaaa{\_\_}$}}}}}

\nc{\simeqdif}{\ovs{\smaaa\simeq}{}}
\declareslashed{}{\text{\tiny dif}}{0}{0.9}{\simeqdif}
\nc{\diffeomorph}{\enmat{\,\slashed{\simeqdif}\,}}

\nc{\simeqiso}{\ovs{\smaaa\simeq}{}}
\declareslashed{}{\text{\tiny iso}}{0}{0.9}{\simeqiso}
\nc{\isomorph}{\enmat{\,\slashed{\simeqiso}\,}}

\nc{\Diffgroup}[2][]{\TOMm{\text{Diff}\unemptynumberone{#1}{^{^{\,}#1}}(#2)}}	

\nc{\smallplus}{{\smaaa+}}
\nc{\smallminus}{{\smaaa-}}
\declareslashed{}{{\smaaa\circ}}{0}{0}{\smallplus}
\declareslashed{}{{\smaaa\circ}}{0}{0}{\smallminus}
\nc{\preseta} {\phantom{i}}
\nc{\premseta}{\phantom{i}}
\declareslashed{}{\slashed{\smallplus}} {0}{-0.2}{\preseta}
\declareslashed{}{\slashed{\smallminus}}{0}{-0.2}{\premseta}
\nc{\seta} {\slashed{\preseta}}
\nc{\mseta}{\slashed{\premseta}}					

\nc{\presplus} {\phantom{i}}
\nc{\presminus}{\phantom{i}}
\declareslashed{}{\smallplus} {0}{-0.2}{\presplus}
\declareslashed{}{\smallminus}{0}{-0.2}{\presminus}
\nc{\splus} {\slashed{\presplus}}
\nc{\sminus}{\slashed{\presminus}}					

\nc{\pn}{\! + \!}
\nc{\mn}{\! - \!}
\nc{\pmn}{\! \pm \!}
\nc{\mpn}{\! \mp \!}
\nc{\pns}{\smaaa{\pn}}
\nc{\mns}{\smaaa{\mn}}
\nc{\pms}{{\smaaa\pm}}
\nc{\mps}{{\smaaa\mp}}
\nc{\len}{\! < \!}
\nc{\gen}{\! > \!}
\nc{\crossn}{\! \cross \!}
\nc{\crosss}{{\smaaa\cross}}
\nc{\crossns}{{\smaaa\crossn}}

\nc{\centerdots}{\phantom{i}}						
\declareslashed{}{{\smaaa\centerdot}}{0.4}{0.36}{\centerdots}
\nc{\dotpro}{\slashed{\centerdots}}
\nc{\dotprol}{\!\slashed{\centerdots}\,}

\nc{\limepszero}{\uns{\lim}{\epsilon \rightarrow 0} \,}
\nc{\limargs}[2]{\uns{\lim}{#1 \rightarrow #2} \,}
\nc{\limzero}[1]{\limargs{#1}{0}}
\nc{\liminfty}[1]{\limargs{#1}{\infty}}
\nc{\limpinfty}[1]{\limargs{#1}{+\infty}}
\nc{\limminfty}[1]{\limargs{#1}{-\infty}}

\nc{\maps}{\TOM{:\;}{:\;\;}}
\rnc{\to}{\TOM{$\,\rightarrow\,$}{\;\rightarrow\;}}
\nc{\lto}{\;\longrightarrow\;}
\nc{\toar}[1]{\;\ovs{\rightarrow}{#1}\;}
\nc{\ltoar}[1]{\;\ovs{\longrightarrow}{#1}\;}
\nc{\toiso}{\ltoa{\text{iso}}}
\nc{\mappedto}{\;\mapsto\;}
\nc{\mappedtoar}[1]{\;\ovs{\mapsto}{#1}\;}
\nc{\lmappedtoar}[1]{\;\ovs{\longmapsto}{#1}\;}
\nc{\krarrow}{\;\rightsquigarrow\;}
\nc{\krarrowar}[1]{\;\ovs{\rightsquigarrow}{#1}\;}

\nc{\mapsn}{:\,}
\nc{\ton}{\rightarrow}
\nc{\lton}{\longrightarrow}
\nc{\toarn}[1]{\ovs{\rightarrow}{#1}}
\nc{\ltoarn}[1]{\ovs{\longrightarrow}{#1}}
\nc{\toison}{\ltoa{\text{iso}}}
\nc{\mappedton}{\mapsto}
\nc{\mappedtoarn}[1]{\ovs{\mapsto}{#1}}
\nc{\lmappedtoarn}[1]{\ovs{\longmapsto}{#1}}
\nc{\krarrown}{\rightsquigarrow}
\nc{\krarrowarn}[1]{\ovs{\rightsquigarrow}{#1}}

\rnc{\implies}{\; \Rightarrow\;}
\nc{\Implies}{\; \Longrightarrow \;}
\nc{\impliesar}[2][]{\; \uns{\ovs{\Rightarrow}{#2}}{#1} \;}
\nc{\Impliesar}[2][]{\; \uns{\ovs{\Longrightarrow}{#2}}{#1} \;}

\rnc{\iff}{\; \Leftrightarrow \;}
\nc{\Iff}{\; \Longleftrightarrow \;}
\nc{\iffar}[2][]{\; \uns{\ovs{\Leftrightarrow}{#2}}{#1} \;}
\nc{\Iffar}[2][]{\; \uns{\ovs{\Longleftrightarrow}{#2}}{#1} \;}

\nc{\included}[1]{ \; \ovs{\hookrightarrow}{#1} \;}

\nc{\order}[1]{\TOMm{O\ar{#1}}}
\nc{\contradiction}{!\text{\ssiz CONTRADICTION}! }

\nc{\difpower}[1]{\unemptynumberone{#1}{ ^{^{_{#1}}} }}
\nc{\dif}[1][]{\TOMm{\text{d}\difpower{#1} }}
\nc{\vol}[2][]{\TOMm{\dif[#1] #2 \;}}
\nc{\volvc}[2][]{\TOMm{\dif[#1] \vc{#2} \;}}
\nc{\fundif}[1]{\TOMm{\mc{D} #1 \;\,}}
\nc{\fundifar}[2]{\TOMm{\mc{D} #1 \ar{#2} \;\,}}
\nc{\intfundif}[1]{\int\!\!\fundif{#1}}
\nc{\intfundifar}[2]{\int\!\!\fundifar{#1}{#2}}

\nc{\hidot}{^{^{_\centerdot}}}

\nc{\toder}[2][]{\TOMm{\fracw{\dif #1}{\dif #2}}}								
\nc{\toders}[2][]{\TOMm{\smaaa{\toder[#1]{#2}}}}								
\nc{\toderat}[3][]{\TOM{$\fracw{\dif #1}{\dif #2} \!\bigr|\ls{#2=#3}$\xspace}		
				   {\fracw{\dif #1}{\dif #2} \!\biiigr|\ls{#2=#3} \,}}		
\nc{\toderats}[3][]{\TOMm{\smaaa{\fracw{\dif #1}{\dif #2}\!} \bigr|\ls{#2=#3} }}

\nc{\todertwo}[2][]{\TOMm{\fracw{\dif[2] #1}{\dif #2\difpower{2}}}}				
\nc{\todertwos}[2][]{\TOMm{\smaaa{\todertwo[#1]{#2}}}}								
\nc{\todertwoat}[3][]{\TOM{$\fracw{\dif[2] #1}{\dif #2\difpower{2}} \!\bigr|\ls{#2=#3}$\xspace}
					  {\fracw{\dif[2] #1}{\dif #2\difpower{2}} \!\biiigr|\ls{#2=#3} \,}}			
\nc{\todertwoats}[3][]{\TOMm{\smaaa{\fracw{\dif[2] #1}{\dif #2\difpower{2}}\!} \bigr|\ls{#2=#3} }}

\nc{\del}[1][]{\partial \unemptynumberone{#1}{\ls{#1}} }
\nc{\dell}{\del\lo}
\nc{\delh}{\del\hi}
\nc{\delco}[2]{\del\lo{#1^{#2}}}
\nc{\vcdel}[1][]{\vc{\del\hs{\!}} \unemptynumberone{#1}{\ls{\vc{#1}}} }
\nc{\vcdelsq}[1][]{\vc{\del\hs{\!}}^{\,2} \unemptynumberone{#1}{\ls{\vc{#1}}} }
\nc{\tivcdel}[1][]{\,\tilde{\!\vcdel} \unemptynumberone{#1}{\ls{\vc{#1}}}{\smaaaa\,}}

\nc{\pader}[2][]{\TOMm{\fracw{\del #1}{\del #2}}}								
\nc{\paders}[2][]{\TOMm{\smaaa{\pader[#1]{#2}}}}
\nc{\paderat}[3][]{\TOM{$\fracw{\del #1}{\del #2} \!\bigr|\ls{#3}$\xspace}
				    {\fracw{\del #1}{\del #2} \!\biiigr|\ls{#3} \,}}
\nc{\paderats}[3][]{\TOMm{\smaaa{\fracw{\del #1}{\del #2}\!} \bigr|\ls{#3} }}

\nc{\paderb}[2]{\TOMm{\biiiglrr{^{\!}\frac{\del #1}{\del #2}^{\!}}} }			
\nc{\paderbs}[2]{\TOMm{\smaaa{\paderb[#1]{#2}}}}
\nc{\paderbat}[3]{\TOM{$\biglrr{^{\!}\frac{\del #1}{\del #2}^{\!}} \ls{\!#3}$\xspace}
					{\biiiglrr{^{\!}\frac{\del #1}{\del #2}^{\!}} \ls{\!#3}\,}}
\nc{\paderbats}[3]{\TOMm{\biglrr{\smaaa{\frac{\del #1}{\del #2}\!}} \ls{\!#3} }}

\nc{\paderB}[2]{\TOMm{\biiiiglrr{\!\frac{\del #1}{\del #2}\!}} }				
\nc{\paderBs}[2]{\TOMm{\smaaa{\paderbb[#1]{#2}}}}
\nc{\paderBat}[3]{\TOM{$\biiglrr{\!\frac{\del #1}{\del #2}\!} \ls{\!#3}$\xspace}
					{\biiiiglrr{\!\frac{\del #1}{\del #2}\!} \ls{\!#3}\,}}
\nc{\paderBats}[3]{\TOMm{\biiglrr{\smaaa{\frac{\del #1}{\del #2}\!}} \ls{\!#3} }}

\nc{\fuder}[2][]{\TOMm{\fracw{\delta #1}{\delta #2}}}							
\nc{\fuders}[2][]{\TOMm{\smaaa{\fuder[#1]{#2}}}}								
\nc{\fuderat}[3][]{\TOMm{\fracw{\delta #1}{\delta #2}\biiiigr|_{#3}}}						

\nc{\covder}{\nabla} 

\nc{\dirdeltaexponent}[1]{\unemptynumberone{#1}{^{^{(\!#1\!)}\!\!}}}		
\nc{\dirdeltaexponenttx}[1]{\unemptynumberone{#1}{^{\smaaa{(\!#1\!)\!}}}}

\nc{\krodelta}[3][]{\TOMm{\delta\dirdeltaexponent{#1}\lo{#2#3}}}				
\nc{\krodeltahi}[3][]{\TOMm{\delta\dirdeltaexponent{#1}\hi{#2,#3}}}				
\nc{\krotensor}[2]{\TOMm{\delta \hilo{#1}{#2}}}								
\nc{\dirdelta}[2][]{\TOM{$\delta\dirdeltaexponenttx{#1} \ar{#2}$\xspace}		
					{ \delta\dirdeltaexponent{#1}   \ar{#2}}}			
\nc{\dirdeltaab}[1]{\TOMm{\delta \ab{#1}\,}}								

\nc{\mb}[1]{\TOMm{\mathbf{#1} }}
\nc{\mbb}[1]{\TOMm{\mathbb{#1} }}
\nc{\mbbm}[1]{\TOMm{\mathbbm{#1} }}
\nc{\mc}[1]{\TOMm{\mathcal{#1} }}
\nc{\mf}[1]{\TOMm{\mathfrak{#1} }}

\nc{\lagdens}{\mc{L}}										
\nc{\lagdensar}[1]{\TOMm{\mc{L} \ar{#1} }}					
\nc{\lagfunc}{\TOMm{L}}										
\nc{\lagfuncar}[1]{\TOMm{L \ar{#1} }}						
\nc{\hamdens}{\mc{H}}										
\nc{\hamdensar}[1]{\TOMm{\mc{H} \ar{#1} }}					
\nc{\hamfunc}{\TOMm{H }}										
\nc{\hamfuncar}[1]{\TOMm{H \ar{#1} }}						

\nc{\emtens}{\TOMm{\mc{T}}}	

\nc{\Sactionar} [2][]{\TOMm{S \ls{#1} \ab{#2} }}
\nc{\SactionArg}[2][]{\TOMm{S \ls{#1} \Argb{#2} }}
\nc{\Saction}   [1][]{\Sactionar[#1]{}}							

\nc{\scatmat}[1][]{\TOMm{\mc{S}\unemptynumberone{#1}{\ls{#1}}}}						

\nc{\timeevol}[1][]{\TOMm{\mc{U}\unemptynumberone{#1}{\lbra{#1}}}}	
\nc{\timeorder}{\opT}										

\nc{\hilb}[1][]{\TOMm{\mathpzc{H} \ls{\,#1} }}					
\nc{\hilbdual}[1][]{\TOMm{\mathpzc{H}^* \ls{\,#1} }}
\nc{\hilbtx}[1]{\hilb[\text{\tiny#1}]}
\nc{\hilbdualtx}[1]{\hilbdual[\text{\tiny#1}]}
\nc{\hilbin}{\hilbtx{in}}
\nc{\hilbdualtin}{\hilbdualtx{in}}
\nc{\hilbout}{\hilbtx{out}}
\nc{\hilbdualtout}{\hilbdualtx{out}}

\nc{\configspace}{\TOMt{C}}

\nc{\rep}[2][]{\TOMm{\op{\mc{R}}\ls{#1} \Arg{#2} }}				

\nc{\laplacian}[1][]{\TOMm{\bigtriangleup \ls{#1}\,\! }}
\nc{\dalembertian}[1][]{\TOMm{\Box \ls{#1}\,\! }}
\nc{\beltrami}[1][]{\TOMm{\Box \ls{#1}\,\! }}

\nc{\wronskian}[1]{\TOMm{W(#1)}}

\nc{\reals}[1][]{\TOMm{\mathbb{R} \hs{#1} }}
\nc{\realspos}{\TOMm{\mathbb{R} \hs{+} }}
\nc{\realsposzero}{\TOMm{\mathbb{R} \hs{+}\ls{0} }}
\nc{\realproj}[1][]{\TOMm{\mathbb{RP} \hs{#1} }}
\nc{\complex}[1][]{\TOMm{\mathbb{C} \hs{#1} }}
\nc{\comproj}[1][]{\TOMm{\mathbb{CP} \hs{#1} }}
\nc{\krecom}[1][]{\TOMm{\mathbb{K} \hs{#1} }}

\nc{\integers}[1][]{\TOMm{\mathbb{Z} \hs{#1} }}
\nc{\naturals}[1][]{\TOMm{\mathbb{N} \hs{#1} }}
\nc{\naturalszero} {\TOMm{\mathbb{N} \ls{0}  }}
\nc{\rationals}[1][]{\TOMm{\mathbb{Q}\hs{#1} }}

\nc{\Zleq}[1]{\integers\,\!\hs{\leq}\!\ar{#1}}
\nc{\Zgeq}[1]{\integers\,\!\hs{\geq}\!\ar{#1}}

\nc{\sphere}[1][]{\TOMm{\mathbb{S} \hs{#1} }}
\nc{\spherel}[2][]{\TOMm{\mathbb{S} \hs{#1}\ls{#2} }}
\nc{\ball}  [1][]{\TOMm{\mathbb{B} \hs{#1} }}
\nc{\balll}  [2][]{\TOMm{\mathbb{B} \hs{#1}\ls{#2} }}
\nc{\disc}  [1][]{\TOMm{\mathbb{D} \hs{#1} }}
\nc{\discl}  [2][]{\TOMm{\mathbb{D} \hs{#1}\ls{#2} }}

\nc{\Cdifar}[2]{\TOMm{C\unemptynumberone{#1}{^{#1}}\unemptynumberone{#2}{(#2)}}}	
\nc{\Cdifarx}[2]{\enmat{C^{#1}(#2)}}										
\nc{\Cdiffar}[2]{\TOMm{C\unemptynumberone{#1}{^{#1}}\AArg{#2}}}
\nc{\Cdif}[1]{\Cdifar{#1}{}}
\nc{\Cdifx}[1]{\Cdifarx{#1}{}}

\nc{\detgdef}[1][g]{{\smaaa{ #1\defeq\det #1_{\mu\nu}}}}

\nc{\flowsymbol}{\mc{F}}
\declareslashed{}{}{0}{0}{\flowsymbol}
\nc{\flow}[2]{\enmat{\slashed{\flowsymbol}_{\!#1}#2\,} }

\nc{\christoffel}[3]{\TOMm{\Ga\hi{#1}\lo{\!#2#3}}}
\nc{\christoffello}[3]{\TOMm{\Ga\looo{#1#2#3}}}

\nc{\riemann}{\TOMm{\mc{R}}}
\nc{\riemannhi}[4]{\TOMm{\riemann\hilo{#1}{#2#3#4}}}
\nc{\riemannlo}[4]{\TOMm{\riemann\lo{#1#2#3#4}}}

\nc{\ricci}{\TOMt{ric}}
\nc{\Ricci}{\TOMt{Ric}}
\nc{\riccisca}{\Ya}

\nc{\GLgroup}[2][]{\TOMm{\txG\txL\unemptynumberone{#1}{_{#1}} \Arg{#2}} }
\nc{\SLgroup}[2][]{\TOMm{\txS\txL\unemptynumberone{#1}{_{#1}} \Arg{#2}} }
\nc{\Ogroup}[2][]{\TOMm{\txO\unemptynumberone{#1}{_{#1}} \Arg{#2}} }
\nc{\SOgroup}[2][]{\TOMm{\txS\txO\unemptynumberone{#1}{_{#1}} \Arg{#2}} }
\nc{\Ugroup}[2][]{\TOMm{\txU\unemptynumberone{#1}{_{#1}} \Arg{2}} }
\nc{\SUgroup}[2][]{\TOMm{\txS\txU\unemptynumberone{#1}{_{#1}} \Arg{2}} }
\nc{\Spgroup}[2][]{\TOMm{\txS\txp\unemptynumberone{#1}{_{#1}} \Arg{#2}} }

\nc{\forrallel}[2]{\forall \; #1 \unemptynumberone{#2}{\elof #2}}			
\nc{\forrallels}[2]{{\smaa{\forall \; #1 \unemptynumberone{#2}{\elofs #2}}}}
\nc{\forrall}[1]{\forrallel{#1}{}}
\nc{\forralls}[1]{\forrallels{#1}{}}
\nc{\forrallelhilo}[4]{^{\forrallels{#1}{#2}}_{\forrallels{#3}{#4}}}	
\nc{\forrallhilo}[2]{^{\forralls{#1}}_{\forralls{#2}}}	

\nc{\notni}{\slashed{\ni}}

\nc{\smallin}{\phantom{\in}}
\nc{\smallins}{\phantom{\in}}
\nc{\smallni}{\phantom{\ni}}
\nc{\smallnis}{\phantom{\ni}}
\nc{\smallnotin}{\phantom{\notin}}
\nc{\smallnotins}{\phantom{\notin}}
\nc{\smallnotni}{\phantom{\notni}}
\nc{\smallnotnis}{\phantom{\notni}}

\nc{\smallsubset}{\phantom{\subset}}
\nc{\smallsupset}{\phantom{\supset}}
\nc{\smallsubseteq}{\phantom{\subseteq}}
\nc{\smallsupseteq}{\phantom{\supseteq}}
\nc{\smallsubsetneq}{\phantom{\subsetneq}}
\nc{\smallsupsetneq}{\phantom{\supsetneq}}

\declareslashed{}{{\smaaa\in}}{0}{0.1}{\smallin}
\declareslashed{}{{\smaaaa\in}}{0}{0.02}{\smallins}
\declareslashed{}{{\smaaa\ni}}{0}{0.1}{\smallni}
\declareslashed{}{{\smaaaa\ni}}{0}{0}{\smallnis}
\declareslashed{}{{\smaaa\notin}}{0}{0.02}{\smallnotin}
\declareslashed{}{{\smaaaa\notin}}{0}{-0.01}{\smallnotins}
\declareslashed{}{{\smaaa\notni}}{0}{0.07}{\smallnotni}
\declareslashed{}{{\smaaaa\notni}}{0}{0.02}{\smallnotnis}

\declareslashed{}{{\smaaa\subset}}{0}{0.2}{\smallsubset}
\declareslashed{}{{\smaaa\supset}}{0}{0.2}{\smallsupset}
\declareslashed{}{{\smaaa\subseteq}}{0}{0.05}{\smallsubseteq}
\declareslashed{}{{\smaaa\supseteq}}{0}{0.05}{\smallsupseteq}
\declareslashed{}{{\smaaa\subsetneq}}{0}{0.05}{\smallsubsetneq}
\declareslashed{}{{\smaaa\supsetneq}}{0}{0.05}{\smallsupsetneq}

\nc{\elof}{\,\slashed{\smallin}\,}							
\nc{\elofs}{\,\slashed{\smallins}\,}
\nc{\fole}{\,\slashed{\smallni}\,}						
\nc{\foles}{\,\slashed{\smallnis}\,}						
\nc{\nelof}{\,\slashed{\smallnotin}\,}						
\nc{\nelofs}{\,\slashed{\smallnotins}\,}						
\nc{\nfole}{\,\slashed{\smallnotni}\,}						
\nc{\nfoles}{\,\slashed{\smallnotnis}\,}						

\nc{\suubset}{\,\slashed{\smallsubset}\,}
\nc{\suupset}{\,\slashed{\smallsupset}\,}
\nc{\suubseteq}{\,\slashed{\smallsubseteq}\,}
\nc{\suupseteq}{\,\slashed{\smallsupseteq}\,}
\nc{\suubsetneq}{\,\slashed{\smallsubsetneq}\,}
\nc{\suupsetneq}{\,\slashed{\smallsupsetneq}\,}

\nc{\Arg}[1]    {\unemptynumberone{#1}{\left( #1 \right)}}
\nc{\AArg}[1]   {\unemptynumberone{#1}{\bigl( #1 \bigr)}}
\nc{\AAArg}[1]  {\unemptynumberone{#1}{\biigl( #1 \biigr)}}
\nc{\AAAArg}[1] {\unemptynumberone{#1}{\biiigl( #1 \biiigr)}}
\nc{\AAAAArg}[1]{\unemptynumberone{#1}{\biiiigl( #1 \biiiigr)}}

\nc{\ar}[1]    {{\smaa{(#1)} }}
\nc{\ars}[1]   {{\unemptynumberone{#1}{\smaaa{(#1)}} }}
\nc{\arvc}[1]    {{\unemptynumberone{#1}{\smaa{(\vc #1)}} }}
\nc{\arvcp}[1]    {{\unemptynumberone{#1}{\smaa{(\vc #1')}} }}
\nc{\arvcm}[1]    {{\unemptynumberone{#1}{\smaa{(-\vc #1)}} }}

\nc{\Argb}[1]{\unemptynumberone{#1}{\left[#1\right]}}
\nc{\ab}[1]{{\unemptynumberone{#1}{\smaa{[#1]}} }}

\nc{\artx}{\ar{t,\vcx}}

\nc{\insertion}[1]{\text{ins}\unemptynumberone{#1}{\Arg{#1}}}

\nc{\sumlim}   [2]{\sum  \limits_{#1}^{#2}}
\nc{\sumliml}  [1]{\sumlim{#1}{}}
\nc{\sumlims}  [2]{\sum  \limits_{\smaaa{#1}}^{\smaaa{#2}}}
\nc{\sumlimls} [1]{\sumlims{#1}{}}
\nc{\prodlim}  [2]{\prod \limits_{#1}^{#2}}
\nc{\prodliml} [1]{\prodlim{#1}{}}
\nc{\prodlims} [2]{\prod  \limits_{\smaaa{#1}}^{\smaaa{#2}}}
\nc{\prodlimls}[1]{\prodlims{#1}{}}

\newlength{\intlimlength}						
\newlength{\intsymlength}
\nc{\intlim} [2]{\settowidth{\intlimlength}{${\smaa{ #1}}$}
			\ifte{\intlimlength > \intsymlength} {\setlength{\intlimlength}{0.5\intlimlength}} {\setlength{\intlimlength}{0.5\intsymlength}}
			\int\limits_{#1}^{#2} \, \hspace{-\intlimlength} }
\nc{\intlims} [2]{\settowidth{\intlimlength}{${\smaaa{#1}}$}				
			\ifte{\intlimlength > \intsymlength} {\setlength{\intlimlength}{0.5\intlimlength}} {\setlength{\intlimlength}{0.5\intsymlength}}
			\int\limits_{\smaaa{#1}}^{\smaaa{#2}} \, \hspace{-\intlimlength} }
\nc{\intliml}[1]{\intlim{#1}{}}
\nc{\intlimls}[1]{\intlims{#1}{}}
\nc{\intvol}[2][]{\int\!\!\vol[#1]{#2}}
			
\declareslashed{\mathop}{\int}{0.05}{0}{\sum}
\nc{\sumint}{\slashed{\sum}}
\nc{\sumintlim}[2]{\slashed{\sum} \limits_{#1}^{#2}}
\nc{\sumintliml}[1]{\sumintlim{#1}{}}
\nc{\sumintlims}[2]{\slashed{\sum} \limits_{\smaaa{#1}}^{\smaaa{#2}}}
\nc{\sumintlimls}[1]{\sumintlims{#1}{}}

\nc{\tansp}[2]{\TOMm{ \text{T}_{\!#1}#2 }}
\nc{\cotsp}[2]{\TOMm{ \text{T}^{*}_{\!#1}#2 }}
\nc{\tanbun}[1]{\TOMm{ \text{T}#1 }}
\nc{\cotbun}[1]{\TOMm{ \text{T}^{*}\text{#1} }}

\nc{\bundle}[3]{#1 \ovs{\longrightarrow}{#2} #3}
\nc{\bundlepi}[2]{\bundle{#1}{\pi}{#2}}
\nc{\tanbunpro}[1]{\bundle{\txT #1}{\tau_{#1}}{#1}} 
\nc{\cotbunpro}[1]{\bundle{\txT^*#1}{\tau^*_{#1}}{#1}} 

\nc{\canpro}[1]{\text{pr}_{#1}}

\nc{\bessel}[2][]{\TOMm{J_{#2} #1 }}						
\nc{\besselar}[3][]{\TOMm{\bessel[#1]{#2} \ar{#3}}}			
\nc{\neumann}[2][]{\TOMm{N_{#2} #1}}
\nc{\neumannar}[3][]{\TOMm{\neumann[#1]{#2} \ar{#3}}}
\nc{\hankel}[2][]{\TOMm{H_{#2} #1}}
\nc{\hankelar}[3][]{\TOMm{\hankel[#1]{#2} \ar{#3}}}

\nc{\spherbessel}[2][]{\TOMm{j_{#2} #1 }}
\nc{\spherbesselar}[3][]{\TOMm{\spherbessel[#1]{#2} \ar{#3}}}
\nc{\spherneumann}[2][]{\TOMm{n_{#2} #1 }}
\nc{\spherneumannar}[3][]{\TOMm{\spherneumann[#1]{#2} \ar{#3}}}
\nc{\spherhankel}[2][]{\TOMm{h_{#2} #1 }}
\nc{\spherhankelar}[3][]{\TOMm{\spherhankel[#1]{#2} \ar{#3}}}

\nc{\asslegendre}[3][]{\TOMm{{P^{#2}_{#3}} #1 }}
\nc{\asslegendrear}[4][]{\TOMm{\asslegendrefunc[#1]{#2}{3} \ar{#4}}}
\nc{\hypergeo}[5][]{\TOMm{F #1 \ar{#2,\,#3;\,#4;\;#5}}}
\nc{\jacobipoly}[4][]{\TOMm{{P^{(#2,#3)}_{#4}} #1 }}
\nc{\jacobipolyar}[5][]{\TOMm{\jacobipoly[#1]{#2}{#3}{#4} \ar{#5}}}
\nc{\spherharmonic}[3][]{\TOMm{{Y^{#2}_{#3}} #1} }
\nc{\spherharmonicar}[4][]{\TOMm{\spherharmonic[#1]{#2}{#3} \ar{#4}}}

\nc{\pochhammer}[2]{\TOMm{(#1)_{#2}\,}}

\nc{\sinn}[1]{\sin^{#1}\!}
\nc{\sinsq}{\sinn{2}}
\nc{\cosn}[1]{\cos^{#1}\!}
\nc{\cossq}{\cosn{2}}
\nc{\tann}[1]{\tan^{#1}\!}
\nc{\tansq}{\tann{2}}
\nc{\cotn}[1]{\cot^{#1}\!}
\nc{\cotsq}{\tann{2}}

\nc{\sinhn}[1]{\sinh^{#1}\!}
\nc{\sinhsq}{\sinhn{2}}
\nc{\coshn}[1]{\cosh^{#1}\!}
\nc{\coshsq}{\coshn{2}}
\nc{\tanhn}[1]{\tanh^{#1}\!}
\nc{\tanhsq}{\tanhn{2}}
\nc{\cothn}[1]{\coth^{#1}\!}
\nc{\cothsq}{\cothn{2}}

\nc{\abs}[1]{\TOMm{\left| #1 \right|}}
\nc{\abss}[1]{\bigl| #1 \bigr|}
\nc{\absss}[1]{\biigl| #1 \biigr|}
\nc{\abssss}[1]{\biiigl| #1 \biiigr|}
\nc{\absssss}[1]{\biiiigl| #1 \biiiigr|}

\nc{\absq}[1]{\TOMm{\left| #1 \right|^{2}}}
\nc{\abssq}[1]{\bigl| #1 \bigr|^{2}}
\nc{\absssq}[1]{\biigl| #1 \biigr|^{2}}
\nc{\abssssq}[1]{\biiigl| #1 \biiigr|^{2}}
\nc{\absssssq}[1]{\biiiigl| #1 \biiiigr|^{2}}

\nc{\norm}[2][]{\TOMm{\left| \! \left| \, #2 \, \right| \! \right|\ls{#1}} }
\nc{\normm}[2][]{\bigl| \! \bigl| \, #2 \, \bigr| \! \bigr|\ls{#1} }
\nc{\normmm}[2][]{\biigl| \! \biigl| \, #2 \, \biigr| \! \biigr|\ls{#1} }
\nc{\normmmm}[2][]{\biiigl| \! \biiigl| \, #2 \, \biiigr| \! \biiigr|\ls{#1} }
\nc{\normmmmm}[2][]{\biiiigl| \! \biiiigl| \, #2 \, \biiiigr| \! \biiiigr|\ls{#1} }

\nc{\normsq}[2][]{\TOMm{\left| \! \left| \, #2 \, \right| \! \right|^{2}\ls{#1} }}
\nc{\normmsq}[2][]{\bigl| \! \bigl| \, #2 \, \bigr| \! \bigr|^{2}\ls{#1} }
\nc{\normmmsq}[2][]{\biigl| \! \biigl| \, #2 \, \biigr| \! \biigr|^{}\ls{#1} }
\nc{\normmmmsq}[2][]{\biiigl| \! \biiigl| \, #2 \, \biiigr| \! \biiigr|^{2}\ls{#1} }
\nc{\normmmmmsq}[2][]{\biiiigl| \! \biiiigl| \, #2 \, \biiiigr| \! \biiiigr|^{2}\ls{#1} }

\newlength{\bralength}
\newlength{\brasymlength}
\newlength{\braasymlength}
\newlength{\braaasymlength}
\newlength{\braaaasymlength}
\newlength{\braaaaasymlength}
\nc{\bra}[2][]{\TOMm{ \settowidth{\bralength}{$\ls{#1}$} \hspace{\bralength}
					\bosym{\langle}\ls{\,\hspace{-\bralength}\hspace{-\brasymlength}{#1}\hspace{\brasymlength}}
					#2 \,\bosym{|}
				}}
\nc{\braa}[2][]{ \settowidth{\bralength}{$\ls{#1}$} \hspace{\bralength}
				\bosym{\bigl\langle}\ls{\hspace{-\bralength}\hspace{-\braasymlength}{#1}\hspace{\braasymlength}}
				 #2 \,\bosym{\bigr|} }
\nc{\braaa}[2][]{ \settowidth{\bralength}{$\ls{#1}$} \hspace{\bralength}
				\bosym{\biigl\langle}\ls{\hspace{-\bralength}\hspace{-\braaasymlength}{#1}\hspace{\braaasymlength}}
				 #2 \,\bosym{\biigr|} }
\nc{\braaaa}[2][]{ \settowidth{\bralength}{$\ls{#1}$} \hspace{\bralength}
				\bosym{\biiigl\langle}\ls{\,\hspace{-\bralength}\hspace{-\braaaasymlength}{#1}\hspace{\braaaasymlength}}
				 #2 \,\bosym{\biiigr|} }
\nc{\braaaaa}[2][]{ \settowidth{\bralength}{$\ls{#1}$} \hspace{\bralength}
				\bosym{\biiiigl\langle}\ls{\:\hspace{-\bralength}\hspace{-\braaaaasymlength}{#1}\hspace{\braaaaasymlength}}
				 #2 \,\bosym{\biiiigr|} }

\nc{\braout}[1]{\bra[\text{out}]{#1}}
\nc{\braaout}[1]{\braa[\text{out}]{#1}}
\nc{\braaaout}[1]{\braaa[\text{out}]{#1}}
\nc{\braaaaout}[1]{\braaaa[\text{out}]{#1}}
\nc{\braaaaaout}[1]{\braaaaa[\text{out}]{#1}}

\nc{\ket}[2][]{\TOMm{\bosym{|}\, #2 \bosym{\rangle}\ls{#1} }}
\nc{\kett}[2][]{\bosym{\bigl|}\, #2 \bosym{\bigr\rangle}\ls{#1} }
\nc{\kettt}[2][]{\bosym{\biigl|}\, #2 \bosym{\biigr\rangle}\ls{\!#1} }
\nc{\ketttt}[2][]{\bosym{\biiigl|}\, #2 \bosym{\biiigr\rangle}\ls{\!\!#1} }
\nc{\kettttt}[2][]{\bosym{\biiiigl|}\, #2 \bosym{\biiiigr\rangle}\ls{\!\!#1} }

\nc{\ketin}[1]{\ket[\text{in}]{#1}}
\nc{\kettin}[1]{\kett[\text{in}]{#1}}
\nc{\ketttin}[1]{\kettt[\text{in}]{#1}}
\nc{\kettttin}[1]{\ketttt[\text{in}]{#1}}
\nc{\ketttttin}[1]{\kettttt[\text{in}]{#1}}

\nc{\ketout}[1]{\ket[\text{out}]{#1}}
\nc{\kettout}[1]{\kett[\text{out}]{#1}}
\nc{\ketttout}[1]{\kettt[\text{out}]{#1}}
\nc{\kettttout}[1]{\ketttt[\text{out}]{#1}}
\nc{\ketttttout}[1]{\kettttt[\text{out}]{#1}}

\nc{\braket}[3][]{\TOMm{\bosym{\langle} #2 \,\bosym{|}\, #3 \bosym{\rangle}\ls{#1} }}
\nc{\brakett}[3][]{\bosym{\bigl\langle} #2 \,\bosym{\bigl|}\, #3 \bosym{\bigr\rangle}\ls{#1} }
\nc{\brakettt}[3][]{\bosym{\biigl\langle} #2 \,\bosym{\biigl|}\, #3 \bosym{\biigr\rangle}\ls{\!#1} }
\nc{\braketttt}[3][]{\bosym{\biiigl\langle} #2 \,\bosym{\biiigl|}\, #3 \bosym{\biiigr\rangle}\ls{\!\!#1} }
\nc{\brakettttt}[3][]{\bosym{\biiiigl\langle} #2 \,\bosym{\biiiigl|}\, #3 \bosym{\biiiigr\rangle}\ls{\!\!#1} }

\nc{\braketl}[4]{\TOMm{ \settowidth{\bralength}{$\ls{#1}$} \hspace{\bralength}
					\bosym{\langle}\ls{\,\hspace{-\bralength}\hspace{-\brasymlength}{#1}\hspace{\brasymlength}} #2 \,
					\bosym{|}\, #3 \bosym{\rangle}\ls{#4} }}
\nc{\brakettl}[4]{\settowidth{\bralength}{$\ls{#1}$} \hspace{\bralength}
				\bosym{\bigl\langle}\ls{\hspace{-\bralength}\hspace{-\braasymlength}{#1}\hspace{\braasymlength}}
				#2 \,\bosym{\bigl|}\, #3 \bosym{\bigr\rangle}\ls{#4} }
\nc{\braketttl}[4]{\settowidth{\bralength}{$\ls{#1}$} \hspace{\bralength}
				\bosym{\biigl\langle}\ls{\hspace{-\bralength}\hspace{-\braaasymlength}{#1}\hspace{\braaasymlength}}
				#2 \,\bosym{\biigl|}\, #3 \bosym{\biigr\rangle}\ls{\!#4} }
\nc{\brakettttl}[4]{\settowidth{\bralength}{$\ls{#1}$} \hspace{\bralength}
				\bosym{\biiigl\langle}\ls{\,\hspace{-\bralength}\hspace{-\braaaasymlength}{#1}\hspace{\braaaasymlength}}
				#2 \,\bosym{\biiigl|}\, #3 \bosym{\biiigr\rangle}\ls{\!\!#4} }
\nc{\braketttttl}[4]{\settowidth{\bralength}{$\ls{#1}$} \hspace{\bralength}
				\bosym{\biiiigl\langle}\ls{\:\hspace{-\bralength}\hspace{-\braaaaasymlength}{#1}\hspace{\braaaaasymlength}}
				#2 \,\bosym{\biiiigl|}\,#3\bosym{\biiiigr\rangle}\ls{\!\!#4} }

\nc{\braketinin}[2]{\braketl{\text{in}}{#1}{#2}{\text{in}}}
\nc{\brakettinin}[2]{\brakettl{\text{in}}{#1}{#2}{\text{in}}}
\nc{\braketttinin}[2]{\braketttl{\text{in}}{#1}{#2}{\text{in}}}
\nc{\brakettttinin}[2]{\brakettttl{\text{in}}{#1}{#2}{\text{in}}}
\nc{\braketttttinin}[2]{\braketttttl{\text{in}}{#1}{#2}{\text{in}}}

\nc{\braketoutout}[2]{\braketl{\text{out}}{#1}{#2}{\text{out}}}
\nc{\brakettoutout}[2]{\brakettl{\text{out}}{#1}{#2}{\text{out}}}
\nc{\braketttoutout}[2]{\braketttl{\text{out}}{#1}{#2}{\text{out}}}
\nc{\brakettttoutout}[2]{\brakettttl{\text{out}}{#1}{#2}{\text{out}}}
\nc{\braketttttoutout}[2]{\braketttttl{\text{out}}{#1}{#2}{\text{out}}}

\nc{\braketoutin}[2]{\braketl{\text{out}}{#1}{#2}{\text{in}}}
\nc{\brakettoutin}[2]{\brakettl{\text{out}}{#1}{#2}{\text{in}}}
\nc{\braketttoutin}[2]{\braketttl{\text{out}}{#1}{#2}{\text{in}}}
\nc{\brakettttoutin}[2]{\brakettttl{\text{out}}{#1}{#2}{\text{in}}}
\nc{\braketttttoutin}[2]{\braketttttl{\text{out}}{#1}{#2}{\text{in}}}

\nc{\braopket}[4][]{\TOMm{\bosym{\langle} #2 \,\bosym{|}\, #3 \,\bosym{|}\, #4 \bosym{\rangle}\ls{#1} }}
\nc{\braopkett}[4][]{\bosym{\bigl\langle} #2 \,\bosym{\bigr|}\, #3 \,\bosym{\bigl|}\, #4 \bosym{\bigr\rangle}\ls{#1}}
\nc{\braopkettt}[4][]{\bosym{\biigl\langle} #2 \,\bosym{\biigr|}\, #3 \,\bosym{\biigl|}\, #4 \bosym{\biigr\rangle}\ls{\!#1}}
\nc{\braopketttt}[4][]{\bosym{\biiigl\langle} #2 \,\bosym{\biiigr|}\, #3 \,\bosym{\biiigl|}\, #4 \bosym{\biiigr\rangle}\ls{\!\!#1}}
\nc{\braopkettttt}[4][]{\bosym{\biiiigl\langle} #2 \,\bosym{\biiiigr|}\, #3 \,\bosym{\biiiigl|}\, #4 \bosym{\biiiigr\rangle}\ls{\!\!#1}}

\nc{\braopketl}[5]{\TOMm{ \settowidth{\bralength}{$\ls{#1}$} \hspace{\bralength}
					\bosym{\langle}\ls{\,\hspace{-\bralength}\hspace{-\brasymlength}{#1}\hspace{\brasymlength}} #2 \,
					\bosym{|}\, #3 \,\bosym{|}\, #4 \bosym{\rangle}\ls{#5} }}
\nc{\braopkettl}[5]{\settowidth{\bralength}{$\ls{#1}$} \hspace{\bralength}
				\bosym{\bigl\langle}\ls{\hspace{-\bralength}\hspace{-\braasymlength}{#1}\hspace{\braasymlength}}
				#2 \,\bosym{\bigl|}\, #3 \,\bosym{\bigl|}\,
				#4 \bosym{\bigr\rangle}\ls{#5} }
\nc{\braopketttl}[5]{\settowidth{\bralength}{$\ls{#1}$} \hspace{\bralength}
				\bosym{\biigl\langle}\ls{\hspace{-\bralength}\hspace{-\braaasymlength}{#1}\hspace{\braaasymlength}}
				#2 \,\bosym{\biigl|}\, #3 \,\bosym{\biigl|}\,
				#4 \bosym{\biigr\rangle}\ls{\!#5} }
\nc{\braopkettttl}[5]{\settowidth{\bralength}{$\ls{#1}$} \hspace{\bralength}
				\bosym{\biiigl\langle}\ls{\,\hspace{-\bralength}\hspace{-\braaaasymlength}{#1}\hspace{\braaaasymlength}}
				#2 \,\bosym{\biiigl|}\, #3 \,\bosym{\biiigl|}\,
				#4 \bosym{\biiigr\rangle}\ls{\!\!#5} }
\nc{\braopketttttl}[5]{\settowidth{\bralength}{$\ls{#1}$} \hspace{\bralength}
				\bosym{\biiiigl\langle}\ls{\:\hspace{-\bralength}\hspace{-\braaaaasymlength}{#1}\hspace{\braaaaasymlength}}
				#2 \,\bosym{\biiiigl|}\, #3 \,\bosym{\biiiigl|}\,
				#4 \bosym{\biiiigr\rangle}\ls{\!\!#5} }

\nc{\braopketoutin}[3]{\braopketl{\text{out}}{#1}{#2}{#3}{\text{in}}}
\nc{\braopkettoutin}[3]{\braopkettl{\text{out}}{#1}{#2}{#3}{\text{in}}}
\nc{\braopketttoutin}[3]{\braopketttl{\text{out}}{#1}{#2}{#3}{\text{in}}}
\nc{\braopkettttoutin}[3]{\braopkettttl{\text{out}}{#1}{#2}{#3}{\text{in}}}
\nc{\braopketttttoutin}[3]{\braopketttttl{\text{out}}{#1}{#2}{#3}{\text{in}}}

\nc{\expval}[2][]{\TOMm{\bosym{\langle} #2 \bosym{\rangle}\ls{#1}} }
\nc{\expvall}[2][]{\bosym{\bigl\langle} #2 \bosym{\bigr\rangle}\ls{#1}}
\nc{\expvalll}[2][]{\bosym{\biigl\langle} #2 \bosym{\biigr\rangle}\ls{\!#1}}
\nc{\expvallll}[2][]{\bosym{\biiigl\langle} #2 \bosym{\biiigr\rangle}\ls{\!#1}}
\nc{\expvalllll}[2][]{\bosym{\biiiigl\langle} #2 \bosym{\biiiigr\rangle}\ls{\!\!#1}}

\nc{\inpro}[3][]{\TOMm{\bosym{\langle} #2 \bosym{,} \; #3 \bosym{\rangle}\ls{#1}} }
\nc{\inproo}[3][]{\bosym{\bigl\langle} #2 \bosym{,} \;\, #3 \bosym{\bigr\rangle}\ls{#1}}
\nc{\inprooo}[3][]{\bosym{\biigl\langle} #2 \bosym{,} \;\; #3 \bosym{\biigr\rangle}\ls{\!#1}}
\nc{\inproooo}[3][]{\bosym{\biiigl\langle} #2 \bosym{,} \;\; #3 \bosym{\biiigr\rangle}\ls{\!#1}}
\nc{\inprooooo}[3][]{\bosym{\biiiigl\langle} #2 \bosym{,} \;\; #3 \bosym{\biiiigr\rangle}\ls{\!\!#1}}

\nc{\pairing}[3][]{\TOMm{\bosym{(} #2 \bosym{,} \, #3 \bosym{)}\ls{#1}}}
\nc{\pairingg}[3][]{\bosym{\bigl(} #2 \bosym{,} \, #3 \bosym{\bigr)}\ls{#1}}
\nc{\pairinggg}[3][]{\bosym{\biigl(} #2 \bosym{,} \, #3 \bosym{\biigr)}\ls{#1}}
\nc{\pairingggg}[3][]{\bosym{\biiigl(} #2 \bosym{,} \, #3 \bosym{\biiigr)}\ls{\!#1}}
\nc{\pairinggggg}[3][]{\bosym{\biiiigl(} #2 \bosym{,} \, #3 \bosym{\biiiigr)}\ls{\!\!#1}}

\nc{\liebra}[2]{\TOMm{\bosym{[} #1 \bosym{,} \; #2 \bosym{]}\ltxs{Lie}}}			
\nc{\liebraa}[2]{\bosym{\bigl[} #1 \bosym{,} \;\, #2 \bosym{\bigr]}\ltxs{Lie}}
\nc{\liebraaa}[2]{\bosym{\biigl[} #1 \bosym{,} \;\; #2 \bosym{\biigr]}\ltxs{Lie}}
\nc{\liebraaaa}[2]{\bosym{\biiigl[} #1 \bosym{,} \;\; #2 \bosym{\biiigr]}\ltxs{Lie}}
\nc{\liebraaaaa}[2]{\bosym{\biiiigl[} #1 \bosym{,} \;\; #2 \bosym{\biiiigr]}\ltxs{Lie}}

\nc{\lieder}[1]{\TOMm{\opL_{#1}} }		
\nc{\liedervc}[1]{\TOMm{\opL\lvc{#1}} }	

\nc{\poibra}[2]{\TOMm{\bosym{\{} #1 \bosym{,} \; #2 \bosym{\}}_{_{\!\textproto{\footnotesize\Adaleth}}}} }				
\nc{\poibraa}[2]{\bosym{\bigl\{} #1 \bosym{,} \;\, #2 \bosym{\bigr\}}_{_{\!\!\textproto{\footnotesize\Adaleth}}} }
\nc{\poibraaa}[2]{\bosym{\biigl\{} #1 \bosym{,} \;\; #2 \bosym{\biigr\}}_{_{\!\!\textproto{\footnotesize\Adaleth}}} }
\nc{\poibraaaa}[2]{\bosym{\biiigl\{} #1 \bosym{,} \;\; #2 \bosym{\biiigr\}}_{\!\!\textproto{\footnotesize\Adaleth}} }
\nc{\poibraaaaa}[2]{\bosym{\biiiigl\{} #1 \bosym{,} \;\; #2 \bosym{\biiiigr\}}_{\!\!\textproto{\footnotesize\Adaleth}} }

\nc{\commu}[2]{\TOMm{\bosym{[} #1 \bosym{,} \; #2 \bosym{]}\ls{-}} }
\nc{\commuu}[2]{\bosym{\bigl[} #1 \bosym{,} \;\, #2 \bosym{\bigr]}\ls{-}}
\nc{\commuuu}[2]{\bosym{\biigl[} #1 \bosym{,} \;\; #2 \bosym{\biigr]}\ls{-}}
\nc{\commuuuu}[2]{\bosym{\biiigl[} #1 \bosym{,} \;\; #2 \bosym{\biiigr]}\ls{-}}
\nc{\commuuuuu}[2]{\bosym{\biiiigl[} #1 \bosym{,} \;\; #2 \bosym{\biiiigr]}\ls{-}}

\nc{\acommu}[2]{\TOMm{\bosym{\{} #1 \bosym{,} \; #2 \bosym{\}}\ls{+}} }
\nc{\acommuu}[2]{\bosym{\bigl\{} #1 \bosym{,} \;\, #2 \bosym{\bigr\}}\ls{+}}
\nc{\acommuuu}[2]{\bosym{\biigl\{} #1 \bosym{,} \;\; #2 \bosym{\biigr\}}\ls{+}}
\nc{\acommuuuu}[2]{\bosym{\biiigl\{} #1 \bosym{,} \;\; #2 \bosym{\biiigr\}}\ls{+}}
\nc{\acommuuuuu}[2]{\bosym{\biiiigl\{} #1 \bosym{,} \;\; #2 \bosym{\biiiigr\}}\ls{+}}

\nc{\set}[2][]{\TOMm{ \{#2\}\ls{#1}}}
\nc{\sett}[2][]{\bigl\{ #2 \bigr\}\ls{#1} }
\nc{\settt}[2][]{\biigl\{ #2 \biigr\}\ls{\!#1} }
\nc{\setttt}[2][]{\biiigl\{ #2 \biiigr\}\ls{\!#1} }
\nc{\settttt}[2][]{\biiiigl\{ #2 \biiiigr\}\ls{\!\!#1} }

\nc{\setc}[2]{\TOMm{\left\{ #1 \; \middle| \; #2 \right\} }}
\nc{\settc}[2]{\bigl\{ #1 \; \bigr| \; #2 \bigr\} }
\nc{\setttc}[2]{\biigl\{ #1 \; \biigr| \;#2 \biigr\} }
\nc{\settttc}[2]{\biiigl\{ #1 \; \biiigr| \; #2 \biiigr\} }
\nc{\setttttc}[2]{\biiiigl\{ #1 \; \biiiigr| \; #2 \biiiigr\} }

\nc{\condprob}[2]{\TOMm{P (#1 | #2) }}
\nc{\condprobb}[2]{P \bigl( #1 \bigr| #2 \bigr) }
\nc{\condprobbb}[2]{P \biigl( #1 \,\biigr|\, #2 \biigr) }
\nc{\condprobbbb}[2]{P \biiigl( #1 \,\biiigr|\, #2 \biiigr) }
\nc{\condprobbbbb}[2]{P \biiiigl( #1 \,\biiiigr|\, #2 \biiiigr) }

\nc{\unit}[1]{\TOMm{\left[#1\right] }}
\nc{\uunit}[1]{\biglrs{#1} }
\nc{\uuunit}[1]{\biiglrs{#1} }
\nc{\uuuunit}[1]{\biiiglrs{#1} }
\nc{\uuuuunit}[1]{\biiiiglrs{#1} }

\nc{\unitt}[1]{\TOMm{\left[\,#1\,\right] }}
\nc{\uunitt}[1]{\biglrs{\,#1\,} }
\nc{\uuunitt}[1]{\biiglrs{\,#1\,} }
\nc{\uuuunitt}[1]{\biiiglrs{\,#1\,} }
\nc{\uuuuunitt}[1]{\biiiiglrs{\,#1\,} }

\nc{\uncertainty}[1]{\enmat{{\smaa{(#1)}}}}
\nc{\tensep}{\enmat{\!\, ^{_{\cross\!}} }}

\nc{\vc}[1]{\TOMm{\unl{#1}} }					
\nc{\lovc}[1]{\lo{\vc{#1}}}
\nc{\hivc}[1]{\hi{\vc{#1}}}

\nc{\multii}[1]{\unl{#1}}						
\nc{\lmultii}[1]{_{\multii{#1}}}					

\nc{\Op}[1]{\TOMm{\widehat{#1}} }				
\nc{\op}[1]{\TOMm{\hat{#1}} }					
\nc{\optx}[1]{\TOMt{\^{#1}} }					

\nc{\fourier}  [1]{\TOMm{\tilde{#1}}}			
\nc{\fourierE} [1]{\TOMm{\check{#1}}}			
\nc{\vcfouE}   [1]{\TOMm{\vc{\fourierE{#1}}}} 	

\nc{\fatstyle}[1]{{\bfseries\textls[12]{#1}}}
\nc{\fat}[1]{\fatstyle{#1}}					
\nc{\emf}[1]{\textit{\textls{#1}}}				

\nc{\itemheadline}[1]{\unl{\textbf{#1}}}		
\nc{\itembo}[1][]{\item[\textbf{#1}]}		
\nc{\itembull}{\item[$\bullet$]}
\nc{\itemBull}{\item[{\Large$\bullet$}]}
\nc{\itembox}{\item[$\scriptstyle\blacksquare$]}
\nc{\itemBox}{\item[$\blacksquare$]}
\nc{\itemtri}{\item[$\scriptstyle\blacktriangleright$]}
\nc{\itemTri}{\item[$\blacktriangleright$]}
\nc{\itemstar}{\item[{\Large$\star$}]}

\nc{\todo}[1]{\indx{\_TODO}\textcolor{red}{\textsc{{\Large ToDo: #1}}}}
\nc{\CHANGED}[1]{\margtx{\greenmedium{\centering\bfseries<= \unl{CHANGED} =>\\ \tiny{#1\\}}}}

\providecommand{\maincheck}[1]{}

\newcounter{privatecounter}
\nc{\private}[1]{\ifte{\value{privatecounter}=1}{#1}{}}

\nc{\ifte}[3]{\ifthenelse{#1}{#2}{#3}}
\nc{\unemptynumberone}[2]{\ifte{\equal{#1}{}}{}{#2}}
\nc{\unemptynumberoneTOMm}[2]{\ifte{\equal{#1}{}} {} {\TOMm{#2}}}
\nc{\emptynumberone}[3]{\ifte{\equal{#1}{}}{#2}{#3}}

\nc{\mlarge}[1]{\text{\large$#1$}}

\nc{\smallheading}[1]{\unl{\bfseries{#1}}}

\newlength{\negphantomlength}
\nc{\negphantom}[1]{\settowidth{\negphantomlength}{{#1}}\hspace{-\negphantomlength}}	

\nc{\indxformat}[1]{\textbf{#1}}									

\nc{\indx}[1]{\index{#1@\indxformat{#1}|textmd}}											
\nc{\indxat}[2]{\index{#1@\indxformat{#2}|textmd}}										
\nc{\indxx}[2]{\index{#1@\indxformat{#1}!#2@\indxformat{#2}|textmd}}							
\nc{\indxxat}[3]{\index{#1@\indxformat{#1}!#2@\indxformat{#3}|textmd}}						
\nc{\indxxatat}[4]{\index{#1@\indxformat{#2}!#3@\indxformat{#4}|textmd}}						
\nc{\indxxx}[3]{\index{#1@\indxformat{#1}!#2@\indxformat{#2}!#3@\indxformat{#3}|textmd}}			
\nc{\indxxxat}[4]{\index{#1@\indxformat{#1}!#2@\indxformat{#2}!#3@\indxformat{#4}|textmd}}		
\nc{\indxxxatatat}[6]{\index{#1@\indxformat{#2}!#3@\indxformat{#4}!#5@\indxformat{#6}|textmd}}			

\nc{\indxbo}[1]{\index{#1@\indxformat{#1}|textbf}}								 
\nc{\indxatbo}[2]{\index{#1@\indxformat{#2}|textbf}}								 
\nc{\indxxbo}[2]{\index{#1@\indxformat{#1}!#2@\indxformat{#2}|textbf}}				 
\nc{\indxxatbo}[3]{\index{#1@\indxformat{#1}!#2@\indxformat{#3}|textbf}}				 
\nc{\indxxatatbo}[4]{\index{#1@\indxformat{#2}!#3@\indxformat{#4}|textbf}}				   	
\nc{\indxxxbo}[3]{\index{#1@\indxformat{#1}!#2@\indxformat{#2}!#3@\indxformat{#3}|textbf}}	
\nc{\indxxxatbo}[4]{\index{#1@\indxformat{#1}!#2@\indxformat{#2}!#3@\indxformat{#4}|textbf}}   
\nc{\indxxxatatatbo}[6]{\index{#1@\indxformat{#2}!#3@\indxformat{#4}!#5@\indxformat{#6}|textbf}}

\nc{\Indx}[1]{#1\index{#1@\indxformat{#1}|textmd}}			
\nc{\Indxbo}[1]{\fat{#1}\index{#1@\indxformat{#1}|textbf}}			

\nc{\igx}{\includegraphics}
\nc{\subfig}[2]{\subfigure[]{\igx[#1]{#2}}}
\nc{\subfigcap}[3]{\subfigure[][#2]{\igx[#1]{#3}}}

\nc{\bfig}[1]{ \ifte{\value{mastereqcounterfigtab}=1} {\setcounter{figure}{\value{equation}}} {}
			\begin{figure}[H]\centering #1 \end{figure} 
			\ifte{\value{mastereqcounterfigtab}=1} {\stepcounter{equation}} {}
			\noi
			}

\nc{\wrapfig}[4]{ \ifte{\value{mastereqcounterfigtab}=1} {\setcounter{figure}{\value{equation}}} {}
		\ifte{\isodd{\value{page}}} {\begin{wrapfigure}{r}{#1}} {\begin{wrapfigure}{l}{#1}}
			\igx[width = #1]{#2}
			\caption{#3}
			\label{#4}
		\end{wrapfigure}
		\ifte{\value{mastereqcounterfigtab}=1} {\stepcounter{equation}} {}
		\noi
		}

\nc{\mathnote}[1]{\smaaa{#1}}

\nc{\margtx}[1]{\marginpar{#1}}
\nc{\margbal}[1]{\marginpar{\begin{align*}#1\end{align*}}}
\nc{\margbals}[1]{\marginpar{\begin{align*}\scriptscriptstyle #1\end{align*}}}

\nc{\inputlabel}[1]{\label{#1}\input{#1}}

\newcounter{prepage}
\newcounter{sucpage}

\nc{\refpmarg}[1]{\margtx{\centering{\ovalbox{\scriptsize$\rightarrow$p.\pageref{#1}}}}} 
\nc{\seesecrefpmarg}[1]{\margtx{\centering{\ovalbox{\scriptsize$\rightarrow$ Sec. \ref{#1} (p.\pageref{#1})}}}}
\nc{\seesecrefmarg}[1]{\margtx{\centering{\ovalbox{\scriptsize$\rightarrow$ Sec. \ref{#1}}}}}
\nc{\checkrefp}[1]{\setcounter{prepage}{\value{page}}\setcounter{sucpage}{\value{page}}%
				\ifte{\isodd{\value{page}}}
					{\addtocounter{prepage}{-3}\addtocounter{sucpage}{+2}}
					{\addtocounter{prepage}{-2}\addtocounter{sucpage}{+3}}%
				\ifte{\value{pageatref}=1 \AND \(\pageref{#1}<\value{prepage} \OR \pageref{#1}>\value{sucpage}\)}
					{\refpmarg{#1}} {}%
				} 
\nc{\checkseesecrefp}[1]{\setcounter{prepage}{\value{page}}\setcounter{sucpage}{\value{page}}%
				\ifte{\isodd{\value{page}}}
					{\addtocounter{prepage}{-3}\addtocounter{sucpage}{+2}}
					{\addtocounter{prepage}{-2}\addtocounter{sucpage}{+3}}%
				\ifte{\value{pageatref}=1 \AND \(\pageref{#1}<\value{prepage} \OR \pageref{#1}>\value{sucpage}\)}
					{\seesecrefpmarg{#1}}
					{\seesecrefmarg{#1}}%
				}

\nc{\refp}[1]{\ref{#1}\checkrefp{#1}}				
\nc{\eqrefp}[1]{equation \eqref{#1}\checkrefp{#1}}	
\nc{\neqrefp}[1]{\eqref{#1}\checkrefp{#1}}			
\nc{\secrefp}[1]{section \ref{#1}\checkrefp{#1}}		
\nc{\seesecrefp}[1]{\checkseesecrefp{#1}}			
\nc{\apprefp}[1]{appendix \ref{#1}\checkrefp{#1}}		

\nc{\citeeq}[2]{\cite{#1}$_{(#2)}$}	
\nc{\citepage}[2]{\cite{#1}$_{\txp.#2}$}	
\nc{\citesec}[2]{\cite{#1}$_{#2}$}	

\nc{\vis}[2]{\visible <#1> {#2}}
\nc{\onl}[2]{\only <#1> {#2}}
\nc{\blockk}[2]{\begin{block}{#1} #2 \end{block}}

\nc{\wideboxsep}{\hspace{1em}}
\nc{\widefbox}     [1]{\fbox{\wideboxsep#1\wideboxsep}}
\nc{\widedoublebox}[1]{\doublebox{\wideboxsep#1\wideboxsep}}
\nc{\wideovalbox}  [1]{\ovalbox{\wideboxsep#1\wideboxsep}}
\nc{\wideOvalbox}  [1]{\Ovalbox{\wideboxsep#1\wideboxsep}}
\nc{\widecolorbox} [2]{\colorbox{#1}{\wideboxsep #2 \wideboxsep}}
\nc{\widefcolorbox}[3]{\fcolorbox{#1}{#2}{\wideboxsep #3 \wideboxsep}}

\nc{\bal}[1]{\begin{align} #1 \end{align}}
\nc{\bals}[1]{\begin{align*} #1 \end{align*}}
\nc{\boxbal}[1]{\begin{empheq}[outerbox=\widefcolorbox{black}{grayveryverylight}]{align}#1\end{empheq}}	

\nc{\bsplit}[1]{\begin{split}#1\end{split}}				
\nc{\balsplit}[1]{\bal{\bsplit{#1}} }		

\nc{\bmult}[1]{\begin{multline}#1\end{multline}}
\nc{\bmults}[1]{\begin{multline*}#1\end{multline*}}

\nc{\bmat}[1]{\begin{matrix} #1 \end{matrix}}
\nc{\bpmat}[1]{\begin{pmatrix} #1 \end{pmatrix}}
\nc{\bdia}[1]{\begin{diagram} #1 \end{diagram}}
\nc{\bcas}[1]{\begin{cases} #1 \end{cases}}

\nc{\bflal}[1]{\begin{flalign} #1 \end{flalign}}
\nc{\bflals}[1]{\begin{flalign*} #1 \end{flalign*}}

\nc{\itemtribflal}[3][0cm]{%
		\vspace{#1}%
		\itemtri%
		\white{x}%
		\vspace{#2}%
		\bflal{#3 &&&}
		}

\nc{\itemtribflals}[3][0cm]{%
		\vspace{#1}%
		\itemtri%
		\white{x}%
		\vspace{#2}%
		\bflals{#3 &&&}
		}

\nc{\binomial}[2]{\bpmat{ #1 \\ #2 }}

\nc{\btab}[1]{ \ifte{\value{mastereqcounterfigtab}=1} 
				{\setcounter{table}{\value{equation}}\stepcounter{equation}}
				{}
			\begin{table}[H] #1 \end{table} 
			}

\nc{\btabu}[2]{\begin{tabular}{#1} #2 \end{tabular}}
\nc{\btabuc}[1]{\btabu{c}{#1}}

\nc{\minpage}[2]{\begin{minipage}{#1} #2 \end{minipage}}
\nc{\minpagel}[2]{\begin{minipage}{#1\linewidth} #2 \end{minipage}}

\newlength{\minpagewidth}
\nc{\minpagewidthof}[2]{\settowidth{\minpagewidth}{#1}\minpage{\minpagewidth}{#2}}

\nc{\formfield}[2][]{\minpagewidthof{\:#2\:}{\center \unlh{\:#2\:}\\$\htxs{#1}$}}

\nc{\bitem}[1]{\begin{itemize} 
					#1 
			\end{itemize}}

\nc{\bitemobo}[1]{\begin{itemize}[<+->]
					#1 
			\end{itemize}}

\nc{\bitemoboa}[1]{\begin{itemize}[<+-| alert@+>]
					#1 
			\end{itemize}}
					
\nc{\numlist}[1]{\begin{enumerate}
			  \setlength{\leftmargin}{2.7cm} \setlength{\labelsep}{0.3cm}
				#1
			  \end{enumerate}}

\nc{\numlistobo}[1]{\begin{enumerate}[<+->]
			  \setlength{\leftmargin}{2.7cm} \setlength{\labelsep}{0.3cm}
				#1
			  \end{enumerate}}

\newcounter{continuednumlistcounter}		
\nc{\keepnumlistcounter}{\setcounter{continuednumlistcounter}{\value{enumi}}}
\nc{\putnumlistcounter}{\setcounter{enumi}{\value{continuednumlistcounter}}}
\nc{\resetcnumlist}{\setcounter{continuednumlistcounter}{0}}
\nc{\cnumlist}[1]{\begin{enumerate}
			   \ifte{\value{continuednumlistcounter}>0}{\putnumlistcounter}{}
			   \setlength{\leftmargin}{2.7cm} \setlength{\labelsep}{0.3cm}
				#1
			   \keepnumlistcounter
			   \end{enumerate}}

\nc{\cnumlistobo}[1]{\begin{enumerate}[<+->]
			   \ifte{\value{continuednumlistcounter}>0}{\putnumlistcounter}{}
			   \setlength{\leftmargin}{2.7cm} \setlength{\labelsep}{0.3cm}
				#1
			   \keepnumlistcounter
			   \end{enumerate}}

\definecolor{bluedark}{rgb}{0.1,0,0.5}\nc{\bluedark}[1]{\textcolor{bluedark}{#1}}

\definecolor{goldunam}{rgb}{0.98,0.89,0.31} \nc{\goldunam}[1]{\textcolor{goldunam}{#1}}

\definecolor{grayveryverylight}{rgb}{0.97,0.97,0.97}\nc{\grayveryverylight}[1]{\textcolor{grayveryverylight}{#1}}

\definecolor{greendark}{rgb}{0,0.57,0}\nc{\greendark}[1]{\textcolor{greendark}{#1}}
\definecolor{greenmedium}{rgb}{0,0.8,0}\nc{\greenmedium}[1]{\textcolor{greenmedium}{#1}}

\definecolor{orangedark}{rgb}{0.9,0.45,0}\nc{\orangedark}[1]{\textcolor{orangedark}{#1}}
\definecolor{orangelight}{rgb}{1,0.85,0.42}\nc{\orangelight}[1]{\textcolor{orangelight}{#1}}

\definecolor{red}{rgb}{1,0,0}\nc{\red}[1]{\textcolor{red}{#1}}
\definecolor{reddark}{rgb}{0.85,0,0}\nc{\reddark}[1]{\textcolor{reddark}{#1}}

\definecolor{violetdark}{rgb}{0.35,0,0.48}\nc{\violetdark}[1]{\textcolor{violetdark}{#1}}

\nc{\green}[1]{\textcolor{green}{#1}}
\nc{\white}[1]{\textcolor{white}{#1}}

\definecolor{yellowdark}{rgb}{0.98,0.97,0}\nc{\yellowdark}[1]{\textcolor{yellowdark}{#1}}

\nc{\al}{\TOMm{\alpha}}
\nc{\be}{\TOMm{\beta}}
\nc{\ga}{\TOMm{\gamma}}
\nc{\de}{\TOMm{\delta}}
\nc{\ep}{\TOMm{\epsilon}}
\nc{\vep}{\TOMm{\varepsilon}}
\nc{\ph}{\TOMm{\phi}}
\nc{\vph}{\TOMm{\varphi}}
\nc{\ps}{\TOMm{\psi}}
\nc{\et}{\TOMm{\eta}}
\nc{\io}{\TOMm{\iota}}
\nc{\ka}{\TOMm{\kappa}}
\nc{\la}{\TOMm{\lambda}}
\nc{\muu}{\TOMm{\mu}}
\nc{\nuu}{\TOMm{\nu}}
\nc{\pii}{\TOMm{\pi}}
\nc{\ro}{\TOMm{\rho}}
\nc{\si}{\TOMm{\sigma}}
\nc{\ta}{\TOMm{\tau}}
\nc{\te}{\TOMm{\theta}}
\nc{\vte}{\TOMm{\vartheta}}
\nc{\om}{\TOMm{\omega}}
\nc{\ki}{\TOMm{\chi}}
\nc{\xii}{\TOMm{\xi}}
\nc{\ze}{\TOMm{\zeta}}

\nc{\Ga}{\TOMm{\Gamma}}
\nc{\De}{\TOMm{\Delta}}
\nc{\Ph}{\TOMm{\Phi}}
\nc{\Ps}{\TOMm{\Psi}}
\nc{\La}{\TOMm{\Lambda}}
\nc{\Pii}{\TOMm{\Pi}}
\nc{\Si}{\TOMm{\Sigma}}
\nc{\Om}{\TOMm{\Omega}}
\nc{\Te}{\TOMm{\Theta}}
\nc{\Up}{\TOMm{\Upsilon}}
\nc{\Xii}{\TOMm{\Xi}}

\nc{\taz}{\TOMm{{\tau_0}}}

\nc{\alar}[2][]{\TOMm{\alpha #1 \ar{#2}}}
\nc{\bear}[2][]{\TOMm{\beta #1 \ar{#2}}}
\nc{\gaar}[2][]{\TOMm{\gamma #1 \ar{#2}}}
\nc{\dear}[2][]{\TOMm{\delta #1 \ar{#2}}}
\nc{\epar}[2][]{\TOMm{\epsilon #1 \ar{#2}}}
\nc{\vepar}[2][]{\TOMm{\varepsilon #1 \ar{#2}}}
\nc{\phar}[2][]{\TOMm{\phi #1 \ar{#2}}}
\nc{\vphar}[2][]{\TOMm{\varphi #1 \ar{#2}}}
\nc{\psar}[2][]{\TOMm{\psi #1 \ar{#2}}}
\nc{\etar}[2][]{\TOMm{\eta #1 \ar{#2}}}
\nc{\ioar}[2][]{\TOMm{\iota #1 \ar{#2}}}
\nc{\kaar}[2][]{\TOMm{\kappa #1 \ar{#2}}}
\nc{\laar}[2][]{\TOMm{\lambda #1 \ar{#2}}}
\nc{\muar}[2][]{\TOMm{\mu #1 \ar{#2}}}
\nc{\nuar}[2][]{\TOMm{\nu #1 \ar{#2}}}
\nc{\piar}[2][]{\TOMm{\pi #1 \ar{#2}}}
\nc{\roar}[2][]{\TOMm{\rho #1 \ar{#2}}}
\nc{\siar}[2][]{\TOMm{\sigma #1 \ar{#2}}}
\nc{\taar}[2][]{\TOMm{\tau #1 \ar{#2}}}
\nc{\tear}[2][]{\TOMm{\theta #1 \ar{#2}}}
\nc{\vtear}[2][]{\TOMm{\vartheta #1 \ar{#2}}}
\nc{\omar}[2][]{\TOMm{\omega #1 \ar{#2}}}
\nc{\kiar}[2][]{\TOMm{\chi #1 \ar{#2}}}
\nc{\xiar}[2][]{\TOMm{\xi #1 \ar{#2}}}
\nc{\zear}[2][]{\TOMm{\zeta #1 \ar{#2}}}

\nc{\Gaar}[2][]{\TOMm{\Gamma #1 \ar{#2}}}
\nc{\Dear}[2][]{\TOMm{\Delta #1 \ar{#2}}}
\nc{\Phar}[2][]{\TOMm{\Phi #1 \ar{#2}}}
\nc{\Psar}[2][]{\TOMm{\Psi #1 \ar{#2}}}
\nc{\Laar}[2][]{\TOMm{\Lambda #1 \ar{#2}}}
\nc{\Piar}[2][]{\TOMm{\Pi #1 \ar{#2}}}
\nc{\Siar}[2][]{\TOMm{\Sigma #1 \ar{#2}}}
\nc{\Omar}[2][]{\TOMm{\Omega #1 \ar{#2}}}
\nc{\Tear}[2][]{\TOMm{\Theta #1 \ar{#2}}}
\nc{\Upar}[2][]{\TOMm{\Upsilon #1 \ar{#2}}}
\nc{\Xiar}[2][]{\TOMm{\Xi #1 \ar{#2}}}

\nc{\alab}[2][]{\TOMm{\alpha #1 \ab{#2}}}
\nc{\beab}[2][]{\TOMm{\beta #1 \ab{#2}}}
\nc{\gaab}[2][]{\TOMm{\gamma #1 \ab{#2}}}
\nc{\deab}[2][]{\TOMm{\delta #1 \ab{#2}}}
\nc{\epab}[2][]{\TOMm{\epsilon #1 \ab{#2}}}
\nc{\vepab}[2][]{\TOMm{\varepsilon #1 \ab{#2}}}
\nc{\phab}[2][]{\TOMm{\phi #1 \ab{#2}}}
\nc{\vphab}[2][]{\TOMm{\varphi #1 \ab{#2}}}
\nc{\psab}[2][]{\TOMm{\psi #1 \ab{#2}}}
\nc{\etab}[2][]{\TOMm{\eta #1 \ab{#2}}}
\nc{\ioab}[2][]{\TOMm{\iota #1 \ab{#2}}}
\nc{\kaab}[2][]{\TOMm{\kappa #1 \ab{#2}}}
\nc{\laab}[2][]{\TOMm{\lambda #1 \ab{#2}}}
\nc{\muab}[2][]{\TOMm{\mu #1 \ab{#2}}}
\nc{\nuab}[2][]{\TOMm{\nu #1 \ab{#2}}}
\nc{\piab}[2][]{\TOMm{\pi #1 \ab{#2}}}
\nc{\roab}[2][]{\TOMm{\rho #1 \ab{#2}}}
\nc{\siab}[2][]{\TOMm{\sigma #1 \ab{#2}}}
\nc{\taab}[2][]{\TOMm{\tau #1 \ab{#2}}}
\nc{\teab}[2][]{\TOMm{\theta #1 \ab{#2}}}
\nc{\vteab}[2][]{\TOMm{\vartheta #1 \ab{#2}}}
\nc{\omab}[2][]{\TOMm{\omega #1 \ab{#2}}}
\nc{\kiab}[2][]{\TOMm{\chi #1 \ab{#2}}}
\nc{\xiab}[2][]{\TOMm{\xi #1 \ab{#2}}}
\nc{\zeab}[2][]{\TOMm{\zeta #1 \ab{#2}}}

\nc{\Gaab}[2][]{\TOMm{\Gamma #1 \ab{#2}}}
\nc{\Deab}[2][]{\TOMm{\Delta #1 \ab{#2}}}
\nc{\Phab}[2][]{\TOMm{\Phi #1 \ab{#2}}}
\nc{\Psab}[2][]{\TOMm{\Psi #1 \ab{#2}}}
\nc{\Laab}[2][]{\TOMm{\Lambda #1 \ab{#2}}}
\nc{\Piab}[2][]{\TOMm{\Pi #1 \ab{#2}}}
\nc{\Siab}[2][]{\TOMm{\Sigma #1 \ab{#2}}}
\nc{\Omab}[2][]{\TOMm{\Omega #1 \ab{#2}}}
\nc{\Teab}[2][]{\TOMm{\Theta #1 \ab{#2}}}
\nc{\Upab}[2][]{\TOMm{\Upsilon #1 \ab{#2}}}
\nc{\Xiab}[2][]{\TOMm{\Xi #1 \ab{#2}}}

\nc{\txa}{\text{a}}
\nc{\txb}{\text{b}}
\nc{\txc}{\text{c}}
\nc{\txd}{\text{d}}
\nc{\txe}{\text{e}}
\nc{\txf}{\text{f}}
\nc{\txg}{\text{g}}
\nc{\txh}{\text{h}}
\nc{\txi}{\text{i}}
\nc{\txj}{\text{j}}
\nc{\txk}{\text{k}}
\nc{\txl}{\text{l}}
\nc{\txm}{\text{m}}
\nc{\txn}{\text{n}}
\nc{\txo}{\text{o}}
\nc{\txp}{\text{p}}
\nc{\txq}{\text{q}}
\nc{\txr}{\text{r}}
\nc{\txs}{\text{s}}
\nc{\txt}{\text{t}}
\nc{\txu}{\text{u}}
\nc{\txv}{\text{v}}
\nc{\txw}{\text{w}}
\nc{\txx}{\text{x}}
\nc{\txy}{\text{y}}
\nc{\txz}{\text{z}}

\nc{\txA}{\text{A}}
\nc{\txB}{\text{B}}
\nc{\txC}{\text{C}}
\nc{\txD}{\text{D}}
\nc{\txE}{\text{E}}
\nc{\txF}{\text{F}}
\nc{\txG}{\text{G}}
\nc{\txH}{\text{H}}
\nc{\txI}{\text{I}}
\nc{\txJ}{\text{J}}
\nc{\txK}{\text{K}}
\nc{\txL}{\text{L}}
\nc{\txM}{\text{M}}
\nc{\txN}{\text{N}}
\nc{\txO}{\text{O}}
\nc{\txP}{\text{P}}
\nc{\txQ}{\text{Q}}
\nc{\txR}{\text{R}}
\nc{\txS}{\text{S}}
\nc{\txT}{\text{T}}
\nc{\txU}{\text{U}}
\nc{\txV}{\text{V}}
\nc{\txW}{\text{W}}
\nc{\txX}{\text{X}}
\nc{\txY}{\text{Y}}
\nc{\txZ}{\text{Z}}

\nc{\txaar}[2][]{\TOMm{\text{a} #1 \ar{#2}}}
\nc{\txbar}[2][]{\TOMm{\text{b} #1 \ar{#2}}}
\nc{\txcar}[2][]{\TOMm{\text{c} #1 \ar{#2}}}
\nc{\txdar}[2][]{\TOMm{\text{d} #1 \ar{#2}}}
\nc{\txear}[2][]{\TOMm{\text{e} #1 \ar{#2}}}
\nc{\txfar}[2][]{\TOMm{\text{f} #1 \ar{#2}}}
\nc{\txgar}[2][]{\TOMm{\text{g} #1 \ar{#2}}}
\nc{\txhar}[2][]{\TOMm{\text{h} #1 \ar{#2}}}
\nc{\txiar}[2][]{\TOMm{\text{i} #1 \ar{#2}}}
\nc{\txjar}[2][]{\TOMm{\text{j} #1 \ar{#2}}}
\nc{\txkar}[2][]{\TOMm{\text{k} #1 \ar{#2}}}
\nc{\txlar}[2][]{\TOMm{\text{l} #1 \ar{#2}}}
\nc{\txmar}[2][]{\TOMm{\text{m} #1 \ar{#2}}}
\nc{\txnar}[2][]{\TOMm{\text{n} #1 \ar{#2}}}
\nc{\txoar}[2][]{\TOMm{\text{o} #1 \ar{#2}}}
\nc{\txpar}[2][]{\TOMm{\text{p} #1 \ar{#2}}}
\nc{\txqar}[2][]{\TOMm{\text{q} #1 \ar{#2}}}
\nc{\txrar}[2][]{\TOMm{\text{r} #1 \ar{#2}}}
\nc{\txsar}[2][]{\TOMm{\text{s} #1 \ar{#2}}}
\nc{\txtar}[2][]{\TOMm{\text{t} #1 \ar{#2}}}
\nc{\txuar}[2][]{\TOMm{\text{u} #1 \ar{#2}}}
\nc{\txvar}[2][]{\TOMm{\text{v} #1 \ar{#2}}}
\nc{\txwar}[2][]{\TOMm{\text{w} #1 \ar{#2}}}
\nc{\txxar}[2][]{\TOMm{\text{x} #1 \ar{#2}}}
\nc{\txyar}[2][]{\TOMm{\text{y} #1 \ar{#2}}}
\nc{\txzar}[2][]{\TOMm{\text{z} #1 \ar{#2}}}

\nc{\txAar}[2][]{\TOMm{\text{A} #1 \ar{#2}}}
\nc{\txBar}[2][]{\TOMm{\text{B} #1 \ar{#2}}}
\nc{\txCar}[2][]{\TOMm{\text{C} #1 \ar{#2}}}
\nc{\txDar}[2][]{\TOMm{\text{D} #1 \ar{#2}}}
\nc{\txEar}[2][]{\TOMm{\text{E} #1 \ar{#2}}}
\nc{\txFar}[2][]{\TOMm{\text{F} #1 \ar{#2}}}
\nc{\txGar}[2][]{\TOMm{\text{G} #1 \ar{#2}}}
\nc{\txHar}[2][]{\TOMm{\text{H} #1 \ar{#2}}}
\nc{\txIar}[2][]{\TOMm{\text{I} #1 \ar{#2}}}
\nc{\txJar}[2][]{\TOMm{\text{J} #1 \ar{#2}}}
\nc{\txKar}[2][]{\TOMm{\text{K} #1 \ar{#2}}}
\nc{\txLar}[2][]{\TOMm{\text{L} #1 \ar{#2}}}
\nc{\txMar}[2][]{\TOMm{\text{M} #1 \ar{#2}}}
\nc{\txNar}[2][]{\TOMm{\text{N} #1 \ar{#2}}}
\nc{\txOar}[2][]{\TOMm{\text{O} #1 \ar{#2}}}
\nc{\txPar}[2][]{\TOMm{\text{P} #1 \ar{#2}}}
\nc{\txQar}[2][]{\TOMm{\text{Q} #1 \ar{#2}}}
\nc{\txRar}[2][]{\TOMm{\text{R} #1 \ar{#2}}}
\nc{\txSar}[2][]{\TOMm{\text{S} #1 \ar{#2}}}
\nc{\txTar}[2][]{\TOMm{\text{T} #1 \ar{#2}}}
\nc{\txUar}[2][]{\TOMm{\text{U} #1 \ar{#2}}}
\nc{\txVar}[2][]{\TOMm{\text{V} #1 \ar{#2}}}
\nc{\txWar}[2][]{\TOMm{\text{W} #1 \ar{#2}}}
\nc{\txXar}[2][]{\TOMm{\text{X} #1 \ar{#2}}}
\nc{\txYar}[2][]{\TOMm{\text{Y} #1 \ar{#2}}}
\nc{\txZar}[2][]{\TOMm{\text{Z} #1 \ar{#2}}}

\nc{\txaab}[2][]{\TOMm{\text{a} #1 \ab{#2}}}
\nc{\txbab}[2][]{\TOMm{\text{b} #1 \ab{#2}}}
\nc{\txcab}[2][]{\TOMm{\text{c} #1 \ab{#2}}}
\nc{\txdab}[2][]{\TOMm{\text{d} #1 \ab{#2}}}
\nc{\txeab}[2][]{\TOMm{\text{e} #1 \ab{#2}}}
\nc{\txfab}[2][]{\TOMm{\text{f} #1 \ab{#2}}}
\nc{\txgab}[2][]{\TOMm{\text{g} #1 \ab{#2}}}
\nc{\txhab}[2][]{\TOMm{\text{h} #1 \ab{#2}}}
\nc{\txiab}[2][]{\TOMm{\text{i} #1 \ab{#2}}}
\nc{\txjab}[2][]{\TOMm{\text{j} #1 \ab{#2}}}
\nc{\txkab}[2][]{\TOMm{\text{k} #1 \ab{#2}}}
\nc{\txlab}[2][]{\TOMm{\text{l} #1 \ab{#2}}}
\nc{\txmab}[2][]{\TOMm{\text{m} #1 \ab{#2}}}
\nc{\txnab}[2][]{\TOMm{\text{n} #1 \ab{#2}}}
\nc{\txoab}[2][]{\TOMm{\text{o} #1 \ab{#2}}}
\nc{\txpab}[2][]{\TOMm{\text{p} #1 \ab{#2}}}
\nc{\txqab}[2][]{\TOMm{\text{q} #1 \ab{#2}}}
\nc{\txrab}[2][]{\TOMm{\text{r} #1 \ab{#2}}}
\nc{\txsab}[2][]{\TOMm{\text{s} #1 \ab{#2}}}
\nc{\txtab}[2][]{\TOMm{\text{t} #1 \ab{#2}}}
\nc{\txuab}[2][]{\TOMm{\text{u} #1 \ab{#2}}}
\nc{\txvab}[2][]{\TOMm{\text{v} #1 \ab{#2}}}
\nc{\txwab}[2][]{\TOMm{\text{w} #1 \ab{#2}}}
\nc{\txxab}[2][]{\TOMm{\text{x} #1 \ab{#2}}}
\nc{\txyab}[2][]{\TOMm{\text{y} #1 \ab{#2}}}
\nc{\txzab}[2][]{\TOMm{\text{z} #1 \ab{#2}}}

\nc{\txAab}[2][]{\TOMm{\text{A} #1 \ab{#2}}}
\nc{\txBab}[2][]{\TOMm{\text{B} #1 \ab{#2}}}
\nc{\txCab}[2][]{\TOMm{\text{C} #1 \ab{#2}}}
\nc{\txDab}[2][]{\TOMm{\text{D} #1 \ab{#2}}}
\nc{\txEab}[2][]{\TOMm{\text{E} #1 \ab{#2}}}
\nc{\txFab}[2][]{\TOMm{\text{F} #1 \ab{#2}}}
\nc{\txGab}[2][]{\TOMm{\text{G} #1 \ab{#2}}}
\nc{\txHab}[2][]{\TOMm{\text{H} #1 \ab{#2}}}
\nc{\txIab}[2][]{\TOMm{\text{I} #1 \ab{#2}}}
\nc{\txJab}[2][]{\TOMm{\text{J} #1 \ab{#2}}}
\nc{\txKab}[2][]{\TOMm{\text{K} #1 \ab{#2}}}
\nc{\txLab}[2][]{\TOMm{\text{L} #1 \ab{#2}}}
\nc{\txMab}[2][]{\TOMm{\text{M} #1 \ab{#2}}}
\nc{\txNab}[2][]{\TOMm{\text{N} #1 \ab{#2}}}
\nc{\txOab}[2][]{\TOMm{\text{O} #1 \ab{#2}}}
\nc{\txPab}[2][]{\TOMm{\text{P} #1 \ab{#2}}}
\nc{\txQab}[2][]{\TOMm{\text{Q} #1 \ab{#2}}}
\nc{\txRab}[2][]{\TOMm{\text{R} #1 \ab{#2}}}
\nc{\txSab}[2][]{\TOMm{\text{S} #1 \ab{#2}}}
\nc{\txTab}[2][]{\TOMm{\text{T} #1 \ab{#2}}}
\nc{\txUab}[2][]{\TOMm{\text{U} #1 \ab{#2}}}
\nc{\txVab}[2][]{\TOMm{\text{V} #1 \ab{#2}}}
\nc{\txWab}[2][]{\TOMm{\text{W} #1 \ab{#2}}}
\nc{\txXab}[2][]{\TOMm{\text{X} #1 \ab{#2}}}
\nc{\txYab}[2][]{\TOMm{\text{Y} #1 \ab{#2}}}
\nc{\txZab}[2][]{\TOMm{\text{Z} #1 \ab{#2}}}

\nc{\vcxcutright}{\!\,\hs{\!}}
\nc{\vcxcutrightd}{\!\,\hs{\!\!}}
\nc{\vcxcutrightt}{\!\,\hs{\!\!\!}}
\nc{\vcxextleft}{\hspace{1pt}}
\nc{\vcxnormright}{\hs{\,}\,\!}
\nc{\vcxnormrightd}{\hs{\,\,}\,\!}
\nc{\vcxnormrightt}{\hs{\,\,\,}\,\!}

\nc{\vca}{\TOMm{\vc{a\vcxcutright}\vcxnormright}}
\nc{\vcb}{\TOMm{\vc{b\vcxcutright}\vcxnormright}}
\nc{\vcc}{\TOMm{\vc{c\vcxcutright}\vcxnormright}}
\nc{\vcd}{\TOMm{\vc{d\vcxcutright}\vcxnormright}}
\nc{\vce}{\TOMm{\vc{e\vcxcutright}\vcxnormright}}
\nc{\vcf}{\TOMm{\vc{f\vcxcutrightd}\vcxnormright}}
\nc{\vcg}{\TOMm{\vc{\vcxextleft g\vcxcutrightd}\vcxnormright}}
\nc{\vch}{\TOMm{\vc{h\vcxcutright}\vcxnormright}}
\nc{\vci}{\TOMm{\vc{i\vcxcutright}\vcxnormright}}
\nc{\vcj}{\TOMm{\vc{j\vcxcutrightd}\vcxnormright}}
\nc{\vck}{\TOMm{\vc{k\vcxcutright}\vcxnormright}}
\nc{\vcl}{\TOMm{\vc{l\vcxcutright}\vcxnormright}}
\nc{\vcm}{\TOMm{\vc{m\vcxcutright}\vcxnormright}}
\nc{\vcn}{\TOMm{\vc{n\vcxcutright}\vcxnormright}}
\nc{\vco}{\TOMm{\vc{o\vcxcutright}\vcxnormright}}
\nc{\vcp}{\TOMm{\vc{\vcxextleft p\vcxcutrightd}\vcxnormright}}
\nc{\vcq}{\TOMm{\vc{q\vcxcutright}\vcxnormright}}
\nc{\vcr}{\TOMm{\vc{r\vcxcutright}\vcxnormright}}
\nc{\vcs}{\TOMm{\vc{s\vcxcutright}\vcxnormright}}
\nc{\vct}{\TOMm{\vc{\vcxextleft t\vcxcutright}\vcxnormright}}
\nc{\vcu}{\TOMm{\vc{u\vcxcutright}\vcxnormright}}
\nc{\vcv}{\TOMm{\vc{v\vcxcutright}\vcxnormright}}
\nc{\vcw}{\TOMm{\vc{w\vcxcutright}\vcxnormright}}
\nc{\vcx}{\TOMm{\vc{x\vcxcutright}\vcxnormright}}
\nc{\vcy}{\TOMm{\vc{y\vcxcutrightd}\vcxnormright}}
\nc{\vcz}{\TOMm{\vc{z\vcxcutrightd}\vcxnormright}}

\nc{\vcA}{\TOMm{\vc{A\vcxcutright}\vcxnormright}}
\nc{\vcB}{\TOMm{\vc{B\vcxcutright}\vcxnormright}}
\nc{\vcC}{\TOMm{\vc{C\vcxcutrightd}\vcxnormrightd}}
\nc{\vcD}{\TOMm{\vc{D\vcxcutrightd}\vcxnormrightd}}
\nc{\vcE}{\TOMm{\vc{E\vcxcutrightd}\vcxnormrightd}}
\nc{\vcF}{\TOMm{\vc{F\vcxcutrightd}\vcxnormrightd}}
\nc{\vcG}{\TOMm{\vc{G\vcxcutrightd}\vcxnormrightd}}
\nc{\vcH}{\TOMm{\vc{H\vcxcutrightd}\vcxnormrightd}}
\nc{\vcI}{\TOMm{\vc{I\vcxcutrightd}\vcxnormright}}
\nc{\vcJ}{\TOMm{\vc{J\vcxcutrightd}\vcxnormright}}
\nc{\vcK}{\TOMm{\vc{K\vcxcutrightd}\vcxnormrightd}}
\nc{\vcL}{\TOMm{\vc{L\vcxcutright}\vcxnormright}}
\nc{\vcM}{\TOMm{\vc{M\vcxcutrightd}\vcxnormrightd}}
\nc{\vcN}{\TOMm{\vc{N\vcxcutrightd}\vcxnormrightd}}
\nc{\vcO}{\TOMm{\vc{O\vcxcutrightd}\vcxnormrightd}}
\nc{\vcP}{\TOMm{\vc{P\vcxcutrightd}\vcxnormright}}
\nc{\vcQ}{\TOMm{\vc{Q\vcxcutrightd}\vcxnormright}}
\nc{\vcR}{\TOMm{\vc{R\vcxcutright}\vcxnormright}}
\nc{\vcS}{\TOMm{\vc{S\vcxcutright}\vcxnormright}}
\nc{\vcT}{\TOMm{\vc{\vcxextleft T\vcxcutrightd}\vcxnormrightd}}
\nc{\vcU}{\TOMm{\vc{U\vcxcutrightd}\vcxnormrightd}}
\nc{\vcV}{\TOMm{\vc{V\vcxcutrightt}\vcxnormrightd}}
\nc{\vcW}{\TOMm{\vc{W\vcxcutrightd}\vcxnormrightd}}
\nc{\vcX}{\TOMm{\vc{X\vcxcutrightd}\vcxnormrightt}}
\nc{\vcY}{\TOMm{\vc{\vcxextleft Y\vcxcutrightt}\vcxnormrightt}}
\nc{\vcZ}{\TOMm{\vc{Z\vcxcutrightd}\vcxnormrightd}}

\nc{\vcal}{\TOMm{\vc\alpha}}
\nc{\vcbe}{\TOMm{\vc\beta}}
\nc{\vcga}{\TOMm{\vc\gamma}}
\nc{\vcde}{\TOMm{\vc\delta}}
\nc{\vcep}{\TOMm{\vc\epsilon}}
\nc{\vcvep}{\TOMm{\vc\varepsilon}}
\nc{\vcph}{\TOMm{\vc\phi}}
\nc{\vcvph}{\TOMm{\vc\varphi}}
\nc{\vcps}{\TOMm{\vc\psi}}
\nc{\vcet}{\TOMm{\vc\eta}}
\nc{\vcio}{\TOMm{\vc\iota}}
\nc{\vcka}{\TOMm{\vc\kappa}}
\nc{\vcla}{\TOMm{\vc\lambda}}
\nc{\vcmu}{\TOMm{\vc\mu}}
\nc{\vcnu}{\TOMm{\vc\nu}}
\nc{\vcpi}{\TOMm{\vc\pi}}
\nc{\vcro}{\TOMm{\vc\rho}}
\nc{\vcsi}{\TOMm{\vc\sigma}}
\nc{\vcta}{\TOMm{\vc\tau}}
\nc{\vcte}{\TOMm{\vc\theta}}
\nc{\vcvte}{\TOMm{\vc\vartheta}}
\nc{\vcom}{\TOMm{\vc\omega}}
\nc{\vcki}{\TOMm{\vc\chi}}
\nc{\vcxi}{\TOMm{\vc\xi}}
\nc{\vcze}{\TOMm{\vc\zeta}}

\nc{\vcGa}{\TOMm{\vc\Gamma}}
\nc{\vcDe}{\TOMm{\vc\Delta}}
\nc{\vcPh}{\TOMm{\vc\Phi}}
\nc{\vcPs}{\TOMm{\vc\Psi}}
\nc{\vcLa}{\TOMm{\vc\Lambda}}
\nc{\vcPi}{\TOMm{\vc\Pi}}
\nc{\vcSi}{\TOMm{\vc\Sigma}}
\nc{\vcOm}{\TOMm{\vc\Omega}}
\nc{\vcTe}{\TOMm{\vc\Theta}}
\nc{\vcUp}{\TOMm{\vc\Upsilon}}
\nc{\vcXi}{\TOMm{\vc\Xi}}

\nc{\vcaar}[2][]{\TOMm{\vc{a} #1 \ar{#2}}}
\nc{\vcbar}[2][]{\TOMm{\vc{b} #1 \ar{#2}}}
\nc{\vccar}[2][]{\TOMm{\vc{c} #1 \ar{#2}}}
\nc{\vcdar}[2][]{\TOMm{\vc{d} #1 \ar{#2}}}
\nc{\vcear}[2][]{\TOMm{\vc{e} #1 \ar{#2}}}
\nc{\vcfar}[2][]{\TOMm{\vc{f} #1 \ar{#2}}}
\nc{\vcgar}[2][]{\TOMm{\vc{g} #1 \ar{#2}}}
\nc{\vchar}[2][]{\TOMm{\vc{h} #1 \ar{#2}}}
\nc{\vciar}[2][]{\TOMm{\vc{i} #1 \ar{#2}}}
\nc{\vcjar}[2][]{\TOMm{\vc{j} #1 \ar{#2}}}
\nc{\vckar}[2][]{\TOMm{\vc{k} #1 \ar{#2}}}
\nc{\vclar}[2][]{\TOMm{\vc{l} #1 \ar{#2}}}
\nc{\vcmar}[2][]{\TOMm{\vc{m} #1 \ar{#2}}}
\nc{\vcnar}[2][]{\TOMm{\vc{n} #1 \ar{#2}}}
\nc{\vcoar}[2][]{\TOMm{\vc{o} #1 \ar{#2}}}
\nc{\vcpar}[2][]{\TOMm{\vc{p} #1 \ar{#2}}}
\nc{\vcqar}[2][]{\TOMm{\vc{q} #1 \ar{#2}}}
\nc{\vcrar}[2][]{\TOMm{\vc{r} #1 \ar{#2}}}
\nc{\vcsar}[2][]{\TOMm{\vc{s} #1 \ar{#2}}}
\nc{\vctar}[2][]{\TOMm{\vc{t} #1 \ar{#2}}}
\nc{\vcuar}[2][]{\TOMm{\vc{u} #1 \ar{#2}}}
\nc{\vcvar}[2][]{\TOMm{\vc{v} #1 \ar{#2}}}
\nc{\vcwar}[2][]{\TOMm{\vc{w} #1 \ar{#2}}}
\nc{\vcxar}[2][]{\TOMm{\vc{x} #1 \ar{#2}}}
\nc{\vcyar}[2][]{\TOMm{\vc{y} #1 \ar{#2}}}
\nc{\vczar}[2][]{\TOMm{\vc{z} #1 \ar{#2}}}

\nc{\vcAar}[2][]{\TOMm{\vc{A} #1 \ar{#2}}}
\nc{\vcBar}[2][]{\TOMm{\vc{B} #1 \ar{#2}}}
\nc{\vcCar}[2][]{\TOMm{\vc{C} #1 \ar{#2}}}
\nc{\vcDar}[2][]{\TOMm{\vc{D} #1 \ar{#2}}}
\nc{\vcEar}[2][]{\TOMm{\vc{E} #1 \ar{#2}}}
\nc{\vcFar}[2][]{\TOMm{\vc{F} #1 \ar{#2}}}
\nc{\vcGar}[2][]{\TOMm{\vc{G} #1 \ar{#2}}}
\nc{\vcHar}[2][]{\TOMm{\vc{H} #1 \ar{#2}}}
\nc{\vcIar}[2][]{\TOMm{\vc{I} #1 \ar{#2}}}
\nc{\vcJar}[2][]{\TOMm{\vc{J} #1 \ar{#2}}}
\nc{\vcKar}[2][]{\TOMm{\vc{K} #1 \ar{#2}}}
\nc{\vcLar}[2][]{\TOMm{\vc{L} #1 \ar{#2}}}
\nc{\vcMar}[2][]{\TOMm{\vc{M} #1 \ar{#2}}}
\nc{\vcNar}[2][]{\TOMm{\vc{N} #1 \ar{#2}}}
\nc{\vcOar}[2][]{\TOMm{\vc{O} #1 \ar{#2}}}
\nc{\vcPar}[2][]{\TOMm{\vc{P} #1 \ar{#2}}}
\nc{\vcQar}[2][]{\TOMm{\vc{Q} #1 \ar{#2}}}
\nc{\vcRar}[2][]{\TOMm{\vc{R} #1 \ar{#2}}}
\nc{\vcSar}[2][]{\TOMm{\vc{S} #1 \ar{#2}}}
\nc{\vcTar}[2][]{\TOMm{\vc{T} #1 \ar{#2}}}
\nc{\vcUar}[2][]{\TOMm{\vc{U} #1 \ar{#2}}}
\nc{\vcVar}[2][]{\TOMm{\vc{V} #1 \ar{#2}}}
\nc{\vcWar}[2][]{\TOMm{\vc{W} #1 \ar{#2}}}
\nc{\vcXar}[2][]{\TOMm{\vc{X} #1 \ar{#2}}}
\nc{\vcYar}[2][]{\TOMm{\vc{Y\!} #1 \ar{#2}}}
\nc{\vcZar}[2][]{\TOMm{\vc{Z} #1 \ar{#2}}}

\nc{\vcalar}[2][]{\TOMm{\vc\alpha #1 \ar{#2}}}
\nc{\vcbear}[2][]{\TOMm{\vc\beta #1 \ar{#2}}}
\nc{\vcgaar}[2][]{\TOMm{\vc\gamma #1 \ar{#2}}}
\nc{\vcdear}[2][]{\TOMm{\vc\delta #1 \ar{#2}}}
\nc{\vcepar}[2][]{\TOMm{\vc\epsilon #1 \ar{#2}}}
\nc{\vcvepar}[2][]{\TOMm{\vc\varepsilon #1 \ar{#2}}}
\nc{\vcphar}[2][]{\TOMm{\vc\phi #1 \ar{#2}}}
\nc{\vcvphar}[2][]{\TOMm{\vc\varphi #1 \ar{#2}}}
\nc{\vcpsar}[2][]{\TOMm{\vc\psi #1 \ar{#2}}}
\nc{\vcetar}[2][]{\TOMm{\vc\eta #1 \ar{#2}}}
\nc{\vcioar}[2][]{\TOMm{\vc\iota #1 \ar{#2}}}
\nc{\vckaar}[2][]{\TOMm{\vc\kappa #1 \ar{#2}}}
\nc{\vclaar}[2][]{\TOMm{\vc\lambda #1 \ar{#2}}}
\nc{\vcmuar}[2][]{\TOMm{\vc\mu #1 \ar{#2}}}
\nc{\vcnuar}[2][]{\TOMm{\vc\nu #1 \ar{#2}}}
\nc{\vcpiar}[2][]{\TOMm{\vc\pi #1 \ar{#2}}}
\nc{\vcroar}[2][]{\TOMm{\vc\rho #1 \ar{#2}}}
\nc{\vcsiar}[2][]{\TOMm{\vc\sigma #1 \ar{#2}}}
\nc{\vctaar}[2][]{\TOMm{\vc\tau #1 \ar{#2}}}
\nc{\vctear}[2][]{\TOMm{\vc\theta #1 \ar{#2}}}
\nc{\vcvtear}[2][]{\TOMm{\vc\vartheta #1 \ar{#2}}}
\nc{\vcomar}[2][]{\TOMm{\vc\omega #1 \ar{#2}}}
\nc{\vckiar}[2][]{\TOMm{\vc\chi #1 \ar{#2}}}
\nc{\vcxiar}[2][]{\TOMm{\vc\xi #1 \ar{#2}}}
\nc{\vczear}[2][]{\TOMm{\vc\zeta #1 \ar{#2}}}

\nc{\vcGaar}[2][]{\TOMm{\vc\Gamma #1 \ar{#2}}}
\nc{\vcDear}[2][]{\TOMm{\vc\Delta #1 \ar{#2}}}
\nc{\vcPhar}[2][]{\TOMm{\vc\Phi #1 \ar{#2}}}
\nc{\vcPsar}[2][]{\TOMm{\vc\Psi #1 \ar{#2}}}
\nc{\vcLaar}[2][]{\TOMm{\vc\Lambda #1 \ar{#2}}}
\nc{\vcPiar}[2][]{\TOMm{\vc\Pi #1 \ar{#2}}}
\nc{\vcSiar}[2][]{\TOMm{\vc\Sigma #1 \ar{#2}}}
\nc{\vcOmar}[2][]{\TOMm{\vc\Omega #1 \ar{#2}}}
\nc{\vcTear}[2][]{\TOMm{\vc\Theta #1 \ar{#2}}}
\nc{\vcUpar}[2][]{\TOMm{\vc\Upsilon #1 \ar{#2}}}
\nc{\vcXiar}[2][]{\TOMm{\vc\Xi #1 \ar{#2}}}

\nc{\Opa}{\TOMm{\Op{a}}}
\nc{\Opb}{\TOMm{\Op{b}}}
\nc{\Opc}{\TOMm{\Op{c}}}
\nc{\Opd}{\TOMm{\Op{d}}}
\nc{\Ope}{\TOMm{\Op{e}}}
\nc{\Opf}{\TOMm{\Op{f}}}
\nc{\Opg}{\TOMm{\Op{g}}}
\nc{\Oph}{\TOMm{\Op{h}}}
\nc{\Opi}{\TOMm{\Op{i}}}
\nc{\Opj}{\TOMm{\Op{j}}}
\nc{\Opk}{\TOMm{\Op{k}}}
\nc{\Opl}{\TOMm{\Op{l}}}
\nc{\Opm}{\TOMm{\Op{m}}}
\nc{\Opn}{\TOMm{\Op{n}}}
\nc{\Opo}{\TOMm{\Op{o}}}
\nc{\Opp}{\TOMm{\Op{p}}}
\nc{\Opq}{\TOMm{\Op{q}}}
\nc{\Opr}{\TOMm{\Op{r}}}
\nc{\Ops}{\TOMm{\Op{s}}}
\nc{\Opt}{\TOMm{\Op{t}}}
\nc{\Opu}{\TOMm{\Op{u}}}
\nc{\Opv}{\TOMm{\Op{v}}}
\nc{\Opw}{\TOMm{\Op{w}}}
\nc{\Opx}{\TOMm{\Op{x}}}
\nc{\Opy}{\TOMm{\Op{y}}}
\nc{\Opz}{\TOMm{\Op{z}}}

\nc{\OpA}{\TOMm{\Op{A}}}
\nc{\OpB}{\TOMm{\Op{B}}}
\nc{\OpC}{\TOMm{\Op{C}}}
\nc{\OpD}{\TOMm{\Op{D}}}
\nc{\OpE}{\TOMm{\Op{E}}}
\nc{\OpF}{\TOMm{\Op{F}}}
\nc{\OpG}{\TOMm{\Op{G}}}
\nc{\OpH}{\TOMm{\Op{H}}}
\nc{\OpI}{\TOMm{\Op{I}}}
\nc{\OpJ}{\TOMm{\Op{J}}}
\nc{\OpK}{\TOMm{\Op{K}}}
\nc{\OpL}{\TOMm{\Op{L}}}
\nc{\OpM}{\TOMm{\Op{M}}}
\nc{\OpN}{\TOMm{\Op{N}}}
\nc{\OpO}{\TOMm{\Op{O}}}
\nc{\OpP}{\TOMm{\Op{P}}}
\nc{\OpQ}{\TOMm{\Op{Q}}}
\nc{\OpR}{\TOMm{\Op{R}}}
\nc{\OpS}{\TOMm{\Op{S}}}
\nc{\OpT}{\TOMm{\Op{T}}}
\nc{\OpU}{\TOMm{\Op{U}}}
\nc{\OpV}{\TOMm{\Op{V}}}
\nc{\OpW}{\TOMm{\Op{W}}}
\nc{\OpX}{\TOMm{\Op{X}}}
\nc{\OpY}{\TOMm{\Op{Y}}}
\nc{\OpZ}{\TOMm{\Op{Z}}}

\nc{\Opal}{\TOMm{\Op\alpha}}
\nc{\Opbe}{\TOMm{\Op\beta}}
\nc{\Opga}{\TOMm{\Op\gamma}}
\nc{\Opde}{\TOMm{\Op\delta}}
\nc{\Opep}{\TOMm{\Op\epsilon}}
\nc{\Opvep}{\TOMm{\Op\varepsilon}}
\nc{\Opph}{\TOMm{\Op\phi}}
\nc{\Opvph}{\TOMm{\Op\varphi}}
\nc{\Opps}{\TOMm{\Op\psi}}
\nc{\Opet}{\TOMm{\Op\eta}}
\nc{\Opio}{\TOMm{\Op\iota}}
\nc{\Opka}{\TOMm{\Op\kappa}}
\nc{\Opla}{\TOMm{\Op\lambda}}
\nc{\Opmu}{\TOMm{\Op\mu}}
\nc{\Opnu}{\TOMm{\Op\nu}}
\nc{\Oppi}{\TOMm{\Op\pi}}
\nc{\Opro}{\TOMm{\Op\rho}}
\nc{\Opsi}{\TOMm{\Op\sigma}}
\nc{\Opta}{\TOMm{\Op\tau}}
\nc{\Opte}{\TOMm{\Op\theta}}
\nc{\Opvte}{\TOMm{\Op\vartheta}}
\nc{\Opom}{\TOMm{\Op\omega}}
\nc{\Opki}{\TOMm{\Op\chi}}
\nc{\Opxi}{\TOMm{\Op\xi}}
\nc{\Opze}{\TOMm{\Op\zeta}}

\nc{\OpGa}{\TOMm{\Op\Gamma}}
\nc{\OpDe}{\TOMm{\Op\Delta}}
\nc{\OpPh}{\TOMm{\Op\Phi}}
\nc{\OpPs}{\TOMm{\Op\Psi}}
\nc{\OpLa}{\TOMm{\Op\Lambda}}
\nc{\OpPi}{\TOMm{\Op\Pi}}
\nc{\OpSi}{\TOMm{\Op\Sigma}}
\nc{\OpOm}{\TOMm{\Op\Omega}}
\nc{\OpTe}{\TOMm{\Op\Theta}}
\nc{\OpUp}{\TOMm{\Op\Upsilon}}
\nc{\OpXi}{\TOMm{\Op\Xi}}

\nc{\Opaar}[2][]{\TOMm{\Op{a} #1 \ar{#2}}}
\nc{\Opbar}[2][]{\TOMm{\Op{b} #1 \ar{#2}}}
\nc{\Opcar}[2][]{\TOMm{\Op{c} #1 \ar{#2}}}
\nc{\Opdar}[2][]{\TOMm{\Op{d} #1 \ar{#2}}}
\nc{\Opear}[2][]{\TOMm{\Op{e} #1 \ar{#2}}}
\nc{\Opfar}[2][]{\TOMm{\Op{f} #1 \ar{#2}}}
\nc{\Opgar}[2][]{\TOMm{\Op{g} #1 \ar{#2}}}
\nc{\Ophar}[2][]{\TOMm{\Op{h} #1 \ar{#2}}}
\nc{\Opiar}[2][]{\TOMm{\Op{i} #1 \ar{#2}}}
\nc{\Opjar}[2][]{\TOMm{\Op{j} #1 \ar{#2}}}
\nc{\Opkar}[2][]{\TOMm{\Op{k} #1 \ar{#2}}}
\nc{\Oplar}[2][]{\TOMm{\Op{l} #1 \ar{#2}}}
\nc{\Opmar}[2][]{\TOMm{\Op{m} #1 \ar{#2}}}
\nc{\Opnar}[2][]{\TOMm{\Op{n} #1 \ar{#2}}}
\nc{\Opoar}[2][]{\TOMm{\Op{o} #1 \ar{#2}}}
\nc{\Oppar}[2][]{\TOMm{\Op{p} #1 \ar{#2}}}
\nc{\Opqar}[2][]{\TOMm{\Op{q} #1 \ar{#2}}}
\nc{\Oprar}[2][]{\TOMm{\Op{r} #1 \ar{#2}}}
\nc{\Opsar}[2][]{\TOMm{\Op{s} #1 \ar{#2}}}
\nc{\Optar}[2][]{\TOMm{\Op{t} #1 \ar{#2}}}
\nc{\Opuar}[2][]{\TOMm{\Op{u} #1 \ar{#2}}}
\nc{\Opvar}[2][]{\TOMm{\Op{v} #1 \ar{#2}}}
\nc{\Opwar}[2][]{\TOMm{\Op{w} #1 \ar{#2}}}
\nc{\Opxar}[2][]{\TOMm{\Op{x} #1 \ar{#2}}}
\nc{\Opyar}[2][]{\TOMm{\Op{y} #1 \ar{#2}}}
\nc{\Opzar}[2][]{\TOMm{\Op{z} #1 \ar{#2}}}

\nc{\OpAar}[2][]{\TOMm{\Op{A} #1 \ar{#2}}}
\nc{\OpBar}[2][]{\TOMm{\Op{B} #1 \ar{#2}}}
\nc{\OpCar}[2][]{\TOMm{\Op{C} #1 \ar{#2}}}
\nc{\OpDar}[2][]{\TOMm{\Op{D} #1 \ar{#2}}}
\nc{\OpEar}[2][]{\TOMm{\Op{E} #1 \ar{#2}}}
\nc{\OpFar}[2][]{\TOMm{\Op{F} #1 \ar{#2}}}
\nc{\OpGar}[2][]{\TOMm{\Op{G} #1 \ar{#2}}}
\nc{\OpHar}[2][]{\TOMm{\Op{H} #1 \ar{#2}}}
\nc{\OpIar}[2][]{\TOMm{\Op{I} #1 \ar{#2}}}
\nc{\OpJar}[2][]{\TOMm{\Op{J} #1 \ar{#2}}}
\nc{\OpKar}[2][]{\TOMm{\Op{K} #1 \ar{#2}}}
\nc{\OpLar}[2][]{\TOMm{\Op{L} #1 \ar{#2}}}
\nc{\OpMar}[2][]{\TOMm{\Op{M} #1 \ar{#2}}}
\nc{\OpNar}[2][]{\TOMm{\Op{N} #1 \ar{#2}}}
\nc{\OpOar}[2][]{\TOMm{\Op{O} #1 \ar{#2}}}
\nc{\OpPar}[2][]{\TOMm{\Op{P} #1 \ar{#2}}}
\nc{\OpQar}[2][]{\TOMm{\Op{Q} #1 \ar{#2}}}
\nc{\OpRar}[2][]{\TOMm{\Op{R} #1 \ar{#2}}}
\nc{\OpSar}[2][]{\TOMm{\Op{S} #1 \ar{#2}}}
\nc{\OpTar}[2][]{\TOMm{\Op{T} #1 \ar{#2}}}
\nc{\OpUar}[2][]{\TOMm{\Op{U} #1 \ar{#2}}}
\nc{\OpVar}[2][]{\TOMm{\Op{V} #1 \ar{#2}}}
\nc{\OpWar}[2][]{\TOMm{\Op{W} #1 \ar{#2}}}
\nc{\OpXar}[2][]{\TOMm{\Op{X} #1 \ar{#2}}}
\nc{\OpYar}[2][]{\TOMm{\Op{Y} #1 \ar{#2}}}
\nc{\OpZar}[2][]{\TOMm{\Op{Z} #1 \ar{#2}}}

\nc{\Opalar}[2][]{\TOMm{\Op\alpha #1 \ar{#2}}}
\nc{\Opbear}[2][]{\TOMm{\Op\beta #1 \ar{#2}}}
\nc{\Opgaar}[2][]{\TOMm{\Op\gamma #1 \ar{#2}}}
\nc{\Opdear}[2][]{\TOMm{\Op\delta #1 \ar{#2}}}
\nc{\Opepar}[2][]{\TOMm{\Op\epsilon #1 \ar{#2}}}
\nc{\Opvepar}[2][]{\TOMm{\Op\varepsilon #1 \ar{#2}}}
\nc{\Opphar}[2][]{\TOMm{\Op\phi #1 \ar{#2}}}
\nc{\Opvphar}[2][]{\TOMm{\Op\varphi #1 \ar{#2}}}
\nc{\Oppsar}[2][]{\TOMm{\Op\psi #1 \ar{#2}}}
\nc{\Opetar}[2][]{\TOMm{\Op\eta #1 \ar{#2}}}
\nc{\Opioar}[2][]{\TOMm{\Op\iota #1 \ar{#2}}}
\nc{\Opkaar}[2][]{\TOMm{\Op\kappa #1 \ar{#2}}}
\nc{\Oplaar}[2][]{\TOMm{\Op\lambda #1 \ar{#2}}}
\nc{\Opmuar}[2][]{\TOMm{\Op\mu #1 \ar{#2}}}
\nc{\Opnuar}[2][]{\TOMm{\Op\nu #1 \ar{#2}}}
\nc{\Oppiar}[2][]{\TOMm{\Op\pi #1 \ar{#2}}}
\nc{\Oproar}[2][]{\TOMm{\Op\rho #1 \ar{#2}}}
\nc{\Opsiar}[2][]{\TOMm{\Op\sigma #1 \ar{#2}}}
\nc{\Optaar}[2][]{\TOMm{\Op\tau #1 \ar{#2}}}
\nc{\Optear}[2][]{\TOMm{\Op\theta #1 \ar{#2}}}
\nc{\Opvtear}[2][]{\TOMm{\Op\vartheta #1 \ar{#2}}}
\nc{\Opomar}[2][]{\TOMm{\Op\omega #1 \ar{#2}}}
\nc{\Opkiar}[2][]{\TOMm{\Op\chi #1 \ar{#2}}}
\nc{\Opxiar}[2][]{\TOMm{\Op\xi #1 \ar{#2}}}
\nc{\Opzear}[2][]{\TOMm{\Op\zeta #1 \ar{#2}}}

\nc{\OpGaar}[2][]{\TOMm{\Op\Gamma #1 \ar{#2}}}
\nc{\OpDear}[2][]{\TOMm{\Op\Delta #1 \ar{#2}}}
\nc{\OpPhar}[2][]{\TOMm{\Op\Phi #1 \ar{#2}}}
\nc{\OpPsar}[2][]{\TOMm{\Op\Psi #1 \ar{#2}}}
\nc{\OpLaar}[2][]{\TOMm{\Op\Lambda #1 \ar{#2}}}
\nc{\OpPiar}[2][]{\TOMm{\Op\Pi #1 \ar{#2}}}
\nc{\OpSiar}[2][]{\TOMm{\Op\Sigma #1 \ar{#2}}}
\nc{\OpOmar}[2][]{\TOMm{\Op\Omega #1 \ar{#2}}}
\nc{\OpTear}[2][]{\TOMm{\Op\Theta #1 \ar{#2}}}
\nc{\OpUpar}[2][]{\TOMm{\Op\Upsilon #1 \ar{#2}}}
\nc{\OpXiar}[2][]{\TOMm{\Op\Xi #1 \ar{#2}}}

\nc{\opa}{\TOMm{\op{a}}}
\nc{\opb}{\TOMm{\op{b}}}
\nc{\opc}{\TOMm{\op{c}}}
\nc{\opd}{\TOMm{\op{d}}}
\nc{\ope}{\TOMm{\op{e}}}
\nc{\opf}{\TOMm{\op{f}}}
\nc{\opg}{\TOMm{\op{g}}}
\nc{\oph}{\TOMm{\op{h}}}
\nc{\opi}{\TOMm{\op{i}}}
\nc{\opj}{\TOMm{\op{j}}}
\nc{\opk}{\TOMm{\op{k}}}
\nc{\opl}{\TOMm{\op{l}}}
\nc{\opm}{\TOMm{\op{m}}}
\nc{\opn}{\TOMm{\op{n}}}
\nc{\opo}{\TOMm{\op{o}}}
\nc{\opp}{\TOMm{\op{p}}}
\nc{\opq}{\TOMm{\op{q}}}
\nc{\opr}{\TOMm{\op{r}}}
\nc{\ops}{\TOMm{\op{s}}}
\nc{\opt}{\TOMm{\op{t}}}
\nc{\opu}{\TOMm{\op{u}}}
\nc{\opv}{\TOMm{\op{v}}}
\nc{\opw}{\TOMm{\op{w}}}
\nc{\opx}{\TOMm{\op{x}}}
\nc{\opy}{\TOMm{\op{y}}}
\nc{\opz}{\TOMm{\op{z}}}

\nc{\opA}{\TOMm{\op{A}}}
\nc{\opB}{\TOMm{\op{B}}}
\nc{\opC}{\TOMm{\op{C}}}
\nc{\opD}{\TOMm{\op{D}}}
\nc{\opE}{\TOMm{\op{E}}}
\nc{\opF}{\TOMm{\op{F}}}
\nc{\opG}{\TOMm{\op{G}}}
\nc{\opH}{\TOMm{\op{H}}}
\nc{\opI}{\TOMm{\op{I}}}
\nc{\opJ}{\TOMm{\op{J}}}
\nc{\opK}{\TOMm{\op{K}}}
\nc{\opL}{\TOMm{\op{L}}}
\nc{\opM}{\TOMm{\op{M}}}
\nc{\opN}{\TOMm{\op{N}}}
\nc{\opO}{\TOMm{\op{O}}}
\nc{\opP}{\TOMm{\op{P}}}
\nc{\opQ}{\TOMm{\op{Q}}}
\nc{\opR}{\TOMm{\op{R}}}
\nc{\opS}{\TOMm{\op{S}}}
\nc{\opT}{\TOMm{\op{T}}}
\nc{\opU}{\TOMm{\op{U}}}
\nc{\opV}{\TOMm{\op{V}}}
\nc{\opW}{\TOMm{\op{W}}}
\nc{\opX}{\TOMm{\op{X}}}
\nc{\opY}{\TOMm{\op{Y}}}
\nc{\opZ}{\TOMm{\op{Z}}}

\nc{\opal}{\TOMm{\op\alpha}}
\nc{\opbe}{\TOMm{\op\beta}}
\nc{\opga}{\TOMm{\op\gamma}}
\nc{\opde}{\TOMm{\op\delta}}
\nc{\opep}{\TOMm{\op\epsilon}}
\nc{\opvep}{\TOMm{\op\varepsilon}}
\nc{\opph}{\TOMm{\op\phi}}
\nc{\opvph}{\TOMm{\op\varphi}}
\nc{\opps}{\TOMm{\op\psi}}
\nc{\opet}{\TOMm{\op\eta}}
\nc{\opio}{\TOMm{\op\iota}}
\nc{\opka}{\TOMm{\op\kappa}}
\nc{\opla}{\TOMm{\op\lambda}}
\nc{\opmu}{\TOMm{\op\mu}}
\nc{\opnu}{\TOMm{\op\nu}}
\nc{\oppi}{\TOMm{\op\pi}}
\nc{\opro}{\TOMm{\op\rho}}
\nc{\opsi}{\TOMm{\op\sigma}}
\nc{\opta}{\TOMm{\op\tau}}
\nc{\opte}{\TOMm{\op\theta}}
\nc{\opvte}{\TOMm{\op\vartheta}}
\nc{\opom}{\TOMm{\op\omega}}
\nc{\opki}{\TOMm{\op\chi}}
\nc{\opxi}{\TOMm{\op\xi}}
\nc{\opze}{\TOMm{\op\zeta}}

\nc{\opGa}{\TOMm{\op\Gamma}}
\nc{\opDe}{\TOMm{\op\Delta}}
\nc{\opPh}{\TOMm{\op\Phi}}
\nc{\opPs}{\TOMm{\op\Psi}}
\nc{\opLa}{\TOMm{\op\Lambda}}
\nc{\opPi}{\TOMm{\op\Pi}}
\nc{\opSi}{\TOMm{\op\Sigma}}
\nc{\opOm}{\TOMm{\op\Omega}}
\nc{\opTe}{\TOMm{\op\Theta}}
\nc{\opUp}{\TOMm{\op\Upsilon}}
\nc{\opXi}{\TOMm{\op\Xi}}

\nc{\opaar}[2][]{\TOMm{\op{a} #1 \ar{#2}}}
\nc{\opbar}[2][]{\TOMm{\op{b} #1 \ar{#2}}}
\nc{\opcar}[2][]{\TOMm{\op{c} #1 \ar{#2}}}
\nc{\opdar}[2][]{\TOMm{\op{d} #1 \ar{#2}}}
\nc{\opear}[2][]{\TOMm{\op{e} #1 \ar{#2}}}
\nc{\opfar}[2][]{\TOMm{\op{f} #1 \ar{#2}}}
\nc{\opgar}[2][]{\TOMm{\op{g} #1 \ar{#2}}}
\nc{\ophar}[2][]{\TOMm{\op{h} #1 \ar{#2}}}
\nc{\opiar}[2][]{\TOMm{\op{i} #1 \ar{#2}}}
\nc{\opjar}[2][]{\TOMm{\op{j} #1 \ar{#2}}}
\nc{\opkar}[2][]{\TOMm{\op{k} #1 \ar{#2}}}
\nc{\oplar}[2][]{\TOMm{\op{l} #1 \ar{#2}}}
\nc{\opmar}[2][]{\TOMm{\op{m} #1 \ar{#2}}}
\nc{\opnar}[2][]{\TOMm{\op{n} #1 \ar{#2}}}
\nc{\opoar}[2][]{\TOMm{\op{o} #1 \ar{#2}}}
\nc{\oppar}[2][]{\TOMm{\op{p} #1 \ar{#2}}}
\nc{\opqar}[2][]{\TOMm{\op{q} #1 \ar{#2}}}
\nc{\oprar}[2][]{\TOMm{\op{r} #1 \ar{#2}}}
\nc{\opsar}[2][]{\TOMm{\op{s} #1 \ar{#2}}}
\nc{\optar}[2][]{\TOMm{\op{t} #1 \ar{#2}}}
\nc{\opuar}[2][]{\TOMm{\op{u} #1 \ar{#2}}}
\nc{\opvar}[2][]{\TOMm{\op{v} #1 \ar{#2}}}
\nc{\opwar}[2][]{\TOMm{\op{w} #1 \ar{#2}}}
\nc{\opxar}[2][]{\TOMm{\op{x} #1 \ar{#2}}}
\nc{\opyar}[2][]{\TOMm{\op{y} #1 \ar{#2}}}
\nc{\opzar}[2][]{\TOMm{\op{z} #1 \ar{#2}}}

\nc{\opAar}[2][]{\TOMm{\op{A} #1 \ar{#2}}}
\nc{\opBar}[2][]{\TOMm{\op{B} #1 \ar{#2}}}
\nc{\opCar}[2][]{\TOMm{\op{C} #1 \ar{#2}}}
\nc{\opDar}[2][]{\TOMm{\op{D} #1 \ar{#2}}}
\nc{\opEar}[2][]{\TOMm{\op{E} #1 \ar{#2}}}
\nc{\opFar}[2][]{\TOMm{\op{F} #1 \ar{#2}}}
\nc{\opGar}[2][]{\TOMm{\op{G} #1 \ar{#2}}}
\nc{\opHar}[2][]{\TOMm{\op{H} #1 \ar{#2}}}
\nc{\opIar}[2][]{\TOMm{\op{I} #1 \ar{#2}}}
\nc{\opJar}[2][]{\TOMm{\op{J} #1 \ar{#2}}}
\nc{\opKar}[2][]{\TOMm{\op{K} #1 \ar{#2}}}
\nc{\opLar}[2][]{\TOMm{\op{L} #1 \ar{#2}}}
\nc{\opMar}[2][]{\TOMm{\op{M} #1 \ar{#2}}}
\nc{\opNar}[2][]{\TOMm{\op{N} #1 \ar{#2}}}
\nc{\opOar}[2][]{\TOMm{\op{O} #1 \ar{#2}}}
\nc{\opPar}[2][]{\TOMm{\op{P} #1 \ar{#2}}}
\nc{\opQar}[2][]{\TOMm{\op{Q} #1 \ar{#2}}}
\nc{\opRar}[2][]{\TOMm{\op{R} #1 \ar{#2}}}
\nc{\opSar}[2][]{\TOMm{\op{S} #1 \ar{#2}}}
\nc{\opTar}[2][]{\TOMm{\op{T} #1 \ar{#2}}}
\nc{\opUar}[2][]{\TOMm{\op{U} #1 \ar{#2}}}
\nc{\opVar}[2][]{\TOMm{\op{V} #1 \ar{#2}}}
\nc{\opWar}[2][]{\TOMm{\op{W} #1 \ar{#2}}}
\nc{\opXar}[2][]{\TOMm{\op{X} #1 \ar{#2}}}
\nc{\opYar}[2][]{\TOMm{\op{Y} #1 \ar{#2}}}
\nc{\opZar}[2][]{\TOMm{\op{Z} #1 \ar{#2}}}

\nc{\opalar}[2][]{\TOMm{\op\alpha #1 \ar{#2}}}
\nc{\opbear}[2][]{\TOMm{\op\beta #1 \ar{#2}}}
\nc{\opgaar}[2][]{\TOMm{\op\gamma #1 \ar{#2}}}
\nc{\opdear}[2][]{\TOMm{\op\delta #1 \ar{#2}}}
\nc{\opepar}[2][]{\TOMm{\op\epsilon #1 \ar{#2}}}
\nc{\opvepar}[2][]{\TOMm{\op\varepsilon #1 \ar{#2}}}
\nc{\opphar}[2][]{\TOMm{\op\phi #1 \ar{#2}}}
\nc{\opvphar}[2][]{\TOMm{\op\varphi #1 \ar{#2}}}
\nc{\oppsar}[2][]{\TOMm{\op\psi #1 \ar{#2}}}
\nc{\opetar}[2][]{\TOMm{\op\eta #1 \ar{#2}}}
\nc{\opioar}[2][]{\TOMm{\op\iota #1 \ar{#2}}}
\nc{\opkaar}[2][]{\TOMm{\op\kappa #1 \ar{#2}}}
\nc{\oplaar}[2][]{\TOMm{\op\lambda #1 \ar{#2}}}
\nc{\opmuar}[2][]{\TOMm{\op\mu #1 \ar{#2}}}
\nc{\opnuar}[2][]{\TOMm{\op\nu #1 \ar{#2}}}
\nc{\oppiar}[2][]{\TOMm{\op\pi #1 \ar{#2}}}
\nc{\oproar}[2][]{\TOMm{\op\rho #1 \ar{#2}}}
\nc{\opsiar}[2][]{\TOMm{\op\sigma #1 \ar{#2}}}
\nc{\optaar}[2][]{\TOMm{\op\tau #1 \ar{#2}}}
\nc{\optear}[2][]{\TOMm{\op\theta #1 \ar{#2}}}
\nc{\opvtear}[2][]{\TOMm{\op\vartheta #1 \ar{#2}}}
\nc{\opomar}[2][]{\TOMm{\op\omega #1 \ar{#2}}}
\nc{\opkiar}[2][]{\TOMm{\op\chi #1 \ar{#2}}}
\nc{\opxiar}[2][]{\TOMm{\op\xi #1 \ar{#2}}}
\nc{\opzear}[2][]{\TOMm{\op\zeta #1 \ar{#2}}}

\nc{\opGaar}[2][]{\TOMm{\op\Gamma #1 \ar{#2}}}
\nc{\opDear}[2][]{\TOMm{\op\Delta #1 \ar{#2}}}
\nc{\opPhar}[2][]{\TOMm{\op\Phi #1 \ar{#2}}}
\nc{\opPsar}[2][]{\TOMm{\op\Psi #1 \ar{#2}}}
\nc{\opLaar}[2][]{\TOMm{\op\Lambda #1 \ar{#2}}}
\nc{\opPiar}[2][]{\TOMm{\op\Pi #1 \ar{#2}}}
\nc{\opSiar}[2][]{\TOMm{\op\Sigma #1 \ar{#2}}}
\nc{\opOmar}[2][]{\TOMm{\op\Omega #1 \ar{#2}}}
\nc{\opTear}[2][]{\TOMm{\op\Theta #1 \ar{#2}}}
\nc{\opUpar}[2][]{\TOMm{\op\Upsilon #1 \ar{#2}}}
\nc{\opXiar}[2][]{\TOMm{\op\Xi #1 \ar{#2}}}

\nc{\dutdef}[2]{\TOMm{{\smaa{\dot{#1} \defeq \frac{\dif#1}{\dif#2}} }}}

\nc{\duta}{\TOMm{\dot{a}}}
\nc{\dutb}{\TOMm{\dot{b}}}
\nc{\dutc}{\TOMm{\dot{c}}}
\nc{\dutd}{\TOMm{\dot{d}}}
\nc{\dute}{\TOMm{\dot{e}}}
\nc{\dutf}{\TOMm{\dot{f}}}
\nc{\dutg}{\TOMm{\dot{g}}}
\nc{\duth}{\TOMm{\dot{h}}}
\nc{\duti}{\TOMm{\dot{i}}}
\nc{\dutj}{\TOMm{\dot{j}}}
\nc{\dutk}{\TOMm{\dot{k}}}
\nc{\dutl}{\TOMm{\dot{l}}}
\nc{\dutm}{\TOMm{\dot{m}}}
\nc{\dutn}{\TOMm{\dot{n}}}
\nc{\duto}{\TOMm{\dot{o}}}
\nc{\dutp}{\TOMm{\dot{p}}}
\nc{\dutq}{\TOMm{\dot{q}}}
\nc{\dutr}{\TOMm{\dot{r}}}
\nc{\duts}{\TOMm{\dot{s}}}
\nc{\dutt}{\TOMm{\dot{t}}}
\nc{\dutu}{\TOMm{\dot{u}}}
\nc{\dutv}{\TOMm{\dot{v}}}
\nc{\dutw}{\TOMm{\dot{w}}}
\nc{\dutx}{\TOMm{\dot{x}}}
\nc{\duty}{\TOMm{\dot{y}}}
\nc{\dutz}{\TOMm{\dot{z}}}

\nc{\dutA}{\TOMm{\dot{A}}}
\nc{\dutB}{\TOMm{\dot{B}}}
\nc{\dutC}{\TOMm{\dot{C}}}
\nc{\dutD}{\TOMm{\dot{D}}}
\nc{\dutE}{\TOMm{\dot{E}}}
\nc{\dutF}{\TOMm{\dot{F}}}
\nc{\dutG}{\TOMm{\dot{G}}}
\nc{\dutH}{\TOMm{\dot{H}}}
\nc{\dutI}{\TOMm{\dot{I}}}
\nc{\dutJ}{\TOMm{\dot{J}}}
\nc{\dutK}{\TOMm{\dot{K}}}
\nc{\dutL}{\TOMm{\dot{L}}}
\nc{\dutM}{\TOMm{\dot{M}}}
\nc{\dutN}{\TOMm{\dot{N}}}
\nc{\dutO}{\TOMm{\dot{O}}}
\nc{\dutP}{\TOMm{\dot{P}}}
\nc{\dutQ}{\TOMm{\dot{Q}}}
\nc{\dutR}{\TOMm{\dot{R}}}
\nc{\dutS}{\TOMm{\dot{S}}}
\nc{\dutT}{\TOMm{\dot{T}}}
\nc{\dutU}{\TOMm{\dot{U}}}
\nc{\dutV}{\TOMm{\dot{V}}}
\nc{\dutW}{\TOMm{\dot{W}}}
\nc{\dutX}{\TOMm{\dot{X}}}
\nc{\dutY}{\TOMm{\dot{Y}}}
\nc{\dutZ}{\TOMm{\dot{Z}}}

\nc{\dutal}{\TOMm{\dot\alpha}}
\nc{\dutbe}{\TOMm{\dot\beta}}
\nc{\dutga}{\TOMm{\dot\gamma}}
\nc{\dutde}{\TOMm{\dot\delta}}
\nc{\dutep}{\TOMm{\dot\epsilon}}
\nc{\dutvep}{\TOMm{\dot\varepsilon}}
\nc{\dutph}{\TOMm{\dot\phi}}
\nc{\dutvph}{\TOMm{\dot\varphi}}
\nc{\dutps}{\TOMm{\dot\psi}}
\nc{\dutet}{\TOMm{\dot\eta}}
\nc{\dutio}{\TOMm{\dot\iota}}
\nc{\dutka}{\TOMm{\dot\kappa}}
\nc{\dutla}{\TOMm{\dot\lambda}}
\nc{\dutmu}{\TOMm{\dot\mu}}
\nc{\dutnu}{\TOMm{\dot\nu}}
\nc{\dutpi}{\TOMm{\dot\pi}}
\nc{\dutro}{\TOMm{\dot\rho}}
\nc{\dutsi}{\TOMm{\dot\sigma}}
\nc{\dutta}{\TOMm{\dot\tau}}
\nc{\dutte}{\TOMm{\dot\theta}}
\nc{\dutvte}{\TOMm{\dot\vartheta}}
\nc{\dutom}{\TOMm{\dot\omega}}
\nc{\dutki}{\TOMm{\dot\chi}}
\nc{\dutxi}{\TOMm{\dot\xi}}
\nc{\dutze}{\TOMm{\dot\zeta}}

\nc{\dutGa}{\TOMm{\dot\Gamma}}
\nc{\dutDe}{\TOMm{\dot\Delta}}
\nc{\dutPh}{\TOMm{\dot\Phi}}
\nc{\dutPs}{\TOMm{\dot\Psi}}
\nc{\dutLa}{\TOMm{\dot\Lambda}}
\nc{\dutPi}{\TOMm{\dot\Pi}}
\nc{\dutSi}{\TOMm{\dot\Sigma}}
\nc{\dutOm}{\TOMm{\dot\Omega}}
\nc{\dutTe}{\TOMm{\dot\Theta}}
\nc{\dutUp}{\TOMm{\dot\Upsilon}}
\nc{\dutXi}{\TOMm{\dot\Xi}}

\nc{\dutaar}[2][]{\TOMm{\dot{a} #1 \ar{#2}}}
\nc{\dutbar}[2][]{\TOMm{\dot{b} #1 \ar{#2}}}
\nc{\dutcar}[2][]{\TOMm{\dot{c} #1 \ar{#2}}}
\nc{\dutdar}[2][]{\TOMm{\dot{d} #1 \ar{#2}}}
\nc{\dutear}[2][]{\TOMm{\dot{e} #1 \ar{#2}}}
\nc{\dutfar}[2][]{\TOMm{\dot{f} #1 \ar{#2}}}
\nc{\dutgar}[2][]{\TOMm{\dot{g} #1 \ar{#2}}}
\nc{\duthar}[2][]{\TOMm{\dot{h} #1 \ar{#2}}}
\nc{\dutiar}[2][]{\TOMm{\dot{i} #1 \ar{#2}}}
\nc{\dutjar}[2][]{\TOMm{\dot{j} #1 \ar{#2}}}
\nc{\dutkar}[2][]{\TOMm{\dot{k} #1 \ar{#2}}}
\nc{\dutlar}[2][]{\TOMm{\dot{l} #1 \ar{#2}}}
\nc{\dutmar}[2][]{\TOMm{\dot{m} #1 \ar{#2}}}
\nc{\dutnar}[2][]{\TOMm{\dot{n} #1 \ar{#2}}}
\nc{\dutoar}[2][]{\TOMm{\dot{o} #1 \ar{#2}}}
\nc{\dutpar}[2][]{\TOMm{\dot{p} #1 \ar{#2}}}
\nc{\dutqar}[2][]{\TOMm{\dot{q} #1 \ar{#2}}}
\nc{\dutrar}[2][]{\TOMm{\dot{r} #1 \ar{#2}}}
\nc{\dutsar}[2][]{\TOMm{\dot{s} #1 \ar{#2}}}
\nc{\duttar}[2][]{\TOMm{\dot{t} #1 \ar{#2}}}
\nc{\dutuar}[2][]{\TOMm{\dot{u} #1 \ar{#2}}}
\nc{\dutvar}[2][]{\TOMm{\dot{v} #1 \ar{#2}}}
\nc{\dutwar}[2][]{\TOMm{\dot{w} #1 \ar{#2}}}
\nc{\dutxar}[2][]{\TOMm{\dot{x} #1 \ar{#2}}}
\nc{\dutyar}[2][]{\TOMm{\dot{y} #1 \ar{#2}}}
\nc{\dutzar}[2][]{\TOMm{\dot{z} #1 \ar{#2}}}

\nc{\dutAar}[2][]{\TOMm{\dot{A} #1 \ar{#2}}}
\nc{\dutBar}[2][]{\TOMm{\dot{B} #1 \ar{#2}}}
\nc{\dutCar}[2][]{\TOMm{\dot{C} #1 \ar{#2}}}
\nc{\dutDar}[2][]{\TOMm{\dot{D} #1 \ar{#2}}}
\nc{\dutEar}[2][]{\TOMm{\dot{E} #1 \ar{#2}}}
\nc{\dutFar}[2][]{\TOMm{\dot{F} #1 \ar{#2}}}
\nc{\dutGar}[2][]{\TOMm{\dot{G} #1 \ar{#2}}}
\nc{\dutHar}[2][]{\TOMm{\dot{H} #1 \ar{#2}}}
\nc{\dutIar}[2][]{\TOMm{\dot{I} #1 \ar{#2}}}
\nc{\dutJar}[2][]{\TOMm{\dot{J} #1 \ar{#2}}}
\nc{\dutKar}[2][]{\TOMm{\dot{K} #1 \ar{#2}}}
\nc{\dutLar}[2][]{\TOMm{\dot{L} #1 \ar{#2}}}
\nc{\dutMar}[2][]{\TOMm{\dot{M} #1 \ar{#2}}}
\nc{\dutNar}[2][]{\TOMm{\dot{N} #1 \ar{#2}}}
\nc{\dutOar}[2][]{\TOMm{\dot{O} #1 \ar{#2}}}
\nc{\dutPar}[2][]{\TOMm{\dot{P} #1 \ar{#2}}}
\nc{\dutQar}[2][]{\TOMm{\dot{Q} #1 \ar{#2}}}
\nc{\dutRar}[2][]{\TOMm{\dot{R} #1 \ar{#2}}}
\nc{\dutSar}[2][]{\TOMm{\dot{S} #1 \ar{#2}}}
\nc{\dutTar}[2][]{\TOMm{\dot{T} #1 \ar{#2}}}
\nc{\dutUar}[2][]{\TOMm{\dot{U} #1 \ar{#2}}}
\nc{\dutVar}[2][]{\TOMm{\dot{V} #1 \ar{#2}}}
\nc{\dutWar}[2][]{\TOMm{\dot{W} #1 \ar{#2}}}
\nc{\dutXar}[2][]{\TOMm{\dot{X} #1 \ar{#2}}}
\nc{\dutYar}[2][]{\TOMm{\dot{Y} #1 \ar{#2}}}
\nc{\dutZar}[2][]{\TOMm{\dot{Z} #1 \ar{#2}}}

\nc{\dutalar}[2][]{\TOMm{\dot\alpha #1 \ar{#2}}}
\nc{\dutbear}[2][]{\TOMm{\dot\beta #1 \ar{#2}}}
\nc{\dutgaar}[2][]{\TOMm{\dot\gamma #1 \ar{#2}}}
\nc{\dutdear}[2][]{\TOMm{\dot\delta #1 \ar{#2}}}
\nc{\dutepar}[2][]{\TOMm{\dot\epsilon #1 \ar{#2}}}
\nc{\dutvepar}[2][]{\TOMm{\dot\varepsilon #1 \ar{#2}}}
\nc{\dutphar}[2][]{\TOMm{\dot\phi #1 \ar{#2}}}
\nc{\dutvphar}[2][]{\TOMm{\dot\varphi #1 \ar{#2}}}
\nc{\dutpsar}[2][]{\TOMm{\dot\psi #1 \ar{#2}}}
\nc{\dutetar}[2][]{\TOMm{\dot\eta #1 \ar{#2}}}
\nc{\dutioar}[2][]{\TOMm{\dot\iota #1 \ar{#2}}}
\nc{\dutkaar}[2][]{\TOMm{\dot\kappa #1 \ar{#2}}}
\nc{\dutlaar}[2][]{\TOMm{\dot\lambda #1 \ar{#2}}}
\nc{\dutmuar}[2][]{\TOMm{\dot\mu #1 \ar{#2}}}
\nc{\dutnuar}[2][]{\TOMm{\dot\nu #1 \ar{#2}}}
\nc{\dutpiar}[2][]{\TOMm{\dot\pi #1 \ar{#2}}}
\nc{\dutroar}[2][]{\TOMm{\dot\rho #1 \ar{#2}}}
\nc{\dutsiar}[2][]{\TOMm{\dot\sigma #1 \ar{#2}}}
\nc{\duttaar}[2][]{\TOMm{\dot\tau #1 \ar{#2}}}
\nc{\duttear}[2][]{\TOMm{\dot\theta #1 \ar{#2}}}
\nc{\dutvtear}[2][]{\TOMm{\dot\vartheta #1 \ar{#2}}}
\nc{\dutomar}[2][]{\TOMm{\dot\omega #1 \ar{#2}}}
\nc{\dutkiar}[2][]{\TOMm{\dot\chi #1 \ar{#2}}}
\nc{\dutxiar}[2][]{\TOMm{\dot\xi #1 \ar{#2}}}
\nc{\dutzear}[2][]{\TOMm{\dot\zeta #1 \ar{#2}}}

\nc{\dutGaar}[2][]{\TOMm{\dot\Gamma #1 \ar{#2}}}
\nc{\dutDear}[2][]{\TOMm{\dot\Delta #1 \ar{#2}}}
\nc{\dutPhar}[2][]{\TOMm{\dot\Phi #1 \ar{#2}}}
\nc{\dutPsar}[2][]{\TOMm{\dot\Psi #1 \ar{#2}}}
\nc{\dutLaar}[2][]{\TOMm{\dot\Lambda #1 \ar{#2}}}
\nc{\dutPiar}[2][]{\TOMm{\dot\Pi #1 \ar{#2}}}
\nc{\dutSiar}[2][]{\TOMm{\dot\Sigma #1 \ar{#2}}}
\nc{\dutOmar}[2][]{\TOMm{\dot\Omega #1 \ar{#2}}}
\nc{\dutTear}[2][]{\TOMm{\dot\Theta #1 \ar{#2}}}
\nc{\dutUpar}[2][]{\TOMm{\dot\Upsilon #1 \ar{#2}}}
\nc{\dutXiar}[2][]{\TOMm{\dot\Xi #1 \ar{#2}}}

\nc{\dduta}{\TOMm{\ddot{a}}}
\nc{\ddutb}{\TOMm{\ddot{b}}}
\nc{\ddutc}{\TOMm{\ddot{c}}}
\nc{\ddutd}{\TOMm{\ddot{d}}}
\nc{\ddute}{\TOMm{\ddot{e}}}
\nc{\ddutf}{\TOMm{\ddot{f}}}
\nc{\ddutg}{\TOMm{\ddot{g}}}
\nc{\dduth}{\TOMm{\ddot{h}}}
\nc{\dduti}{\TOMm{\ddot{i}}}
\nc{\ddutj}{\TOMm{\ddot{j}}}
\nc{\ddutk}{\TOMm{\ddot{k}}}
\nc{\ddutl}{\TOMm{\ddot{l}}}
\nc{\ddutm}{\TOMm{\ddot{m}}}
\nc{\ddutn}{\TOMm{\ddot{n}}}
\nc{\dduto}{\TOMm{\ddot{o}}}
\nc{\ddutp}{\TOMm{\ddot{p}}}
\nc{\ddutq}{\TOMm{\ddot{q}}}
\nc{\ddutr}{\TOMm{\ddot{r}}}
\nc{\dduts}{\TOMm{\ddot{s}}}
\nc{\ddutt}{\TOMm{\ddot{t}}}
\nc{\ddutu}{\TOMm{\ddot{u}}}
\nc{\ddutv}{\TOMm{\ddot{v}}}
\nc{\ddutw}{\TOMm{\ddot{w}}}
\nc{\ddutx}{\TOMm{\ddot{x}}}
\nc{\dduty}{\TOMm{\ddot{y}}}
\nc{\ddutz}{\TOMm{\ddot{z}}}

\nc{\ddutA}{\TOMm{\ddot{A}}}
\nc{\ddutB}{\TOMm{\ddot{B}}}
\nc{\ddutC}{\TOMm{\ddot{C}}}
\nc{\ddutD}{\TOMm{\ddot{D}}}
\nc{\ddutE}{\TOMm{\ddot{E}}}
\nc{\ddutF}{\TOMm{\ddot{F}}}
\nc{\ddutG}{\TOMm{\ddot{G}}}
\nc{\ddutH}{\TOMm{\ddot{H}}}
\nc{\ddutI}{\TOMm{\ddot{I}}}
\nc{\ddutJ}{\TOMm{\ddot{J}}}
\nc{\ddutK}{\TOMm{\ddot{K}}}
\nc{\ddutL}{\TOMm{\ddot{L}}}
\nc{\ddutM}{\TOMm{\ddot{M}}}
\nc{\ddutN}{\TOMm{\ddot{N}}}
\nc{\ddutO}{\TOMm{\ddot{O}}}
\nc{\ddutP}{\TOMm{\ddot{P}}}
\nc{\ddutQ}{\TOMm{\ddot{Q}}}
\nc{\ddutR}{\TOMm{\ddot{R}}}
\nc{\ddutS}{\TOMm{\ddot{S}}}
\nc{\ddutT}{\TOMm{\ddot{T}}}
\nc{\ddutU}{\TOMm{\ddot{U}}}
\nc{\ddutV}{\TOMm{\ddot{V}}}
\nc{\ddutW}{\TOMm{\ddot{W}}}
\nc{\ddutX}{\TOMm{\ddot{X}}}
\nc{\ddutY}{\TOMm{\ddot{Y}}}
\nc{\ddutZ}{\TOMm{\ddot{Z}}}

\nc{\ddutal}{\TOMm{\ddot\alpha}}
\nc{\ddutbe}{\TOMm{\ddot\beta}}
\nc{\ddutga}{\TOMm{\ddot\gamma}}
\nc{\ddutde}{\TOMm{\ddot\delta}}
\nc{\ddutep}{\TOMm{\ddot\epsilon}}
\nc{\ddutvep}{\TOMm{\ddot\varepsilon}}
\nc{\ddutph}{\TOMm{\ddot\phi}}
\nc{\ddutvph}{\TOMm{\ddot\varphi}}
\nc{\ddutps}{\TOMm{\ddot\psi}}
\nc{\ddutet}{\TOMm{\ddot\eta}}
\nc{\ddutio}{\TOMm{\ddot\iota}}
\nc{\ddutka}{\TOMm{\ddot\kappa}}
\nc{\ddutla}{\TOMm{\ddot\lambda}}
\nc{\ddutmu}{\TOMm{\ddot\mu}}
\nc{\ddutnu}{\TOMm{\ddot\nu}}
\nc{\ddutpi}{\TOMm{\ddot\pi}}
\nc{\ddutro}{\TOMm{\ddot\rho}}
\nc{\ddutsi}{\TOMm{\ddot\sigma}}
\nc{\ddutta}{\TOMm{\ddot\tau}}
\nc{\ddutte}{\TOMm{\ddot\theta}}
\nc{\ddutvte}{\TOMm{\ddot\vartheta}}
\nc{\ddutom}{\TOMm{\ddot\omega}}
\nc{\ddutki}{\TOMm{\ddot\chi}}
\nc{\ddutxi}{\TOMm{\ddot\xi}}
\nc{\ddutze}{\TOMm{\ddot\zeta}}

\nc{\ddutGa}{\TOMm{\ddot\Gamma}}
\nc{\ddutDe}{\TOMm{\ddot\Delta}}
\nc{\ddutPh}{\TOMm{\ddot\Phi}}
\nc{\ddutPs}{\TOMm{\ddot\Psi}}
\nc{\ddutLa}{\TOMm{\ddot\Lambda}}
\nc{\ddutPi}{\TOMm{\ddot\Pi}}
\nc{\ddutSi}{\TOMm{\ddot\Sigma}}
\nc{\ddutOm}{\TOMm{\ddot\Omega}}
\nc{\ddutTe}{\TOMm{\ddot\Theta}}
\nc{\ddutUp}{\TOMm{\ddot\Upsilon}}
\nc{\ddutXi}{\TOMm{\ddot\Xi}}

\nc{\ddutaar}[2][]{\TOMm{\ddot{a} #1 \ar{#2}}}
\nc{\ddutbar}[2][]{\TOMm{\ddot{b} #1 \ar{#2}}}
\nc{\ddutcar}[2][]{\TOMm{\ddot{c} #1 \ar{#2}}}
\nc{\ddutdar}[2][]{\TOMm{\ddot{d} #1 \ar{#2}}}
\nc{\ddutear}[2][]{\TOMm{\ddot{e} #1 \ar{#2}}}
\nc{\ddutfar}[2][]{\TOMm{\ddot{f} #1 \ar{#2}}}
\nc{\ddutgar}[2][]{\TOMm{\ddot{g} #1 \ar{#2}}}
\nc{\dduthar}[2][]{\TOMm{\ddot{h} #1 \ar{#2}}}
\nc{\ddutiar}[2][]{\TOMm{\ddot{i} #1 \ar{#2}}}
\nc{\ddutjar}[2][]{\TOMm{\ddot{j} #1 \ar{#2}}}
\nc{\ddutkar}[2][]{\TOMm{\ddot{k} #1 \ar{#2}}}
\nc{\ddutlar}[2][]{\TOMm{\ddot{l} #1 \ar{#2}}}
\nc{\ddutmar}[2][]{\TOMm{\ddot{m} #1 \ar{#2}}}
\nc{\ddutnar}[2][]{\TOMm{\ddot{n} #1 \ar{#2}}}
\nc{\ddutoar}[2][]{\TOMm{\ddot{o} #1 \ar{#2}}}
\nc{\ddutpar}[2][]{\TOMm{\ddot{p} #1 \ar{#2}}}
\nc{\ddutqar}[2][]{\TOMm{\ddot{q} #1 \ar{#2}}}
\nc{\ddutrar}[2][]{\TOMm{\ddot{r} #1 \ar{#2}}}
\nc{\ddutsar}[2][]{\TOMm{\ddot{s} #1 \ar{#2}}}
\nc{\dduttar}[2][]{\TOMm{\ddot{t} #1 \ar{#2}}}
\nc{\ddutuar}[2][]{\TOMm{\ddot{u} #1 \ar{#2}}}
\nc{\ddutvar}[2][]{\TOMm{\ddot{v} #1 \ar{#2}}}
\nc{\ddutwar}[2][]{\TOMm{\ddot{w} #1 \ar{#2}}}
\nc{\ddutxar}[2][]{\TOMm{\ddot{x} #1 \ar{#2}}}
\nc{\ddutyar}[2][]{\TOMm{\ddot{y} #1 \ar{#2}}}
\nc{\ddutzar}[2][]{\TOMm{\ddot{z} #1 \ar{#2}}}

\nc{\ddutAar}[2][]{\TOMm{\ddot{A} #1 \ar{#2}}}
\nc{\ddutBar}[2][]{\TOMm{\ddot{B} #1 \ar{#2}}}
\nc{\ddutCar}[2][]{\TOMm{\ddot{C} #1 \ar{#2}}}
\nc{\ddutDar}[2][]{\TOMm{\ddot{D} #1 \ar{#2}}}
\nc{\ddutEar}[2][]{\TOMm{\ddot{E} #1 \ar{#2}}}
\nc{\ddutFar}[2][]{\TOMm{\ddot{F} #1 \ar{#2}}}
\nc{\ddutGar}[2][]{\TOMm{\ddot{G} #1 \ar{#2}}}
\nc{\ddutHar}[2][]{\TOMm{\ddot{H} #1 \ar{#2}}}
\nc{\ddutIar}[2][]{\TOMm{\ddot{I} #1 \ar{#2}}}
\nc{\ddutJar}[2][]{\TOMm{\ddot{J} #1 \ar{#2}}}
\nc{\ddutKar}[2][]{\TOMm{\ddot{K} #1 \ar{#2}}}
\nc{\ddutLar}[2][]{\TOMm{\ddot{L} #1 \ar{#2}}}
\nc{\ddutMar}[2][]{\TOMm{\ddot{M} #1 \ar{#2}}}
\nc{\ddutNar}[2][]{\TOMm{\ddot{N} #1 \ar{#2}}}
\nc{\ddutOar}[2][]{\TOMm{\ddot{O} #1 \ar{#2}}}
\nc{\ddutPar}[2][]{\TOMm{\ddot{P} #1 \ar{#2}}}
\nc{\ddutQar}[2][]{\TOMm{\ddot{Q} #1 \ar{#2}}}
\nc{\ddutRar}[2][]{\TOMm{\ddot{R} #1 \ar{#2}}}
\nc{\ddutSar}[2][]{\TOMm{\ddot{S} #1 \ar{#2}}}
\nc{\ddutTar}[2][]{\TOMm{\ddot{T} #1 \ar{#2}}}
\nc{\ddutUar}[2][]{\TOMm{\ddot{U} #1 \ar{#2}}}
\nc{\ddutVar}[2][]{\TOMm{\ddot{V} #1 \ar{#2}}}
\nc{\ddutWar}[2][]{\TOMm{\ddot{W} #1 \ar{#2}}}
\nc{\ddutXar}[2][]{\TOMm{\ddot{X} #1 \ar{#2}}}
\nc{\ddutYar}[2][]{\TOMm{\ddot{Y} #1 \ar{#2}}}
\nc{\ddutZar}[2][]{\TOMm{\ddot{Z} #1 \ar{#2}}}

\nc{\ddutalar}[2][]{\TOMm{\ddot\alpha #1 \ar{#2}}}
\nc{\ddutbear}[2][]{\TOMm{\ddot\beta #1 \ar{#2}}}
\nc{\ddutgaar}[2][]{\TOMm{\ddot\gamma #1 \ar{#2}}}
\nc{\ddutdear}[2][]{\TOMm{\ddot\delta #1 \ar{#2}}}
\nc{\ddutepar}[2][]{\TOMm{\ddot\epsilon #1 \ar{#2}}}
\nc{\ddutvepar}[2][]{\TOMm{\ddot\varepsilon #1 \ar{#2}}}
\nc{\ddutphar}[2][]{\TOMm{\ddot\phi #1 \ar{#2}}}
\nc{\ddutvphar}[2][]{\TOMm{\ddot\varphi #1 \ar{#2}}}
\nc{\ddutpsar}[2][]{\TOMm{\ddot\psi #1 \ar{#2}}}
\nc{\ddutetar}[2][]{\TOMm{\ddot\eta #1 \ar{#2}}}
\nc{\ddutioar}[2][]{\TOMm{\ddot\iota #1 \ar{#2}}}
\nc{\ddutkaar}[2][]{\TOMm{\ddot\kappa #1 \ar{#2}}}
\nc{\ddutlaar}[2][]{\TOMm{\ddot\lambda #1 \ar{#2}}}
\nc{\ddutmuar}[2][]{\TOMm{\ddot\mu #1 \ar{#2}}}
\nc{\ddutnuar}[2][]{\TOMm{\ddot\nu #1 \ar{#2}}}
\nc{\ddutpiar}[2][]{\TOMm{\ddot\pi #1 \ar{#2}}}
\nc{\ddutroar}[2][]{\TOMm{\ddot\rho #1 \ar{#2}}}
\nc{\ddutsiar}[2][]{\TOMm{\ddot\sigma #1 \ar{#2}}}
\nc{\dduttaar}[2][]{\TOMm{\ddot\tau #1 \ar{#2}}}
\nc{\dduttear}[2][]{\TOMm{\ddot\theta #1 \ar{#2}}}
\nc{\ddutvtear}[2][]{\TOMm{\ddot\vartheta #1 \ar{#2}}}
\nc{\ddutomar}[2][]{\TOMm{\ddot\omega #1 \ar{#2}}}
\nc{\ddutkiar}[2][]{\TOMm{\ddot\chi #1 \ar{#2}}}
\nc{\ddutxiar}[2][]{\TOMm{\ddot\xi #1 \ar{#2}}}
\nc{\ddutzear}[2][]{\TOMm{\ddot\zeta #1 \ar{#2}}}

\nc{\ddutGaar}[2][]{\TOMm{\ddot\Gamma #1 \ar{#2}}}
\nc{\ddutDear}[2][]{\TOMm{\ddot\Delta #1 \ar{#2}}}
\nc{\ddutPhar}[2][]{\TOMm{\ddot\Phi #1 \ar{#2}}}
\nc{\ddutPsar}[2][]{\TOMm{\ddot\Psi #1 \ar{#2}}}
\nc{\ddutLaar}[2][]{\TOMm{\ddot\Lambda #1 \ar{#2}}}
\nc{\ddutPiar}[2][]{\TOMm{\ddot\Pi #1 \ar{#2}}}
\nc{\ddutSiar}[2][]{\TOMm{\ddot\Sigma #1 \ar{#2}}}
\nc{\ddutOmar}[2][]{\TOMm{\ddot\Omega #1 \ar{#2}}}
\nc{\ddutTear}[2][]{\TOMm{\ddot\Theta #1 \ar{#2}}}
\nc{\ddutUpar}[2][]{\TOMm{\ddot\Upsilon #1 \ar{#2}}}
\nc{\ddutXiar}[2][]{\TOMm{\ddot\Xi #1 \ar{#2}}}

\nc{\Tia}{\TOMm{\Ti{a}}}
\nc{\Tib}{\TOMm{\Ti{b}}}
\nc{\Tic}{\TOMm{\Ti{c}}}
\nc{\Tid}{\TOMm{\Ti{d}}}
\nc{\Tie}{\TOMm{\Ti{e}}}
\nc{\Tif}{\TOMm{\Ti{f}}}
\nc{\Tig}{\TOMm{\Ti{g}}}
\nc{\Tih}{\TOMm{\Ti{h}}}
\nc{\Tii}{\TOMm{\Ti{i}}}
\nc{\Tij}{\TOMm{\Ti{j}}}
\nc{\Tik}{\TOMm{\Ti{k}}}
\nc{\Til}{\TOMm{\Ti{l}}}
\nc{\Tim}{\TOMm{\Ti{m}}}
\nc{\Tin}{\TOMm{\Ti{n}}}
\nc{\Tio}{\TOMm{\Ti{o}}}
\nc{\Tip}{\TOMm{\Ti{p}}}
\nc{\Tiq}{\TOMm{\Ti{q}}}
\nc{\Tir}{\TOMm{\Ti{r}}}
\nc{\Tis}{\TOMm{\Ti{s}}}
\nc{\Tit}{\TOMm{\Ti{t}}}
\nc{\Tiu}{\TOMm{\Ti{u}}}
\nc{\Tiv}{\TOMm{\Ti{v}}}
\nc{\Tiw}{\TOMm{\Ti{w}}}
\nc{\Tix}{\TOMm{\Ti{x}}}
\nc{\Tiy}{\TOMm{\Ti{y}}}
\nc{\Tiz}{\TOMm{\Ti{z}}}

\nc{\TiA}{\TOMm{\Ti{A}}}
\nc{\TiB}{\TOMm{\Ti{B}}}
\nc{\TiC}{\TOMm{\Ti{C}}}
\nc{\TiD}{\TOMm{\Ti{D}}}
\nc{\TiE}{\TOMm{\Ti{E}}}
\nc{\TiF}{\TOMm{\Ti{F}}}
\nc{\TiG}{\TOMm{\Ti{G}}}
\nc{\TiH}{\TOMm{\Ti{H}}}
\nc{\TiI}{\TOMm{\Ti{I}}}
\nc{\TiJ}{\TOMm{\Ti{J}}}
\nc{\TiK}{\TOMm{\Ti{K}}}
\nc{\TiL}{\TOMm{\Ti{L}}}
\nc{\TiM}{\TOMm{\Ti{M}}}
\nc{\TiN}{\TOMm{\Ti{N}}}
\nc{\TiO}{\TOMm{\Ti{O}}}
\nc{\TiP}{\TOMm{\Ti{P}}}
\nc{\TiQ}{\TOMm{\Ti{Q}}}
\nc{\TiR}{\TOMm{\Ti{R}}}
\nc{\TiS}{\TOMm{\Ti{S}}}
\nc{\TiT}{\TOMm{\Ti{T}}}
\nc{\TiU}{\TOMm{\Ti{U}}}
\nc{\TiV}{\TOMm{\Ti{V}}}
\nc{\TiW}{\TOMm{\Ti{W}}}
\nc{\TiX}{\TOMm{\Ti{X}}}
\nc{\TiY}{\TOMm{\Ti{Y}}}
\nc{\TiZ}{\TOMm{\Ti{Z}}}

\nc{\Tial}{\TOMm{\Ti\alpha}}
\nc{\Tibe}{\TOMm{\Ti\beta}}
\nc{\Tiga}{\TOMm{\Ti\gamma}}
\nc{\Tide}{\TOMm{\Ti\delta}}
\nc{\Tiep}{\TOMm{\Ti\epsilon}}
\nc{\Tivep}{\TOMm{\Ti\varepsilon}}
\nc{\Tiph}{\TOMm{\Ti\phi}}
\nc{\Tivph}{\TOMm{\Ti\varphi}}
\nc{\Tips}{\TOMm{\Ti\psi}}
\nc{\Tiet}{\TOMm{\Ti\eta}}
\nc{\Tiio}{\TOMm{\Ti\iota}}
\nc{\Tika}{\TOMm{\Ti\kappa}}
\nc{\Tila}{\TOMm{\Ti\lambda}}
\nc{\Timu}{\TOMm{\Ti\mu}}
\nc{\Tinu}{\TOMm{\Ti\nu}}
\nc{\Tipi}{\TOMm{\Ti\pi}}
\nc{\Tiro}{\TOMm{\Ti\rho}}
\nc{\Tisi}{\TOMm{\Ti\sigma}}
\nc{\Tita}{\TOMm{\Ti\tau}}
\nc{\Tite}{\TOMm{\Ti\theta}}
\nc{\Tivte}{\TOMm{\Ti\vartheta}}
\nc{\Tiom}{\TOMm{\Ti\omega}}
\nc{\Tiki}{\TOMm{\Ti\chi}}
\nc{\Tixi}{\TOMm{\Ti\xi}}
\nc{\Tize}{\TOMm{\Ti\zeta}}

\nc{\TiGa}{\TOMm{\Ti\Gamma}}
\nc{\TiDe}{\TOMm{\Ti\Delta}}
\nc{\TiPh}{\TOMm{\Ti\Phi}}
\nc{\TiPs}{\TOMm{\Ti\Psi}}
\nc{\TiLa}{\TOMm{\Ti\Lambda}}
\nc{\TiPi}{\TOMm{\Ti\Pi}}
\nc{\TiSi}{\TOMm{\Ti\Sigma}}
\nc{\TiOm}{\TOMm{\Ti\Omega}}
\nc{\TiTe}{\TOMm{\Ti\Theta}}
\nc{\TiUp}{\TOMm{\Ti\Upsilon}}
\nc{\TiXi}{\TOMm{\Ti\Xi}}

\nc{\Tiaar}[2][]{\TOMm{\Ti{a} #1 \ar{#2}}}
\nc{\Tibar}[2][]{\TOMm{\Ti{b} #1 \ar{#2}}}
\nc{\Ticar}[2][]{\TOMm{\Ti{c} #1 \ar{#2}}}
\nc{\Tidar}[2][]{\TOMm{\Ti{d} #1 \ar{#2}}}
\nc{\Tiear}[2][]{\TOMm{\Ti{e} #1 \ar{#2}}}
\nc{\Tifar}[2][]{\TOMm{\Ti{f} #1 \ar{#2}}}
\nc{\Tigar}[2][]{\TOMm{\Ti{g} #1 \ar{#2}}}
\nc{\Tihar}[2][]{\TOMm{\Ti{h} #1 \ar{#2}}}
\nc{\Tiiar}[2][]{\TOMm{\Ti{i} #1 \ar{#2}}}
\nc{\Tijar}[2][]{\TOMm{\Ti{j} #1 \ar{#2}}}
\nc{\Tikar}[2][]{\TOMm{\Ti{k} #1 \ar{#2}}}
\nc{\Tilar}[2][]{\TOMm{\Ti{l} #1 \ar{#2}}}
\nc{\Timar}[2][]{\TOMm{\Ti{m} #1 \ar{#2}}}
\nc{\Tinar}[2][]{\TOMm{\Ti{n} #1 \ar{#2}}}
\nc{\Tioar}[2][]{\TOMm{\Ti{o} #1 \ar{#2}}}
\nc{\Tipar}[2][]{\TOMm{\Ti{p} #1 \ar{#2}}}
\nc{\Tiqar}[2][]{\TOMm{\Ti{q} #1 \ar{#2}}}
\nc{\Tirar}[2][]{\TOMm{\Ti{r} #1 \ar{#2}}}
\nc{\Tisar}[2][]{\TOMm{\Ti{s} #1 \ar{#2}}}
\nc{\Titar}[2][]{\TOMm{\Ti{t} #1 \ar{#2}}}
\nc{\Tiuar}[2][]{\TOMm{\Ti{u} #1 \ar{#2}}}
\nc{\Tivar}[2][]{\TOMm{\Ti{v} #1 \ar{#2}}}
\nc{\Tiwar}[2][]{\TOMm{\Ti{w} #1 \ar{#2}}}
\nc{\Tixar}[2][]{\TOMm{\Ti{x} #1 \ar{#2}}}
\nc{\Tiyar}[2][]{\TOMm{\Ti{y} #1 \ar{#2}}}
\nc{\Tizar}[2][]{\TOMm{\Ti{z} #1 \ar{#2}}}

\nc{\TiAar}[2][]{\TOMm{\Ti{A} #1 \ar{#2}}}
\nc{\TiBar}[2][]{\TOMm{\Ti{B} #1 \ar{#2}}}
\nc{\TiCar}[2][]{\TOMm{\Ti{C} #1 \ar{#2}}}
\nc{\TiDar}[2][]{\TOMm{\Ti{D} #1 \ar{#2}}}
\nc{\TiEar}[2][]{\TOMm{\Ti{E} #1 \ar{#2}}}
\nc{\TiFar}[2][]{\TOMm{\Ti{F} #1 \ar{#2}}}
\nc{\TiGar}[2][]{\TOMm{\Ti{G} #1 \ar{#2}}}
\nc{\TiHar}[2][]{\TOMm{\Ti{H} #1 \ar{#2}}}
\nc{\TiIar}[2][]{\TOMm{\Ti{I} #1 \ar{#2}}}
\nc{\TiJar}[2][]{\TOMm{\Ti{J} #1 \ar{#2}}}
\nc{\TiKar}[2][]{\TOMm{\Ti{K} #1 \ar{#2}}}
\nc{\TiLar}[2][]{\TOMm{\Ti{L} #1 \ar{#2}}}
\nc{\TiMar}[2][]{\TOMm{\Ti{M} #1 \ar{#2}}}
\nc{\TiNar}[2][]{\TOMm{\Ti{N} #1 \ar{#2}}}
\nc{\TiOar}[2][]{\TOMm{\Ti{O} #1 \ar{#2}}}
\nc{\TiPar}[2][]{\TOMm{\Ti{P} #1 \ar{#2}}}
\nc{\TiQar}[2][]{\TOMm{\Ti{Q} #1 \ar{#2}}}
\nc{\TiRar}[2][]{\TOMm{\Ti{R} #1 \ar{#2}}}
\nc{\TiSar}[2][]{\TOMm{\Ti{S} #1 \ar{#2}}}
\nc{\TiTar}[2][]{\TOMm{\Ti{T} #1 \ar{#2}}}
\nc{\TiUar}[2][]{\TOMm{\Ti{U} #1 \ar{#2}}}
\nc{\TiVar}[2][]{\TOMm{\Ti{V} #1 \ar{#2}}}
\nc{\TiWar}[2][]{\TOMm{\Ti{W} #1 \ar{#2}}}
\nc{\TiXar}[2][]{\TOMm{\Ti{X} #1 \ar{#2}}}
\nc{\TiYar}[2][]{\TOMm{\Ti{Y} #1 \ar{#2}}}
\nc{\TiZar}[2][]{\TOMm{\Ti{Z} #1 \ar{#2}}}

\nc{\Tialar}[2][]{\TOMm{\Ti\alpha #1 \ar{#2}}}
\nc{\Tibear}[2][]{\TOMm{\Ti\beta #1 \ar{#2}}}
\nc{\Tigaar}[2][]{\TOMm{\Ti\gamma #1 \ar{#2}}}
\nc{\Tidear}[2][]{\TOMm{\Ti\delta #1 \ar{#2}}}
\nc{\Tiepar}[2][]{\TOMm{\Ti\epsilon #1 \ar{#2}}}
\nc{\Tivepar}[2][]{\TOMm{\Ti\varepsilon #1 \ar{#2}}}
\nc{\Tiphar}[2][]{\TOMm{\Ti\phi #1 \ar{#2}}}
\nc{\Tivphar}[2][]{\TOMm{\Ti\varphi #1 \ar{#2}}}
\nc{\Tipsar}[2][]{\TOMm{\Ti\psi #1 \ar{#2}}}
\nc{\Tietar}[2][]{\TOMm{\Ti\eta #1 \ar{#2}}}
\nc{\Tiioar}[2][]{\TOMm{\Ti\iota #1 \ar{#2}}}
\nc{\Tikaar}[2][]{\TOMm{\Ti\kappa #1 \ar{#2}}}
\nc{\Tilaar}[2][]{\TOMm{\Ti\lambda #1 \ar{#2}}}
\nc{\Timuar}[2][]{\TOMm{\Ti\mu #1 \ar{#2}}}
\nc{\Tinuar}[2][]{\TOMm{\Ti\nu #1 \ar{#2}}}
\nc{\Tipiar}[2][]{\TOMm{\Ti\pi #1 \ar{#2}}}
\nc{\Tiroar}[2][]{\TOMm{\Ti\rho #1 \ar{#2}}}
\nc{\Tisiar}[2][]{\TOMm{\Ti\sigma #1 \ar{#2}}}
\nc{\Titaar}[2][]{\TOMm{\Ti\tau #1 \ar{#2}}}
\nc{\Titear}[2][]{\TOMm{\Ti\theta #1 \ar{#2}}}
\nc{\Tivtear}[2][]{\TOMm{\Ti\vartheta #1 \ar{#2}}}
\nc{\Tiomar}[2][]{\TOMm{\Ti\omega #1 \ar{#2}}}
\nc{\Tikiar}[2][]{\TOMm{\Ti\chi #1 \ar{#2}}}
\nc{\Tixiar}[2][]{\TOMm{\Ti\xi #1 \ar{#2}}}
\nc{\Tizear}[2][]{\TOMm{\Ti\zeta #1 \ar{#2}}}

\nc{\TiGaar}[2][]{\TOMm{\Ti\Gamma #1 \ar{#2}}}
\nc{\TiDear}[2][]{\TOMm{\Ti\Delta #1 \ar{#2}}}
\nc{\TiPhar}[2][]{\TOMm{\Ti\Phi #1 \ar{#2}}}
\nc{\TiPsar}[2][]{\TOMm{\Ti\Psi #1 \ar{#2}}}
\nc{\TiLaar}[2][]{\TOMm{\Ti\Lambda #1 \ar{#2}}}
\nc{\TiPiar}[2][]{\TOMm{\Ti\Pi #1 \ar{#2}}}
\nc{\TiSiar}[2][]{\TOMm{\Ti\Sigma #1 \ar{#2}}}
\nc{\TiOmar}[2][]{\TOMm{\Ti\Omega #1 \ar{#2}}}
\nc{\TiTear}[2][]{\TOMm{\Ti\Theta #1 \ar{#2}}}
\nc{\TiUpar}[2][]{\TOMm{\Ti\Upsilon #1 \ar{#2}}}
\nc{\TiXiar}[2][]{\TOMm{\Ti\Xi #1 \ar{#2}}}

\nc{\tia}{\TOMm{\ti{a}}}
\nc{\tib}{\TOMm{\ti{b}}}
\nc{\tic}{\TOMm{\ti{c}}}
\nc{\tid}{\TOMm{\ti{d}}}
\nc{\tie}{\TOMm{\ti{e}}}
\nc{\tif}{\TOMm{\ti{f}}}
\nc{\tig}{\TOMm{\ti{g}}}
\nc{\tih}{\TOMm{\ti{h}}}
\nc{\tii}{\TOMm{\ti{i}}}
\nc{\tij}{\TOMm{\ti{j}}}
\nc{\tik}{\TOMm{\ti{k}}}
\nc{\til}{\TOMm{\ti{l}}}
\nc{\tim}{\TOMm{\ti{m}}}
\nc{\tin}{\TOMm{\ti{n}}}
\nc{\tio}{\TOMm{\ti{o}}}
\nc{\tip}{\TOMm{\ti{p}}}
\nc{\tiq}{\TOMm{\ti{q}}}
\nc{\tir}{\TOMm{\ti{r}}}
\nc{\tis}{\TOMm{\ti{s}}}
\nc{\tit}{\TOMm{\ti{t}}}
\nc{\tiu}{\TOMm{\ti{u}}}
\nc{\tiv}{\TOMm{\ti{v}}}
\nc{\tiw}{\TOMm{\ti{w}}}
\nc{\tix}{\TOMm{\ti{x}}}
\nc{\tiy}{\TOMm{\ti{y}}}
\nc{\tiz}{\TOMm{\ti{z}}}

\nc{\tiA}{\TOMm{\ti{A}}}
\nc{\tiB}{\TOMm{\ti{B}}}
\nc{\tiC}{\TOMm{\ti{C}}}
\nc{\tiD}{\TOMm{\ti{D}}}
\nc{\tiE}{\TOMm{\ti{E}}}
\nc{\tiF}{\TOMm{\ti{F}}}
\nc{\tiG}{\TOMm{\ti{G}}}
\nc{\tiH}{\TOMm{\ti{H}}}
\nc{\tiI}{\TOMm{\ti{I}}}
\nc{\tiJ}{\TOMm{\ti{J}}}
\nc{\tiK}{\TOMm{\ti{K}}}
\nc{\tiL}{\TOMm{\ti{L}}}
\nc{\tiM}{\TOMm{\ti{M}}}
\nc{\tiN}{\TOMm{\ti{N}}}
\nc{\tiO}{\TOMm{\ti{O}}}
\nc{\tiP}{\TOMm{\ti{P}}}
\nc{\tiQ}{\TOMm{\ti{Q}}}
\nc{\tiR}{\TOMm{\ti{R}}}
\nc{\tiS}{\TOMm{\ti{S}}}
\nc{\tiT}{\TOMm{\ti{T}}}
\nc{\tiU}{\TOMm{\ti{U}}}
\nc{\tiV}{\TOMm{\ti{V}}}
\nc{\tiW}{\TOMm{\ti{W}}}
\nc{\tiX}{\TOMm{\ti{X}}}
\nc{\tiY}{\TOMm{\ti{Y}}}
\nc{\tiZ}{\TOMm{\ti{Z}}}

\nc{\tial}{\TOMm{\ti\alpha}}
\nc{\tibe}{\TOMm{\ti\beta}}
\nc{\tiga}{\TOMm{\ti\gamma}}
\nc{\tide}{\TOMm{\ti\delta}}
\nc{\tiep}{\TOMm{\ti\epsilon}}
\nc{\tivep}{\TOMm{\ti\varepsilon}}
\nc{\tiph}{\TOMm{\ti\phi}}
\nc{\tivph}{\TOMm{\ti\varphi}}
\nc{\tips}{\TOMm{\ti\psi}}
\nc{\tiet}{\TOMm{\ti\eta}}
\nc{\tiio}{\TOMm{\ti\iota}}
\nc{\tika}{\TOMm{\ti\kappa}}
\nc{\tila}{\TOMm{\ti\lambda}}
\nc{\timu}{\TOMm{\ti\mu}}
\nc{\tinu}{\TOMm{\ti\nu}}
\nc{\tipi}{\TOMm{\ti\pi}}
\nc{\tiro}{\TOMm{\ti\rho}}
\nc{\tisi}{\TOMm{\ti\sigma}}
\nc{\tita}{\TOMm{\ti\tau}}
\nc{\tite}{\TOMm{\ti\theta}}
\nc{\tivte}{\TOMm{\ti\vartheta}}
\nc{\tiom}{\TOMm{\ti\omega}}
\nc{\tiki}{\TOMm{\ti\chi}}
\nc{\tixi}{\TOMm{\ti\xi}}
\nc{\tize}{\TOMm{\ti\zeta}}

\nc{\tiGa}{\TOMm{\ti\Gamma}}
\nc{\tiDe}{\TOMm{\ti\Delta}}
\nc{\tiPh}{\TOMm{\ti\Phi}}
\nc{\tiPs}{\TOMm{\ti\Psi}}
\nc{\tiLa}{\TOMm{\ti\Lambda}}
\nc{\tiPi}{\TOMm{\ti\Pi}}
\nc{\tiSi}{\TOMm{\ti\Sigma}}
\nc{\tiOm}{\TOMm{\ti\Omega}}
\nc{\tiTe}{\TOMm{\ti\Theta}}
\nc{\tiUp}{\TOMm{\ti\psilon}}
\nc{\tiXi}{\TOMm{\ti\Xi}}

\nc{\tiaar}[2][]{\TOMm{\ti{a} #1 \ar{#2}}}
\nc{\tibar}[2][]{\TOMm{\ti{b} #1 \ar{#2}}}
\nc{\ticar}[2][]{\TOMm{\ti{c} #1 \ar{#2}}}
\nc{\tidar}[2][]{\TOMm{\ti{d} #1 \ar{#2}}}
\nc{\tiear}[2][]{\TOMm{\ti{e} #1 \ar{#2}}}
\nc{\tifar}[2][]{\TOMm{\ti{f} #1 \ar{#2}}}
\nc{\tigar}[2][]{\TOMm{\ti{g} #1 \ar{#2}}}
\nc{\tihar}[2][]{\TOMm{\ti{h} #1 \ar{#2}}}
\nc{\tiiar}[2][]{\TOMm{\ti{i} #1 \ar{#2}}}
\nc{\tijar}[2][]{\TOMm{\ti{j} #1 \ar{#2}}}
\nc{\tikar}[2][]{\TOMm{\ti{k} #1 \ar{#2}}}
\nc{\tilar}[2][]{\TOMm{\ti{l} #1 \ar{#2}}}
\nc{\timar}[2][]{\TOMm{\ti{m} #1 \ar{#2}}}
\nc{\tinar}[2][]{\TOMm{\ti{n} #1 \ar{#2}}}
\nc{\tioar}[2][]{\TOMm{\ti{o} #1 \ar{#2}}}
\nc{\tipar}[2][]{\TOMm{\ti{p} #1 \ar{#2}}}
\nc{\tiqar}[2][]{\TOMm{\ti{q} #1 \ar{#2}}}
\nc{\tirar}[2][]{\TOMm{\ti{r} #1 \ar{#2}}}
\nc{\tisar}[2][]{\TOMm{\ti{s} #1 \ar{#2}}}
\nc{\titar}[2][]{\TOMm{\ti{t} #1 \ar{#2}}}
\nc{\tiuar}[2][]{\TOMm{\ti{u} #1 \ar{#2}}}
\nc{\tivar}[2][]{\TOMm{\ti{v} #1 \ar{#2}}}
\nc{\tiwar}[2][]{\TOMm{\ti{w} #1 \ar{#2}}}
\nc{\tixar}[2][]{\TOMm{\ti{x} #1 \ar{#2}}}
\nc{\tiyar}[2][]{\TOMm{\ti{y} #1 \ar{#2}}}
\nc{\tizar}[2][]{\TOMm{\ti{z} #1 \ar{#2}}}

\nc{\tiAar}[2][]{\TOMm{\ti{A} #1 \ar{#2}}}
\nc{\tiBar}[2][]{\TOMm{\ti{B} #1 \ar{#2}}}
\nc{\tiCar}[2][]{\TOMm{\ti{C} #1 \ar{#2}}}
\nc{\tiDar}[2][]{\TOMm{\ti{D} #1 \ar{#2}}}
\nc{\tiEar}[2][]{\TOMm{\ti{E} #1 \ar{#2}}}
\nc{\tiFar}[2][]{\TOMm{\ti{F} #1 \ar{#2}}}
\nc{\tiGar}[2][]{\TOMm{\ti{G} #1 \ar{#2}}}
\nc{\tiHar}[2][]{\TOMm{\ti{H} #1 \ar{#2}}}
\nc{\tiIar}[2][]{\TOMm{\ti{I} #1 \ar{#2}}}
\nc{\tiJar}[2][]{\TOMm{\ti{J} #1 \ar{#2}}}
\nc{\tiKar}[2][]{\TOMm{\ti{K} #1 \ar{#2}}}
\nc{\tiLar}[2][]{\TOMm{\ti{L} #1 \ar{#2}}}
\nc{\tiMar}[2][]{\TOMm{\ti{M} #1 \ar{#2}}}
\nc{\tiNar}[2][]{\TOMm{\ti{N} #1 \ar{#2}}}
\nc{\tiOar}[2][]{\TOMm{\ti{O} #1 \ar{#2}}}
\nc{\tiPar}[2][]{\TOMm{\ti{P} #1 \ar{#2}}}
\nc{\tiQar}[2][]{\TOMm{\ti{Q} #1 \ar{#2}}}
\nc{\tiRar}[2][]{\TOMm{\ti{R} #1 \ar{#2}}}
\nc{\tiSar}[2][]{\TOMm{\ti{S} #1 \ar{#2}}}
\nc{\tiTar}[2][]{\TOMm{\ti{T} #1 \ar{#2}}}
\nc{\tiUar}[2][]{\TOMm{\ti{U} #1 \ar{#2}}}
\nc{\tiVar}[2][]{\TOMm{\ti{V} #1 \ar{#2}}}
\nc{\tiWar}[2][]{\TOMm{\ti{W} #1 \ar{#2}}}
\nc{\tiXar}[2][]{\TOMm{\ti{X} #1 \ar{#2}}}
\nc{\tiYar}[2][]{\TOMm{\ti{Y} #1 \ar{#2}}}
\nc{\tiZar}[2][]{\TOMm{\ti{Z} #1 \ar{#2}}}

\nc{\tialar}[2][]{\TOMm{\ti\alpha #1 \ar{#2}}}
\nc{\tibear}[2][]{\TOMm{\ti\beta #1 \ar{#2}}}
\nc{\tigaar}[2][]{\TOMm{\ti\gamma #1 \ar{#2}}}
\nc{\tidear}[2][]{\TOMm{\ti\delta #1 \ar{#2}}}
\nc{\tiepar}[2][]{\TOMm{\ti\epsilon #1 \ar{#2}}}
\nc{\tivepar}[2][]{\TOMm{\ti\varepsilon #1 \ar{#2}}}
\nc{\tiphar}[2][]{\TOMm{\ti\phi #1 \ar{#2}}}
\nc{\tivphar}[2][]{\TOMm{\ti\varphi #1 \ar{#2}}}
\nc{\tipsar}[2][]{\TOMm{\ti\psi #1 \ar{#2}}}
\nc{\tietar}[2][]{\TOMm{\ti\eta #1 \ar{#2}}}
\nc{\tiioar}[2][]{\TOMm{\ti\iota #1 \ar{#2}}}
\nc{\tikaar}[2][]{\TOMm{\ti\kappa #1 \ar{#2}}}
\nc{\tilaar}[2][]{\TOMm{\ti\lambda #1 \ar{#2}}}
\nc{\timuar}[2][]{\TOMm{\ti\mu #1 \ar{#2}}}
\nc{\tinuar}[2][]{\TOMm{\ti\nu #1 \ar{#2}}}
\nc{\tipiar}[2][]{\TOMm{\ti\pi #1 \ar{#2}}}
\nc{\tiroar}[2][]{\TOMm{\ti\rho #1 \ar{#2}}}
\nc{\tisiar}[2][]{\TOMm{\ti\sigma #1 \ar{#2}}}
\nc{\titaar}[2][]{\TOMm{\ti\tau #1 \ar{#2}}}
\nc{\titear}[2][]{\TOMm{\ti\theta #1 \ar{#2}}}
\nc{\tivtear}[2][]{\TOMm{\ti\vartheta #1 \ar{#2}}}
\nc{\tiomar}[2][]{\TOMm{\ti\omega #1 \ar{#2}}}
\nc{\tikiar}[2][]{\TOMm{\ti\chi #1 \ar{#2}}}
\nc{\tixiar}[2][]{\TOMm{\ti\xi #1 \ar{#2}}}
\nc{\tizear}[2][]{\TOMm{\ti\zeta #1 \ar{#2}}}

\nc{\tiGaar}[2][]{\TOMm{\ti\Gamma #1 \ar{#2}}}
\nc{\tiDear}[2][]{\TOMm{\ti\Delta #1 \ar{#2}}}
\nc{\tiPhar}[2][]{\TOMm{\ti\Phi #1 \ar{#2}}}
\nc{\tiPsar}[2][]{\TOMm{\ti\Psi #1 \ar{#2}}}
\nc{\tiLaar}[2][]{\TOMm{\ti\Lambda #1 \ar{#2}}}
\nc{\tiPiar}[2][]{\TOMm{\ti\Pi #1 \ar{#2}}}
\nc{\tiSiar}[2][]{\TOMm{\ti\Sigma #1 \ar{#2}}}
\nc{\tiOmar}[2][]{\TOMm{\ti\Omega #1 \ar{#2}}}
\nc{\tiTear}[2][]{\TOMm{\ti\Theta #1 \ar{#2}}}
\nc{\tiUpar}[2][]{\TOMm{\ti\Upsilon #1 \ar{#2}}}
\nc{\tiXiar}[2][]{\TOMm{\ti\Xi #1 \ar{#2}}}

\nc{\vola}[1][]{\TOMm{\emptynumberone{#1} {\vol{a}} {\vol[#1\!\!]{a}} } }
\nc{\volb}[1][]{\TOMm{\emptynumberone{#1} {\vol{b}} {\vol[#1\!\!]{b}} } }
\nc{\volc}[1][]{\TOMm{\emptynumberone{#1} {\vol{c}} {\vol[#1\!\!]{c}} } }
\nc{\vold}[1][]{\TOMm{\emptynumberone{#1} {\vol{d}} {\vol[#1\!\!]{d}} } }
\nc{\vole}[1][]{\TOMm{\emptynumberone{#1} {\vol{e}} {\vol[#1\!\!]{e}} } }
\nc{\volf}[1][]{\TOMm{\emptynumberone{#1} {\vol{\!f}} {\vol[#1\!\!\!]{f}} } }
\nc{\volg}[1][]{\TOMm{\emptynumberone{#1} {\vol{g}} {\vol[#1\!\!]{g}} } }
\nc{\volh}[1][]{\TOMm{\emptynumberone{#1} {\vol{h}} {\vol[#1\!\!]{h}} } }
\nc{\voli}[1][]{\TOMm{\emptynumberone{#1} {\vol{i}} {\vol[#1\!]{i}} } }
\nc{\volj}[1][]{\TOMm{\emptynumberone{#1} {\vol{j}} {\vol[#1\!\!]{j}} } }
\nc{\volk}[1][]{\TOMm{\emptynumberone{#1} {\vol{k}} {\vol[#1\!\!]{k}} } }
\nc{\voll}[1][]{\TOMm{\emptynumberone{#1} {\vol{l}} {\vol[#1\!]{l}} } }
\nc{\volm}[1][]{\TOMm{\emptynumberone{#1} {\vol{m}} {\vol[#1\!\!]{m}} } }
\nc{\voln}[1][]{\TOMm{\emptynumberone{#1} {\vol{n}} {\vol[#1\!\!]{n}} } }
\nc{\volo}[1][]{\TOMm{\emptynumberone{#1} {\vol{o}} {\vol[#1\!\!\!]{o}} } }
\nc{\volp}[1][]{\TOMm{\emptynumberone{#1} {\vol{p}} {\vol[#1\!\!]{p}} } }
\nc{\volq}[1][]{\TOMm{\emptynumberone{#1} {\vol{q}} {\vol[#1\!\!]{q}} } }
\nc{\volr}[1][]{\TOMm{\emptynumberone{#1} {\vol{r}} {\vol[#1\!\!]{r}} } }
\nc{\vols}[1][]{\TOMm{\emptynumberone{#1} {\vol{s}} {\vol[#1\!\!]{s}} } }
\nc{\volt}[1][]{\TOMm{\emptynumberone{#1} {\vol{t}} {\vol[#1\!\!]{t}} } }
\nc{\volu}[1][]{\TOMm{\emptynumberone{#1} {\vol{u}} {\vol[#1\!\!]{u}} } }
\nc{\volv}[1][]{\TOMm{\emptynumberone{#1} {\vol{v}} {\vol[#1\!\!]{v}} } }
\nc{\volw}[1][]{\TOMm{\emptynumberone{#1} {\vol{w}} {\vol[#1\!\!]{w}} } }
\nc{\volx}[1][]{\TOMm{\emptynumberone{#1} {\vol{x}} {\vol[#1\!\!]{x}} } }
\nc{\voly}[1][]{\TOMm{\emptynumberone{#1} {\vol{y}} {\vol[#1\!\!]{y}} } }
\nc{\volz}[1][]{\TOMm{\emptynumberone{#1} {\vol{z}} {\vol[#1\!\!]{z}} } }

\nc{\volA}[1][]{\TOMm{\emptynumberone{#1} {\vol{A}} {\vol[#1\!\!\!]{A}} } }
\nc{\volB}[1][]{\TOMm{\emptynumberone{#1} {\vol{B}} {\vol[#1\!\!\!]{B}} } }
\nc{\volC}[1][]{\TOMm{\emptynumberone{#1} {\vol{C}} {\vol[#1\!\!]{C}} } }
\nc{\volD}[1][]{\TOMm{\emptynumberone{#1} {\vol{D}} {\vol[#1\!\!]{D}} } }
\nc{\volE}[1][]{\TOMm{\emptynumberone{#1} {\vol{E}} {\vol[#1\!\!]{E}} } }
\nc{\volF}[1][]{\TOMm{\emptynumberone{#1} {\vol{F}} {\vol[#1\!\!]{F}} } }
\nc{\volG}[1][]{\TOMm{\emptynumberone{#1} {\vol{G}} {\vol[#1\!\!]{G}} } }
\nc{\volH}[1][]{\TOMm{\emptynumberone{#1} {\vol{H}} {\vol[#1\!\!]{H}} } }
\nc{\volI}[1][]{\TOMm{\emptynumberone{#1} {\vol{I}} {\vol[#1\!\!]{I}} } }
\nc{\volJ}[1][]{\TOMm{\emptynumberone{#1} {\vol{J}} {\vol[#1\!\!\!]{J}} } }
\nc{\volK}[1][]{\TOMm{\emptynumberone{#1} {\vol{K}} {\vol[#1\!\!]{K}} } }
\nc{\volL}[1][]{\TOMm{\emptynumberone{#1} {\vol{L}} {\vol[#1\!\!]{L}} } }
\nc{\volM}[1][]{\TOMm{\emptynumberone{#1} {\vol{M}} {\vol[#1\!\!]{M}} } }
\nc{\volN}[1][]{\TOMm{\emptynumberone{#1} {\vol{N}} {\vol[#1\!\!]{N}} } }
\nc{\volO}[1][]{\TOMm{\emptynumberone{#1} {\vol{O}} {\vol[#1\!\!\!]{O}} } }
\nc{\volP}[1][]{\TOMm{\emptynumberone{#1} {\vol{P}} {\vol[#1\!\!]{P}} } }
\nc{\volQ}[1][]{\TOMm{\emptynumberone{#1} {\vol{Q}} {\vol[#1\!\!]{Q}} } }
\nc{\volR}[1][]{\TOMm{\emptynumberone{#1} {\vol{R}} {\vol[#1\!\!]{R}} } }
\nc{\volS}[1][]{\TOMm{\emptynumberone{#1} {\vol{S}} {\vol[#1\!\!]{S}} } }
\nc{\volT}[1][]{\TOMm{\emptynumberone{#1} {\vol{T}} {\vol[#1\!]{T}} } }
\nc{\volU}[1][]{\TOMm{\emptynumberone{#1} {\vol{U}} {\vol[#1\!\!]{U}} } }
\nc{\volV}[1][]{\TOMm{\emptynumberone{#1} {\vol{V}} {\vol[#1\!]{V}} } }
\nc{\volW}[1][]{\TOMm{\emptynumberone{#1} {\vol{W}} {\vol[#1\!\!]{W}} } }
\nc{\volX}[1][]{\TOMm{\emptynumberone{#1} {\vol{X}} {\vol[#1\!\!]{X}} } }
\nc{\volY}[1][]{\TOMm{\emptynumberone{#1} {\vol{Y}} {\vol[#1\!]{Y}} } }
\nc{\volZ}[1][]{\TOMm{\emptynumberone{#1} {\vol{Z}} {\vol[#1\!\!]{Z}} } }

\nc{\volal}[1][]{\TOMm{\emptynumberone{#1} {\vol{\alpha}} {\vol[#1\!\!]{\alpha}} } }
\nc{\volbe}[1][]{\TOMm{\emptynumberone{#1} {\vol{\beta}} {\vol[#1\!\!]{\beta}} } }
\nc{\volga}[1][]{\TOMm{\emptynumberone{#1} {\vol{\gamma}} {\vol[#1\!\!]{\gamma}} } }
\nc{\volde}[1][]{\TOMm{\emptynumberone{#1} {\vol{\delta}} {\vol[#1\!\!]{\delta}} } }
\nc{\volep}[1][]{\TOMm{\emptynumberone{#1} {\vol{\ep}} {\vol[#1\!\!\!]{\ep}} } }
\nc{\volvep}[1][]{\TOMm{\emptynumberone{#1} {\vol{\vep}} {\vol[#1\!\!\!]{\vep}} } }
\nc{\volze}[1][]{\TOMm{\emptynumberone{#1} {\vol{\zeta}} {\vol[#1\!\!]{\zeta}} } }
\nc{\volet}[1][]{\TOMm{\emptynumberone{#1} {\vol{\eta}} {\vol[#1\!\!]{\eta}} } }
\nc{\volte}[1][]{\TOMm{\emptynumberone{#1} {\vol{\theta}} {\vol[#1\!\!]{\theta}} } }
\nc{\volvte}[1][]{\TOMm{\emptynumberone{#1} {\vol{\vth}} {\vol[#1\!\!]{\vth}} } }
\nc{\volka}[1][]{\TOMm{\emptynumberone{#1} {\vol{\kappa}} {\vol[#1\!\!\!]{\kappa}} } }
\nc{\volla}[1][]{\TOMm{\emptynumberone{#1} {\vol{\lambda}} {\vol[#1\!\!]{\lambda}} } }
\nc{\volmu}[1][]{\TOMm{\emptynumberone{#1} {\vol{\mu}} {\vol[#1\!\!]{\mu}} } }
\nc{\volnu}[1][]{\TOMm{\emptynumberone{#1} {\vol{\nu}} {\vol[#1\!\!]{\nu}} } }
\nc{\volki}[1][]{\TOMm{\emptynumberone{#1} {\vol{\chi}} {\vol[#1\!\!]{\chi}} } }
\nc{\volxi}[1][]{\TOMm{\emptynumberone{#1} {\vol{\xi}} {\vol[#1\!\!]{\xi}} } }
\nc{\volps}[1][]{\TOMm{\emptynumberone{#1} {\vol{\psi}} {\vol[#1\!\!]{\psi}} } }
\nc{\volph}[1][]{\TOMm{\emptynumberone{#1} {\vol{\phi}} {\vol[#1\!\!]{\phi}} } }
\nc{\volvph}[1][]{\TOMm{\emptynumberone{#1} {\vol{\vph}} {\vol[#1\!\!]{\vph}} } }
\nc{\volpi}[1][]{\TOMm{\emptynumberone{#1} {\vol{\pi}} {\vol[#1\!\!]{\pi}} } }
\nc{\volro}[1][]{\TOMm{\emptynumberone{#1} {\vol[\!]{\rho}} {\vol[#1\!\!\!]{\rho}} } }
\nc{\volsi}[1][]{\TOMm{\emptynumberone{#1} {\vol{\sigma}} {\vol[#1\!\!]{\sigma}} } }
\nc{\volta}[1][]{\TOMm{\emptynumberone{#1} {\vol{\tau}} {\vol[#1\!\!]{\tau}} } }
\nc{\volom}[1][]{\TOMm{\emptynumberone{#1} {\vol{\omega}} {\vol[#1\!\!]{\omega}} } }

\nc{\volGa}[1][]{\TOMm{\emptynumberone{#1} {\vol{\Gamma}} {\vol[#1\!]{\Gamma}} } }
\nc{\volDe}[1][]{\TOMm{\emptynumberone{#1} {\vol{\Delta}} {\vol[#1\!\!]{\Delta}} } }
\nc{\volTe}[1][]{\TOMm{\emptynumberone{#1} {\vol{\Theta}} {\vol[#1\!]{\Theta}} } }
\nc{\volLa}[1][]{\TOMm{\emptynumberone{#1} {\vol{\Lambda}} {\vol[#1\!\!]{\Lambda}} } }
\nc{\volXi}[1][]{\TOMm{\emptynumberone{#1} {\vol{\Xi}} {\vol[#1\!]{\Xi}} } }
\nc{\volPs}[1][]{\TOMm{\emptynumberone{#1} {\vol{\Psi}} {\vol[#1\!]{\Psi}} } }
\nc{\volPh}[1][]{\TOMm{\emptynumberone{#1} {\vol{\Phi}} {\vol[#1\!]{\Phi}} } }
\nc{\volSi}[1][]{\TOMm{\emptynumberone{#1} {\vol[\!]{\Sigma}} {\vol[#1\!]{\Sigma}} } }
\nc{\volUp}[1][]{\TOMm{\emptynumberone{#1} {\vol{\Upsilon}} {\vol[#1\!]{\Upsilon}} } }
\nc{\volOm}[1][]{\TOMm{\emptynumberone{#1} {\vol{\Omega}} {\vol[#1\!]{\Omega}} } }

\nc{\voltwopi}[2][]{\fracw{\dif[#1]#2}{(2\piu)^{#1}} }
\nc{\voltwopiE}[2][]{\fracw{\dif[#1]#2}{(2\piu)^{#1}2E\unemptynumberone{#2}{\lvc{#2}}} }

\nc{\voltwopik}[1][]{\voltwopi[#1]{k}}
\nc{\voltwopiEk}[1][]{\voltwopiE[#1]{\!k}}

\nc{\voltwopip}[1][]{\voltwopi[#1]{p}}
\nc{\voltwopiEp}[1][]{\voltwopiE[#1]{p}}

\nc{\voltwopiq}[1][]{\voltwopi[#1]{q}}
\nc{\voltwopiEq}[1][]{\voltwopiE[#1]{q}}

\nc{\todert}[1][]{\toder[#1]{t}}											
\nc{\toderatt}[2][]{\toderat[#1]{t}{#2}}
\nc{\toderattxt}[2][]{\toderattx[#1]{t}{#2}}
\nc{\toderst}[1][]{\toders[#1]{t}}
\nc{\toderatst}[2][]{\toderats[#1]{t}{#2}}

\nc{\padert}[1][]{\pader[#1]{t}}
\nc{\paderatt}[2][]{\paderat[#1]{t}{#2}}
\nc{\paderattxt}[2][]{\paderattx[#1]{t}{#2}}
\nc{\paderst}[1][]{\paders[#1]{t}}
\nc{\paderatst}[2][]{\paderats[#1]{t}{#2}}

\nc{\delslash}[1][]{\TOMm{\slashed{\del} \ls{#1} }}
\nc{\kslash}{\TOMm{\slashed{k}}}
\nc{\pslash}{\TOMm{\slashed{p}}}
\nc{\qslash}{\TOMm{\slashed{q}}}
\nc{\xslash}{\TOMm{\slashed{x}}}
\nc{\yslash}{\TOMm{\slashed{y}}}

\nc{\Aslash}{\TOMm{\slashed{A}}}

\nc{\delrest}[1][]{\TOMm{\widetilde{\delvc\,}\ls{#1} }}
\nc{\krest}{\TOMm{\widetilde{\vc{k}\,} } }
\nc{\prest}{\TOMm{\widetilde{\vc{p}\,} } }
\nc{\qrest}{\TOMm{\widetilde{\vc{q}\,} } }
\nc{\xrest}{\TOMm{\widetilde{\vc{x}\,} } }
\nc{\yrest}{\TOMm{\widetilde{\vc{y}\,} } }
\nc{\Arest}{\TOMm{\widetilde{\vc{A}\,} } }

%% file: __new_commands_special.tex
\nc{\cococab}{\coco{c_a}c_b \mn c_a\coco{c_b}}

\nc{\Sibar}{{\ovl\Si}}
\nc{\hsita}{^{\Si_\ta}}

\nc{\kgsol}[1]{\TOMt{S\smaa{OL}$(#1)$}}
\nc{\kgsolC}[1]{\TOMt{S\smaa{OL}$_{\complex}(#1)$}}

\nc{\solinpro}[3]{\TOMm{\left\{#1,\,#2\right\}^{#3}}}
\nc{\solinproi}[3]{\TOMm{\bigl\{#1,\,#2\bigr\}^{#3}}}
\nc{\solinproii}[3]{\TOMm{\biigl\{#1,\,#2\biigr\}^{#3}}}
\nc{\solinproiii}[3]{\TOMm{\biiigl\{#1,\,#2\biiigr\}^{#3}}}
\nc{\solinproiiii}[3]{\TOMm{\biiiigl\{#1,\,#2\biiiigr\}^{#3}}}

\nc{\confman}[3]{\txM^{(#1,#2)}_{#3}}

\nc{\AdS}[1]{\TOMt{AdS$_{1,#1}$}}
\nc{\Rads}{\TOMm{R\lads}}
\nc{\lads}{_{\smaaaa{\text{AdS}}}}
\nc{\hads}{^{\smaaaa{\text{AdS}}}}
\nc{\lmink}{_{\smaaaa{\text{Mink}}}}
\nc{\hmink}{^{\smaaaa{\text{Mink}}}}

\nc{\omag}[3]{\TOMm{\om^{\smaaaa{#1}}_{#2,#3}}}
\nc{\omsubmag}[3]{\TOMm{\si^{#1}_{#2,#3}}}

\nc{\Norm}[3]{\TOMm{N^{#1}_{#2,#3}}}
\nc{\opNorm}[3]{\TOMm{\opN^{#1}_{#2,#3}}}
\nc{\RNorm}[1][]{\biiglrrn{\Rads^{d\mn1}\Norm\pm nl}{#1}}
\nc{\opRNorm}[1][]{\biiglrrn{\Rads^{d\mn1}\opNorm\pm\opn\opl}{#1}}

\nc{\artrOm}{\ar{t,r,\Om}}
\nc{\artroOm}{\ar{t,\ro,\Om}}
\nc{\artOm}{\ar{t,\Om}}
\nc{\arroOm}{\ar{\ro,\Om}}

\nc{\artax}{\ar{\ta,\vc x}}

\nc{\voltom}{\volt\!\volOm}
\nc{\intvoltom}[1][]{\int\!\!\volt\!\volOm[#1]}
\nc{\intvoltaom}[1][]{\int\!\!\volta\!\volOm[#1]}
\nc{\intvolrom}[1][]{\int\!\!\volr\!\volOm[#1]}
\nc{\intvolroom}[1][]{\int\!\!\volro\!\volOm[#1]}
\nc{\intesumlm}{\intlim{}{}\volE\!\sumliml{l,m_l}\,}
\nc{\intesumlvcm}{\intlim{}{}\volE\!\sumliml{l,\vc m}\,}
\nc{\intemsumlm}{\intlim{E>m}{}\,\volE\!\!\sumliml{l,m_l}\,}
\nc{\intpsumlm}{\intlim{0}{\infty}\,\volp\!\!\sumliml{l,m_l}\,}
\nc{\intomsumlvcm}{\intlim{}{}\volom\!\sumliml{l,\vc m}\,}
\nc{\inttiomsumlvcm}{\intlim{}{}\vol{\tiom}\!\sumliml{l,\vc m}\,}

\nc{\titiph}{\tilde{\tiph}}

\nc{\opft}{\TOMm{\opf\ar t}}
\nc{\opfta}{\TOMm{\opf\ar \ta}}
\nc{\opdtx}{\TOMm{\opd\ar{t,\vc x}}}
\nc{\opdtax}{\TOMm{\opd\ar{\ta,\vc x}}}

\nc{\gaC}{\TOMm{\ga^{\txC}}}
\nc{\gaS}{\TOMm{\ga^{\txS}}}

\nc{\alp}{\TOMm{\al_+}}
\nc{\alm}{\TOMm{\al_-}}
\nc{\alpm}{\TOMm{\al_\pm}}
\nc{\almp}{\TOMm{\al_\mp}}
\nc{\bep}{\TOMm{\be_+}}
\nc{\bem}{\TOMm{\be_-}}
\nc{\bepm}{\TOMm{\be_\pm}}
\nc{\bemp}{\TOMm{\be_\mp}}
\nc{\gaCp}{\TOMm{\gaC_+}}
\nc{\gaCm}{\TOMm{\gaC_-}}
\nc{\gaCpm}{\TOMm{\gaC_\pm}}
\nc{\gaCmp}{\TOMm{\gaC_\mp}}
\nc{\Timp}{\TOMm{\Tim_+}}
\nc{\Timm}{\TOMm{\Tim_-}}
\nc{\Timpm}{\TOMm{\Tim_\pm}}
\nc{\Timmp}{\TOMm{\Tim_\mp}}
\nc{\timp}{\TOMm{\tim_+}}
\nc{\timm}{\TOMm{\tim_-}}
\nc{\timpm}{\TOMm{\tim_\pm}}
\nc{\timmp}{\TOMm{\tim_\mp}}

\nc{\SCfun}[4]{\TOMm{\,\!\hs{#1\!}#2\htxs{#3}_{#4}}}
\nc{\oS}[1]{\SCfun 1 S {} {#1}}
\nc{\tSodd}[1]{\SCfun 2 S {odd} {#1}}
\nc{\tSeve}[1]{\SCfun 2 S {eve} {#1}}
\nc{\oC}[1]{\SCfun 1 C {} {#1}}
\nc{\tCnon}[1]{\SCfun 2 C {non} {#1}}
\nc{\tCint}[1]{\SCfun 2 C {int} {#1}}

\nc{\cSCfun}[5]{\TOMm{\,\!\ls{#2}\hs{\,#1\!}#3\htxs{#4}_{#5}}}
\nc{\tcSeve}[2]{\cSCfun 2 {#1} S {eve} {#2}}
\nc{\tcCint}[2]{\cSCfun 2 {#1} C {int} {#2}}

\nc{\Jac}[4][]{\TOMm{J#1\,\!^{#2}_{#3,#4}}}
\nc{\Jacar}[5][]{\TOMm{\Jac[#1]{#2}{#3}{#4}\ar{#5}}}

\nc{\msqBF}{m^2\ltx{BF}} 

\nc{\gthree}{\tig}
\nc{\ruutabsg}{\ruutabs g}
\nc{\ruutabsgt}[1]{\ruutabs{\gthree\ls{#1}}}
\nc{\ruutabsgtt}{\ruutabs{g^{tt}\!}}
\nc{\ruutabsggtt}[1]{\ruutabs{\gthree \ls{#1} g^{tt}\!}}
\nc{\ruutabsgtata}{\ruutabs{g^{\ta\ta}\!}}
\nc{\ruutabsggtata}[1]{\ruutabs{\gthree \ls{#1} g^{\ta\ta}\!}}
\nc{\ruutabsgroro}{\ruutabs{g^{\ro\ro}\!}}
\nc{\ruutabsggroro}[1]{\ruutabs{\gthree \ls{#1} g^{\ro\ro}\!}}

\nc{\tiphp}[1]{\TOMm{\tiph^+_{#1}}}
\nc{\tiphm}[1]{\TOMm{\tiph^-_{#1}}}
\nc{\tiphpm}[1]{\TOMm{\tiph^\pm_{#1}}}
\nc{\tiphR}[1]{\TOMm{\tiph^{\,\mathbb{R}}_{#1}}}
\nc{\tiphI}[1]{\TOMm{\tiph^{\,\mathbb{I}}_{#1}}}

\nc{\tietp}[1]{\TOMm{\tiet^+_{#1}}}
\nc{\tietm}[1]{\TOMm{\tiet^-_{#1}}}

\nc{\tizep}[1]{\TOMm{\tize^+_{#1}}}
\nc{\tizem}[1]{\TOMm{\tize^-_{#1}}}

\nc{\oneph}{\TOMm{\,\!\hs{1\!}\ph}}
\nc{\twoph}{\TOMm{\,\!\hs{2\!}\ph}}
\nc{\onetiph}{\TOMm{\,\!\hs{1\!}\tiph}}
\nc{\twotiph}{\TOMm{\,\!\hs{2\!}\tiph}}

\nc{\onetiphp}[1]{\TOMm{\onetiph^+_{#1}}}
\nc{\onetiphm}[1]{\TOMm{\onetiph^-_{#1}}}
\nc{\twotiphp}[1]{\TOMm{\twotiph^+_{#1}}}
\nc{\twotiphm}[1]{\TOMm{\twotiph^-_{#1}}}

\nc{\onetitiphp}[1]{\TOMm{\,\!\hs{1\!}\tilde{\tiph}^+_{#1}}}
\nc{\onetitiphm}[1]{\TOMm{\,\!\hs{1\!}\tilde{\tiph}^-_{#1}}}
\nc{\twotitiphp}[1]{\TOMm{\,\!\hs{2\!}\tilde{\tiph}^+_{#1}}}
\nc{\twotitiphm}[1]{\TOMm{\,\!\hs{2\!}\tilde{\tiph}^-_{#1}}}

\nc{\cinpro}[3][]{\TOMm{\{ #2 {,} \; #3 \}\ls{#1}} }
\nc{\cinproo}[3][]{{\biigl\{} #2 {,} \;\, #3 {\biigr\}}\ls{#1}}
\nc{\cinprooo}[3][]{{\biigl\{} #2 {,} \;\; #3 {\biigr\}}\ls{\!#1}}
\nc{\cinproooo}[3][]{{\biiigl\{} #2 {,} \;\; #3 {\biiigr\}}\ls{\!#1}}
\nc{\cinprooooo}[3][]{{\biiiigl\{} #2 {,} \;\; #3 {\biiiigr\}}\ls{\!\!#1}}

%% file: zzzzz_title_iop.tex
\title{S-Matrix for AdS from General Boundary QFT}
\vspace{-1mm}
\author{Daniele Colosi, Max Dohse and Robert Oeckl}
\address{Centro de Ciencias Matem\'aticas (CCM), UNAM Campus Morelia,
		58190 Morelia, Mexico}
\vspace{-1mm}
\ead{colosi/max/robert@matmor.unam.mx}
%
%
\vspace{-2mm}
\begin{abstract}
The General Boundary Formulation (GBF) is a new framework for studying quantum theories.
After concise overviews of the GBF and Schrödinger-Feynman quantization
we apply the GBF to resolve a well known problem
on Anti-deSitter spacetime where
due to the lack of temporally asymptotic free states
the usual S-matrix cannot be defined.
We construct a different type of S-matrix
plus propagators for free and interacting real Klein-Gordon theory. 
\end{abstract}
%